\documentclass{aa}\newcommand{\ratio}{0.57}\newcommand{\ratioo}{0.66}\newcommand{\ratiooo}{0.6}\newcommand{\ratioA}{0.8}\newcommand{\ratioB}{0.99}

\usepackage{graphicx}
\usepackage{txfonts}
\usepackage{enumerate}

\usepackage{multicol}
\usepackage{xcolor}
\let\oldtextbf\textbf
\renewcommand{\textbf}[1]{\textcolor{black}{{#1}}}
\let\ftextbf\textbf
\renewcommand{\ftextbf}[1]{\textcolor{black}{{#1}}}

\usepackage[linkcolor=forestgreen,citecolor=royalblue,filecolor=darkmagenta,urlcolor=royalblue]{hyperref}
\usepackage{amstext}

\begin{document}

   \title{Bonn Optimized Stellar Tracks (BoOST)}

   \subtitle{Simulated populations of massive and very assive stars for astrophysical applications}

   \author{Dorottya Szécsi
          \inst{1,2}\thanks{The BoOST data (stellar model grids, interpolated tracks and synthetic populations) are all available online: \url{http://boost.asu.cas.cz}. Email: dorottya.szecsi@gmail.com}
          \and
          Poojan Agrawal\inst{3,4,5}
          \and
          Richard Wünsch\inst{6}
          \and
          Norbert Langer\inst{7}
          \fnmsep
          }

   \institute{Institute of Astronomy, Faculty of Physics, Astronomy and Informatics, Nicolaus Copernicus University, Grudziądzka 5, 87-100 Toruń, Poland -- \email{dorottya.szecsi@gmail.com}
         \and
             I. Physikalisches Institut, Universität zu Köln, Zülpicher-Strasse 77, D-50937 Cologne, Germany
        \and
            Center for Astrophysics and Supercomputing, Swinburne University of Technology, Hawthorn, Victoria 3122, Australia
        \and
            OzGrav: The ARC Center of Excellence for Gravitational Wave Discovery
        \and 
                McWilliams Center for Cosmology, Department of Physics, Carnegie Mellon University, Pittsburgh, PA 15213, USA
        \and
                        Astronomical Institute of the Czech Academy of Sciences, Bo\v{c}n\'{i} II 1401, 141 00 Prague 4, Czech Republic            
        \and
            Argelander-Institut f\"ur Astronomie der Universit\"at Bonn, Auf dem H\"ugel 71, 53121 Bonn, Germany
    }
   \date{Accepted: 07/12/2021.}

 
  \abstract
{Massive and very massive stars can play important roles in stellar populations by ejecting strong stellar winds and exploding in energetic phenomena. It is therefore imperative that their behavior be properly accounted for in synthetic model populations. 

We present nine grids of stellar evolutionary model sequences, together with finely resolved interpolated sequences and synthetic populations, of stars with 9--500\,M$_{\odot}$ and with metallicities ranging from Galactic metallicity down to 1/250\,Z$_{\odot}$. The stellar models were computed with the Bonn evolutionary code with consistent physical ingredients, and {covering core hydrogen- and core helium-burning phases}. 
The interpolation and population synthesis were performed with our newly developed routine \textsc{synStars}. Eight~of the grids represent slowly rotating massive stars with a normal or classical \textbf{evolutionary path}, while one grid represents fast-rotating, chemically homogeneously evolving models. The grids contain data on stellar wind properties such as estimated wind velocity and kinetic energy of the wind, as well as common stellar parameters such as mass, radius, surface temperature, luminosity, mass-loss rate, and surface abundances of 34 isotopes. We also provide estimates of the helium and carbon-oxygen core mass for calculating the mass of stellar remnants. 

The Bonn Optimized Stellar Tracks (BoOST) project is published as simple tables that include stellar models, interpolated tracks, and synthetic populations. Covering the broadest mass and metallicity range of any published massive star evolutionary model sets to date, BoOST is ideal for further scientific applications such as star formation studies in both low- and high-redshift galaxies. 
}


   \maketitle
%

\section{Introduction}\label{sec:intro}

Stellar evolutionary model sequences provide the basis for several astrophysical investigations. These investigations include \textbf{the simulation of galaxies close \citep[e.g.,][]{Gatto:2017} and far} \citep[e.g.,][]{Rosdahl:2018}, obtaining mass and age of observed stars \citep[e.g.,][]{Schneider:2014b,Grin:2017,RamirezAgudelo:2017}, studying the formation of ancient globular clusters \citep[e.g.,][]{deMink:2009,Szecsi:2019}, and predicting the outcome of binary populations in terms of gravitational wave event rates \citep[e.g.,][]{Kruckow:2018}. 
Because massive ($>$9~M$_{\odot}$) and very massive ($>$100~M$_{\odot}$) stars can play important roles in stellar populations by ejecting strong stellar winds and exploding in energetic phenomena, it is very important that their behaviour is properly accounted for in synthetic populations \citep{Agrawal:2020}. For example if the early Universe is to be studied, very massive stars are key: the first few generations of galaxies at cosmic dawn might have formed them in larger numbers and with higher initial mass than what is typical today.

\textbf{The Bonn code\footnote{also known as the Binary Evolutionary Code}} has been used in the past decades to compute stellar evolutionary model sequences with the most recent input physics. {Amongst other things, it is especially suited to simulating massive stars due to the large nuclear reaction network and the high spacial resolution it applies. Stellar grids of massive stars} with various chemical composition and various rotational properties have been created and analyzed \citep{Yoon:2006,Yoon:2012,Brott:2011,Koehler:2015,Szecsi:2015}, occasionally including detailed binary models \citep{deMink:2009,deMink:2009b,Yoon:2015}.

\begin{table*}\raggedleft
        \caption{Nine BoOST grids published here, and their compositions in various units.
        }\centering
        \begin{tabular}{ c|c|c|c|c|c|c|c|c||c } 
                \hline
                & \oldtextbf{MW}$^\mathrm{a}$ & \oldtextbf{LMC}$^\mathrm{b}$ & \oldtextbf{SMC}$^\mathrm{a}$ & \oldtextbf{dwarfA} & \oldtextbf{dwarfB} & \oldtextbf{IZw18}$^\mathrm{c}$ & \oldtextbf{dwarfD} & \oldtextbf{dwarfE} & \oldtextbf{IZw18-CHE}$^\mathrm{c}$ \\
                &  &  &  & SMC/2 & SMC/5 & SMC/10 & SMC/20 & SMC/50 & SMC/10 \\\hline
                Z$_{\rm MW}$ & 1 & $\sim$1/2.5 & $\sim$1/5 & $\sim$1/10 & $\sim$1/25 & $\sim$1/50 & $\sim$1/100 & $\sim$1/250 & $\sim$1/50 \\
                Z$_{\rm SMC}$ & $\sim$5 & $\sim$2 & 1 & 0.5 & 0.2 & 0.1 & 0.05 & 0.02 & 0.1 \\
                $[$Fe/H$]$ & $\lesssim$\,0.0 & $\sim$\,$-$0.4 & $-$0.7 & $-$1.0 & $-$1.4 & $-$1.7 & $-$2.0 & $-$2.4 & $-$1.7 \\
                Z & 0.0088 & 0.0047 & 0.0021 & 0.00105 & 0.00042 & 0.00021 & 0.00011 & 0.00005 & 0.00021 \\\hline
        \end{tabular}\\ \vspace{5pt}
        \raggedright{\footnotesize 
                The values Z$_{\rm MW}$~$=$~0.0088, Z$_{\rm LMC}$~$=$~0.0047, and Z$_{\rm SMC}$~$=$~0.0021 were defined by \citet[][see their Tables~1 and~2]{Brott:2011}. For dwarf galaxy grids dwarfA...E (including IZw18 and IZw18-CHE), the initial metallicity is scaled down from that of the SMC by a factor given in the table header.\\               
                $^\mathrm{a}$~Models between 9-60~M$_{\odot}$ were published  in \citet{Brott:2011}. 
                \\ $^\mathrm{b}$~Models between 9-60~M$_{\odot}$ were published in \citet{Brott:2011}, and the main-sequence phase of the models between 60-500~M$_{\odot}$ in \citet{Koehler:2015}. 
                \\ $^\mathrm{c}$~The main-sequence phases of the models between 9-300~M$_{\odot}$ were published in \citet{Szecsi:2015}.
}
        \label{tab:Z}
\end{table*}

Nonetheless, the models published so far can be further optimized. Consistency in the initial parameter space, as well as a reliable treatment of the late phases of evolution, is necessary in order to make these models applicable within a wide range of astrophysical studies. The Bonn code has been used to create and publish four grids of rotating single-star models. Their initial compositions correspond to the Milky Way \citep[MW;][]{Brott:2011}, the Large and Small Magellanic Clouds \citep[LMC and SMC;][]{Brott:2011,Koehler:2015}, and the dwarf galaxy I~Zwicky~18 \citep[IZw18;][]{Szecsi:2015}. In the current project, we extend the mass range of all published Bonn grids up to very massive stars. The metallicity range of the models is also extended down to the lowest metallicities observed in ancient globular clusters.

We publish stellar evolutionary model predictions as well as synthetic populations based on published and unpublished results of the Bonn code \ftextbf{in a physically consistent way}. That is, the Bonn Optimized Stellar Tracks (BoOST) project provides the following three types of published data: 
\begin{enumerate}[a.)]
        \item \textit{Grids:} Nine~grids of stellar models with initial masses of 9$-$500~M$_{\odot}$ each. The metallicities of the grids are equally distributed between Galactic\footnote{By Galactic metallicity, we mean the metallicity of MW stars as defined and applied to stellar models by \citet{Brott:2011}, cf. their Tables~1 and~2: Z$_{\rm MW}$~$=$~0.0088. This is somewhat lower than the solar metallicity, which is Z$_{\odot}\sim$~0.012. See \citet{Brott:2011} for a discussion of this definition.} ([Fe/H]~$\lesssim$~0.0) and very low ([Fe/H]~$=$~$-$2.4) metallicities. These grids are in part based on published stellar models (MW, LMC, SMC, IZw18, and IZw18-CHE), while four new grids have been computed for the present work and cover various sub-SMC metallicities. These are typical for dwarf galaxies and the early Universe. The models have been post-processed into equivalent evolutionary phase format.
        
        \item \textit{Tracks:} Interpolated sets of tracks\footnote{We use the term `stellar model' to mean a detailed evolutionary model sequence computed in a stellar evolution code from first principles. In turn, we use the term `track' to mean synthetically created (here: interpolated) evolutionary models. Similarly, we consistently use the term `grid' to mean a grid of stellar models, and the term `set' to mean a set of interpolated tracks.} corresponding to all nine~grids, which can serve as the basis of synthetic stellar populations. We publish them as simple tables. 
        
        \item \textit{Populations:} After weighting the interpolated sets of tracks with a Salpeter initial mass function, the result are synthetic populations of massive and very massive stars for an instantaneous starburst episode at age~0. The total mass of the populations is set to 10$^7$~M$_{\odot}$. The populations are published as tables, up to $\sim$25~Myr (when the 9~M$_{\odot}$ stars die). 
\end{enumerate}

This is the first time that stellar evoluionary predictions for massive and very massive stars in such a broad metallicity range are published. 
The models in eight~of the grids (a.) have been computed with moderate rotation rates leading to normal or classical stellar evolution (i.e.,~initial rotational velocity of 100~km~s$^{-1}$ uniformly), while one of the grids corresponds to extreme rotation rates leading to chemically homogeneous evolution (CHE;~initial rotational velocity of 500~km~s$^{-1}$). In addition to the usual surface properties (mass, temperature, luminosity, mass-loss rate, etc.),
detailed information about the chemical composition (yields of 34 isotopes) and kinetic energy of the stellar winds is provided at various metallicities as a function of time. Additionally, estimates for the helium and carbon-oxygen core masses are provided to facilitate predictions for the mass of the remnant. 

Moreover, we present the newly developed stellar population synthesis tool \textsc{synStars}, which performs spline-based interpolation on the precomputed stellar models based on the initial mass of the star. In addition to performing the interpolation, \textsc{synStars} also creates the time-dependent synthetic populations by weighting the tracks with a Salpeter initial mass function and integrating over the stellar feedback (total mass in the wind, total kinetic energy in the wind, etc.). To facilitate reproducibility, the current version of \textsc{synStars} is also available along with the model data. 

With their broad range in mass and metallicity, the BoOST model populations are suitable for applications in star formation studies, for instance.
They can also be used to simulate the formation and evolution of young clusters, dwarf galaxies, and high-redshift galaxies, where the feedback from massive stellar winds play a crucial and metallicity dependent, role. 

This paper is organized as follows.
In Sect.~\ref{sec:stellarmodels} we present nine grids of stellar models computed with the Bonn code. 
In Sect.~\ref{sec:EEPs} we describe how we identified the equivalent evolutionary phases (EEPs), while in Sects.~\ref{sec:interpolation}--\ref{sec:popsyn}, we present our new tool \textsc{synStars} and use it to interpolate between the stellar models and perform population synthesis.
In Sect.~\ref{sec:extrapol} we discuss the models that did not converge due to numerical instabilities and our novel solution for including their contributions into the populations. 
In Sect.~\ref{sec:core} we explain our method for defining the mass of the final stellar cores, which is a proxy for the mass of the compact object remnant. 
In Sect.~\ref{sec:discussion} we discuss similarities to previous projects, suggest possible astrophysical applications, and describe future plans. In Sect.~\ref{sec:summary} we conclude.

\section{Grids of stellar models}\label{sec:stellarmodels}

\subsection{Physical ingredients}\label{sec:ingredients}

The BoOST project relies on stellar evolutionary models created with the Bonn code. Some of these models have been published earlier (\citealt{Brott:2011a,Koehler:2015} and \citealt{Szecsi:2015}), but most are newly computed.
A description of the input physics implemented in the version of the Bonn code we use here was given by \citet{Heger:2000a,Heger:2000b,Petrovic:2005,Brott:2011,Yoon:2012,Szecsi:2015} and references therein. {To summarize, the models follow the prescription from \citet{Vink:2000} for hot wind-driven mass loss of OB~stars and \citet{Nieuwenhuijzen:1990} for cool dust-driven mass loss of supergiants. As for the naked helium star phase -- i.e., Wolf--Rayet stars or Transparent Wind UV-Intense (TWUIN) stars; cf. \citealt{Kubatova:2019} --, they follow the rates from \citet{Hamann:1995} reduced by a factor of ten \citep{Yoon:2005}. A metallicity dependence of $\sim$Z$^{0.86}$ \citep{Vink:2001} was applied.}
{Convective mixing was included based on the mixing-length theory approach \citep{Boehm:1958} with a mixing length parameter $\alpha_{\rm MLT}$~$=$~1.5. For convective overshooting, step overshooting was employed with a parameter $\alpha_{\rm over}$~$=$~0.335 as calibrated by \citet{Brott:2011a} for massive stars in the LMC. Convective boundaries were determined using the Ledoux criterion with a semiconvective mixing parameter of $\alpha_{\rm sc}$~$=$~1.0.} 
{Rotationally induced mixing of chemical elements was treated with an efficiency parameter $f_c=0.0228$ (\citealt{Heger:2000a,Heger:2000b}, calibrated by \citealt{Brott:2011a}). Furthermore, transport of angular momentum by magnetic fields due to the Spruit--Taylor dynamo \citep{Spruit:2002,Heger:2005} was included. }

Table~\ref{tab:Z} provides a summary of all the BoOST grids and their initial compositions. {The already published and the newly created models were computed with the same version of the code, that is, the same input physics as above were applied consistently.} As for the previous publications,
\citet{Brott:2011a} published stellar models with an MW, LMC, and SMC composition between 9$-$60~M$_{\odot}$. \citet{Koehler:2015} extended the LMC grid with main-sequence models up to 500~M$_{\odot}$. \citet{Szecsi:2015} published models with the much lower IZw18 metallicity (corresponding to 0.1$\times$Z$_{\rm SMC}$) in the mass range of 9$-$300~M$_{\odot}$, also on the main sequence. {All other models, as well as the post-main-sequence phases when needed, were newly computed by us.}

We consistently used the moderate initial rotational rate of 100~km~s$^{-1}$ because this rotational rate is representative of nonrotating or slowly rotating massive stars. Additionally, one of our grids was composed of chemically homogeneous models, which have a very fast initial rotation rate, 500~km~s$^{-1}$. 

\begin{figure*}\centering\vspace{-10pt}
        \includegraphics[width=\ratio\columnwidth,page=1]{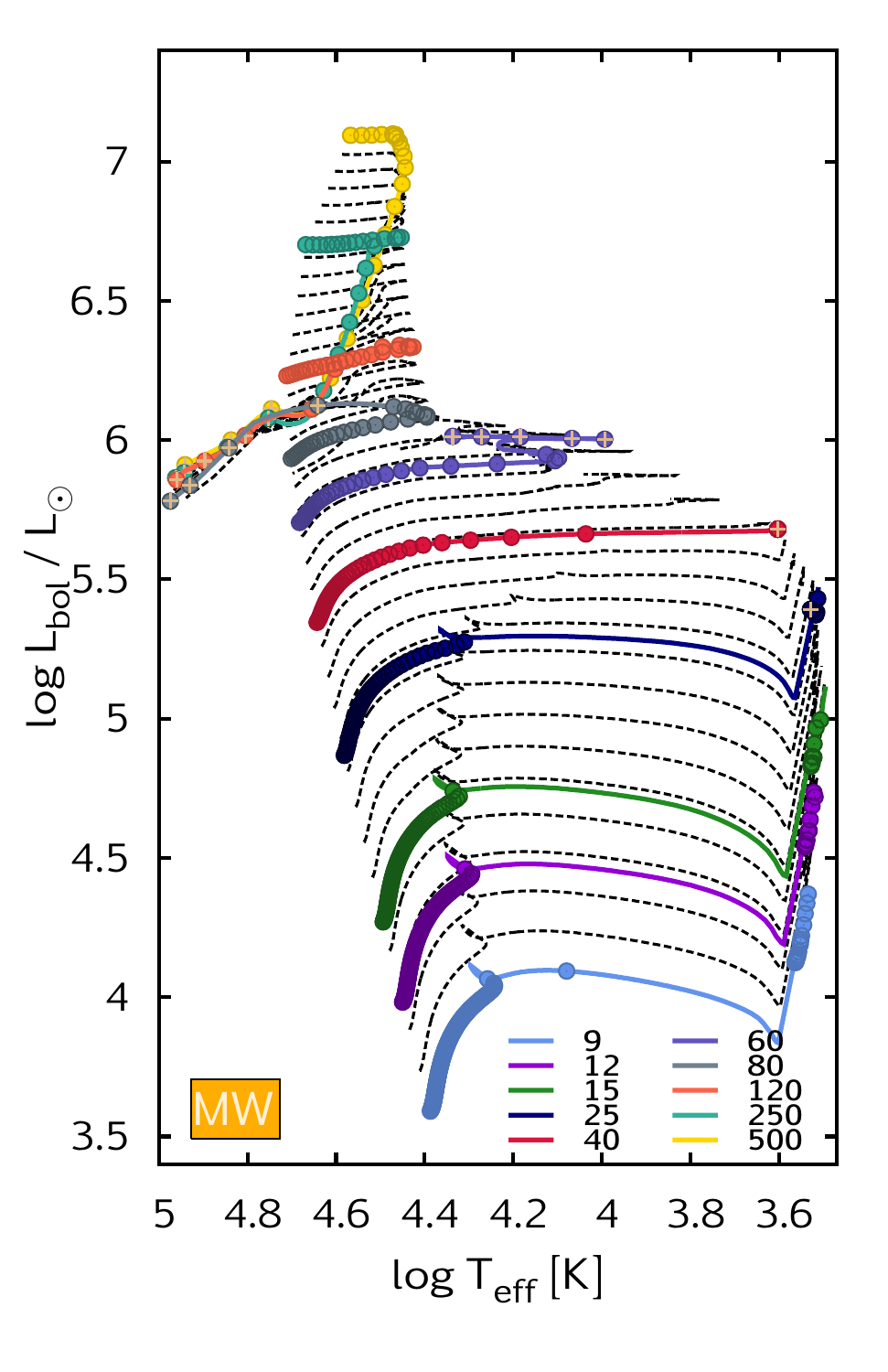}\hspace{0pt}
        \includegraphics[width=\ratio\columnwidth,page=1]{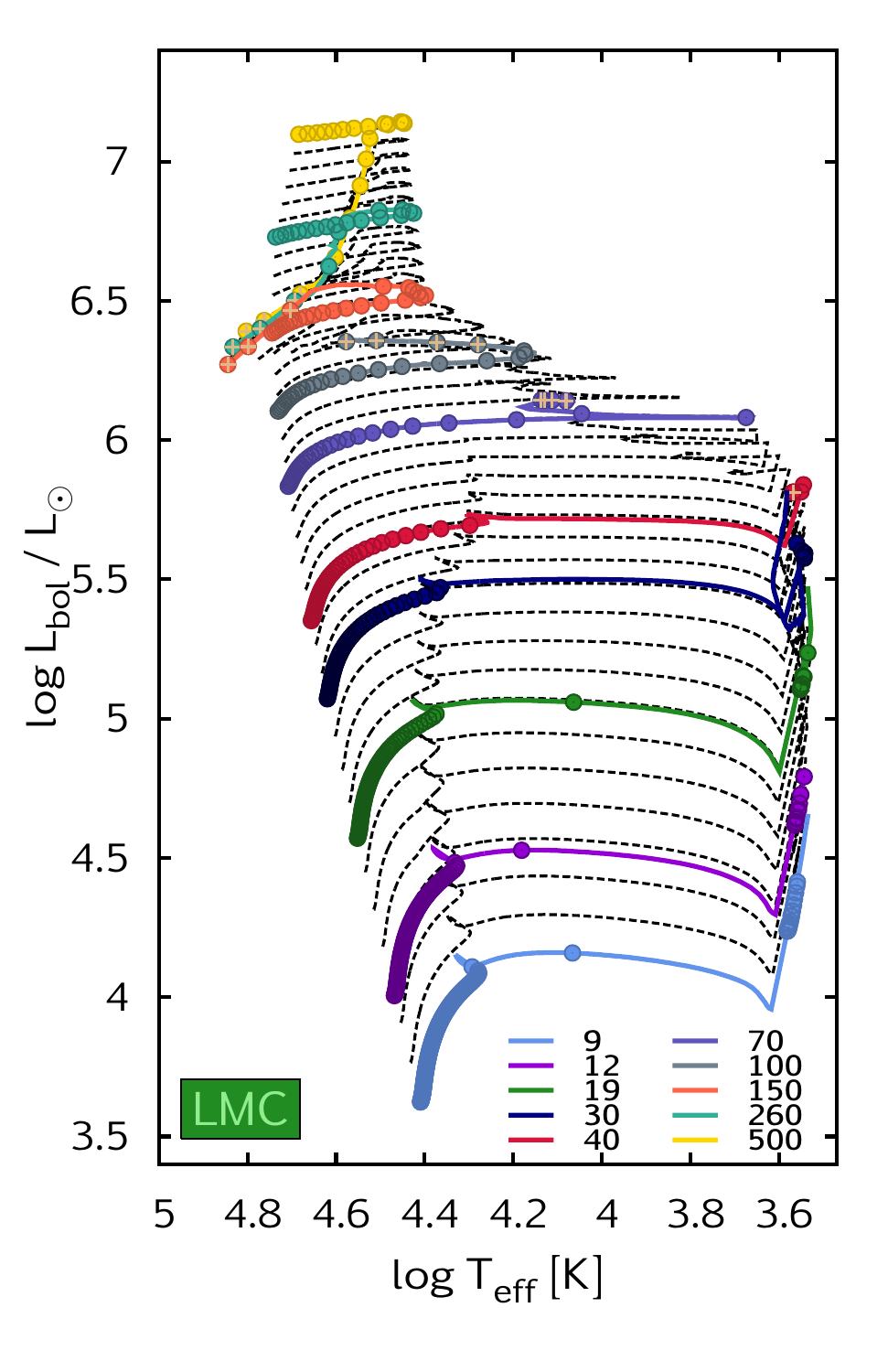}\hspace{0pt}
        \includegraphics[width=\ratio\columnwidth,page=1]{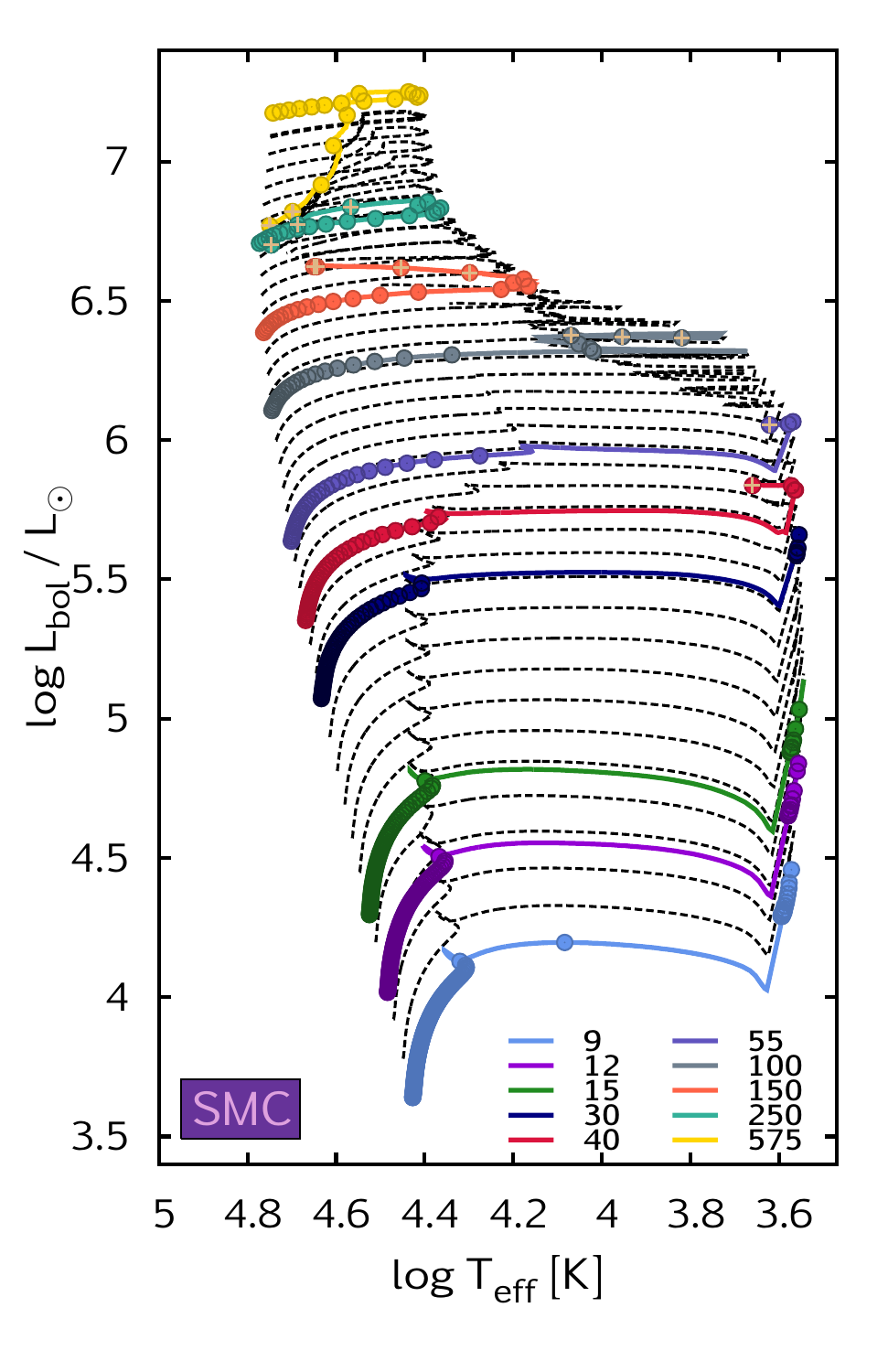}\vspace{-10pt} \\
        \includegraphics[width=\ratio\columnwidth,page=1]{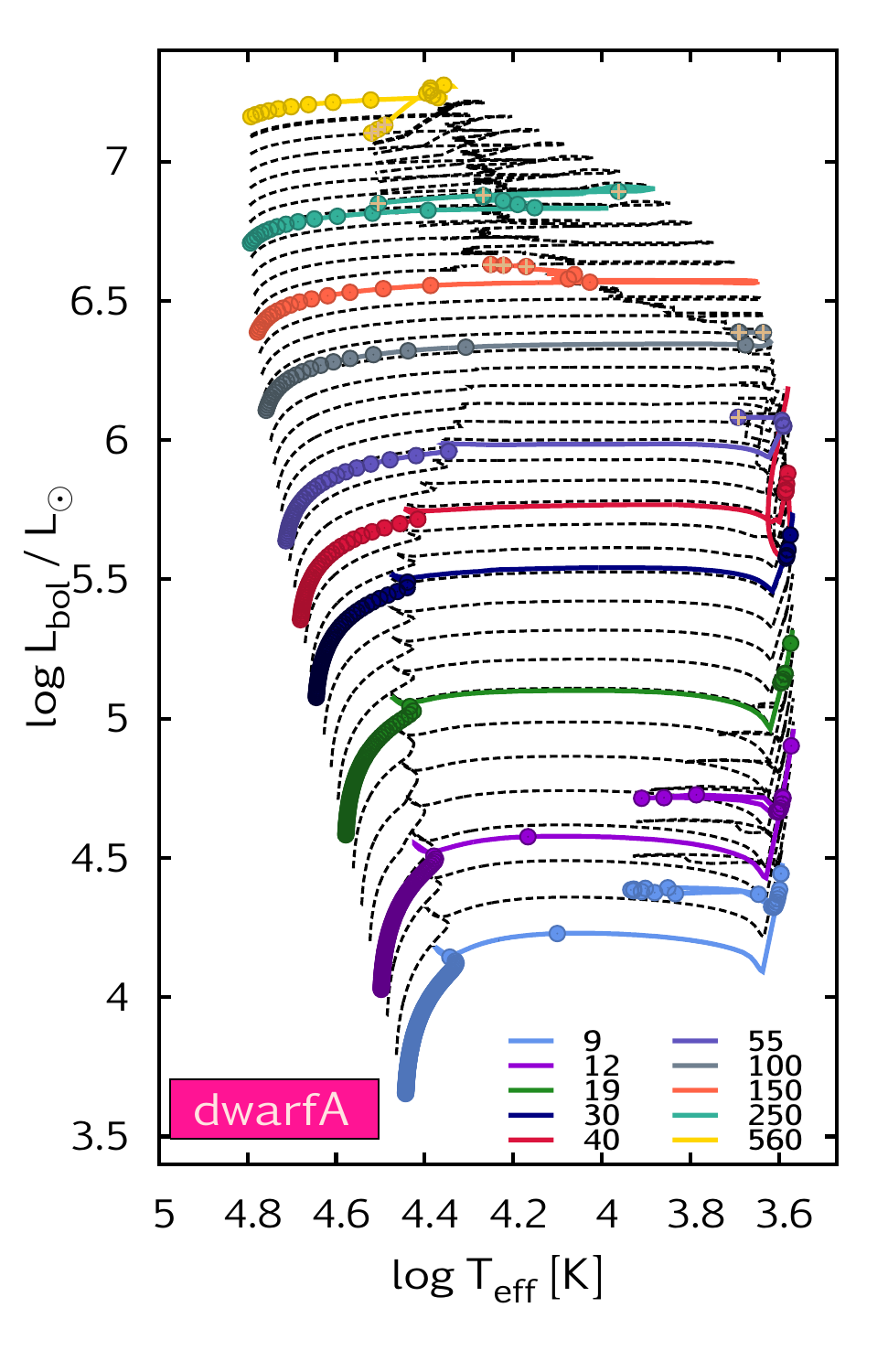}\hspace{0pt}
        \includegraphics[width=\ratio\columnwidth,page=1]{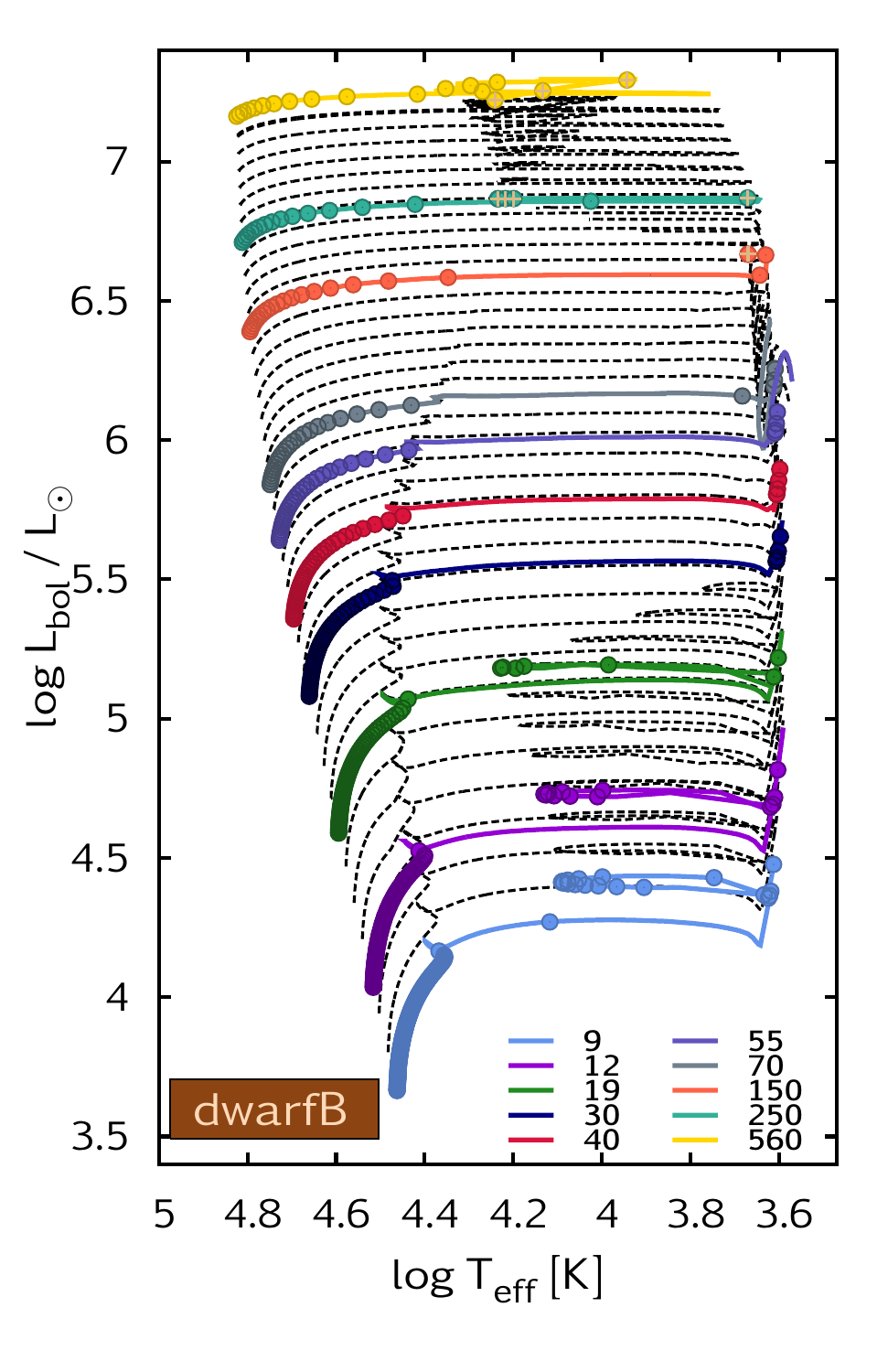}\hspace{0pt}
        \includegraphics[width=\ratio\columnwidth,page=1]{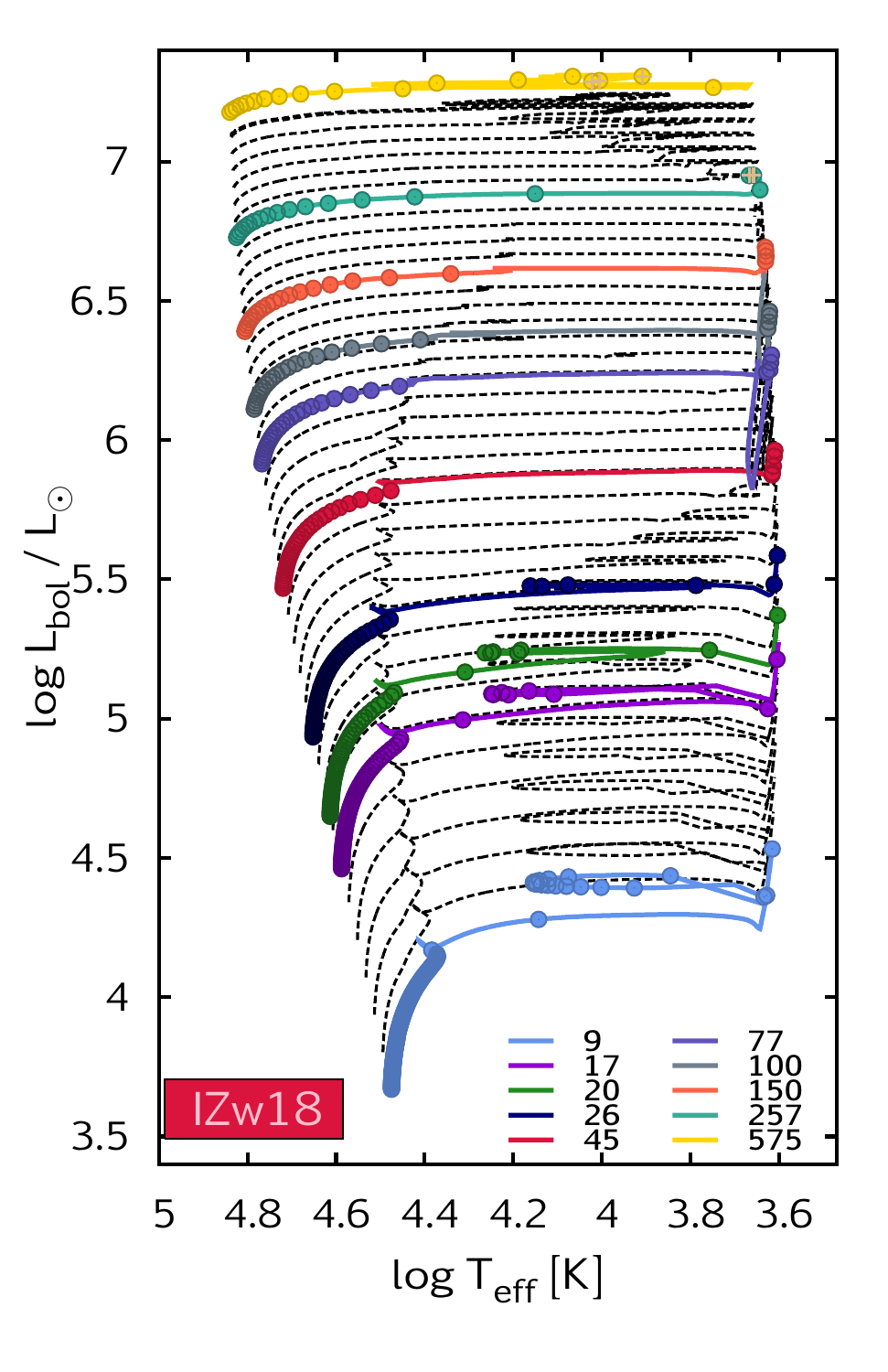}\vspace{-10pt} \\
        \includegraphics[width=\ratio\columnwidth,page=1]{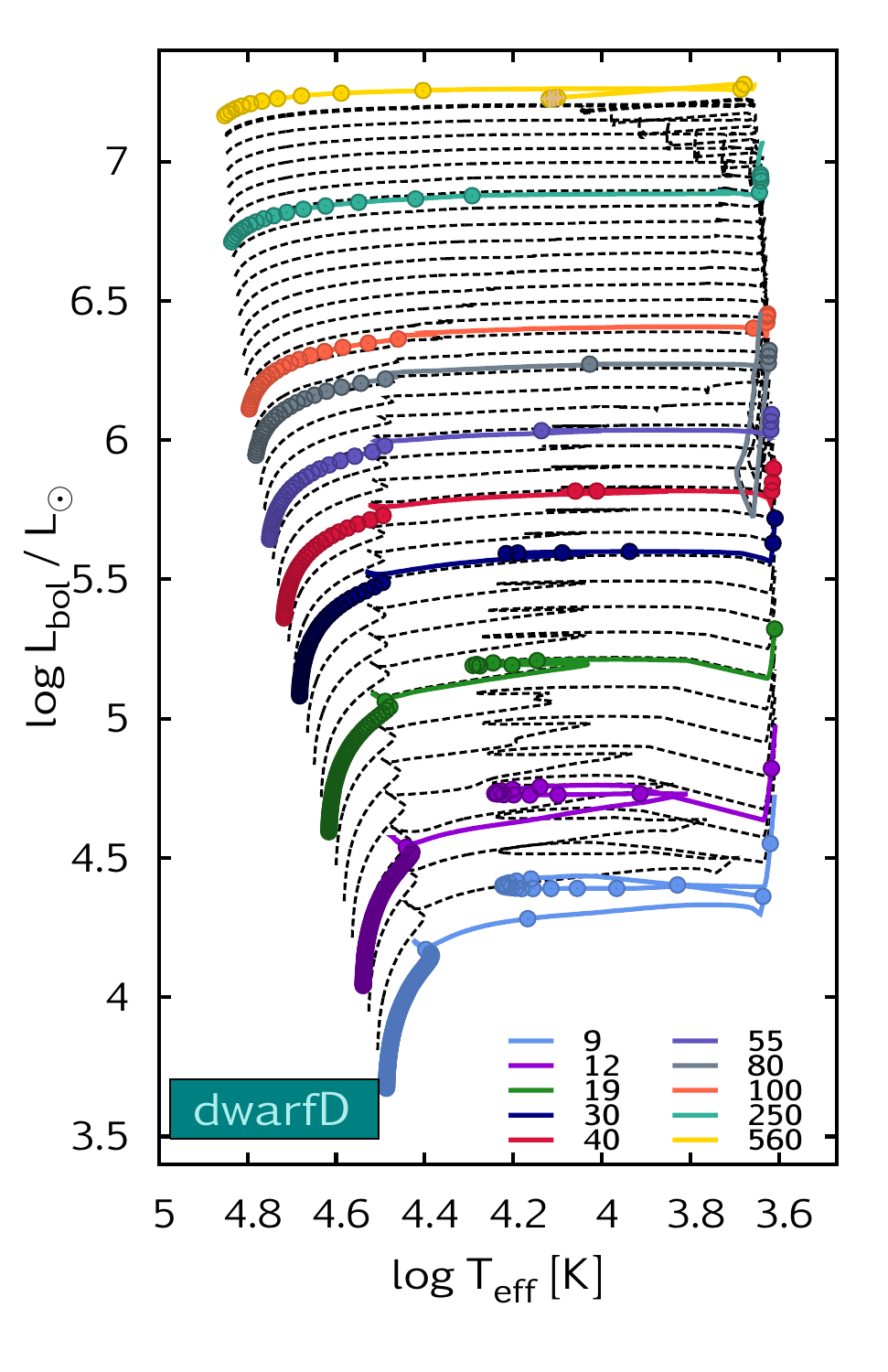}\hspace{0pt}
        \includegraphics[width=\ratio\columnwidth,page=1]{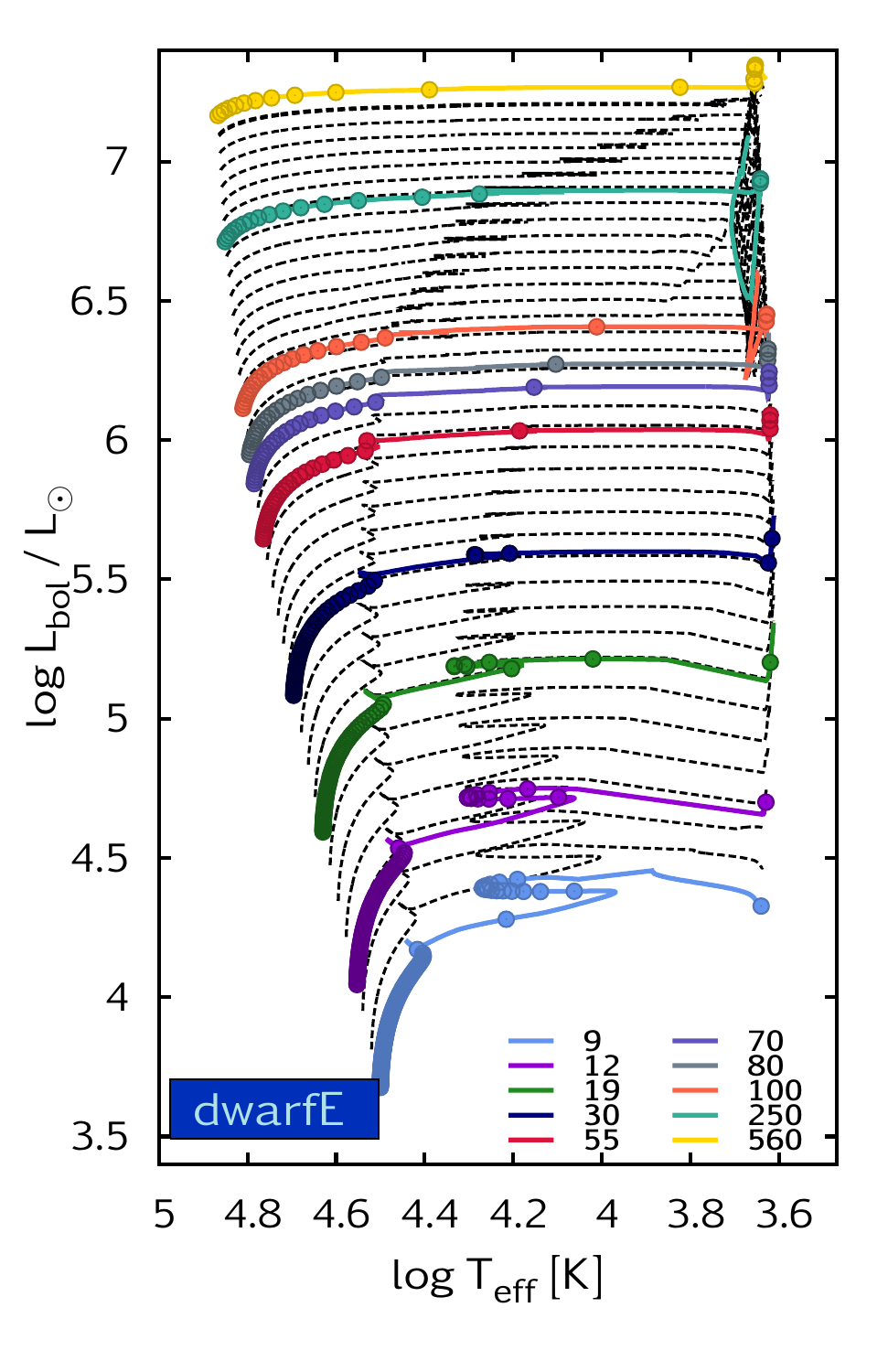}\hspace{0pt}
        \includegraphics[width=\ratio\columnwidth,page=1]{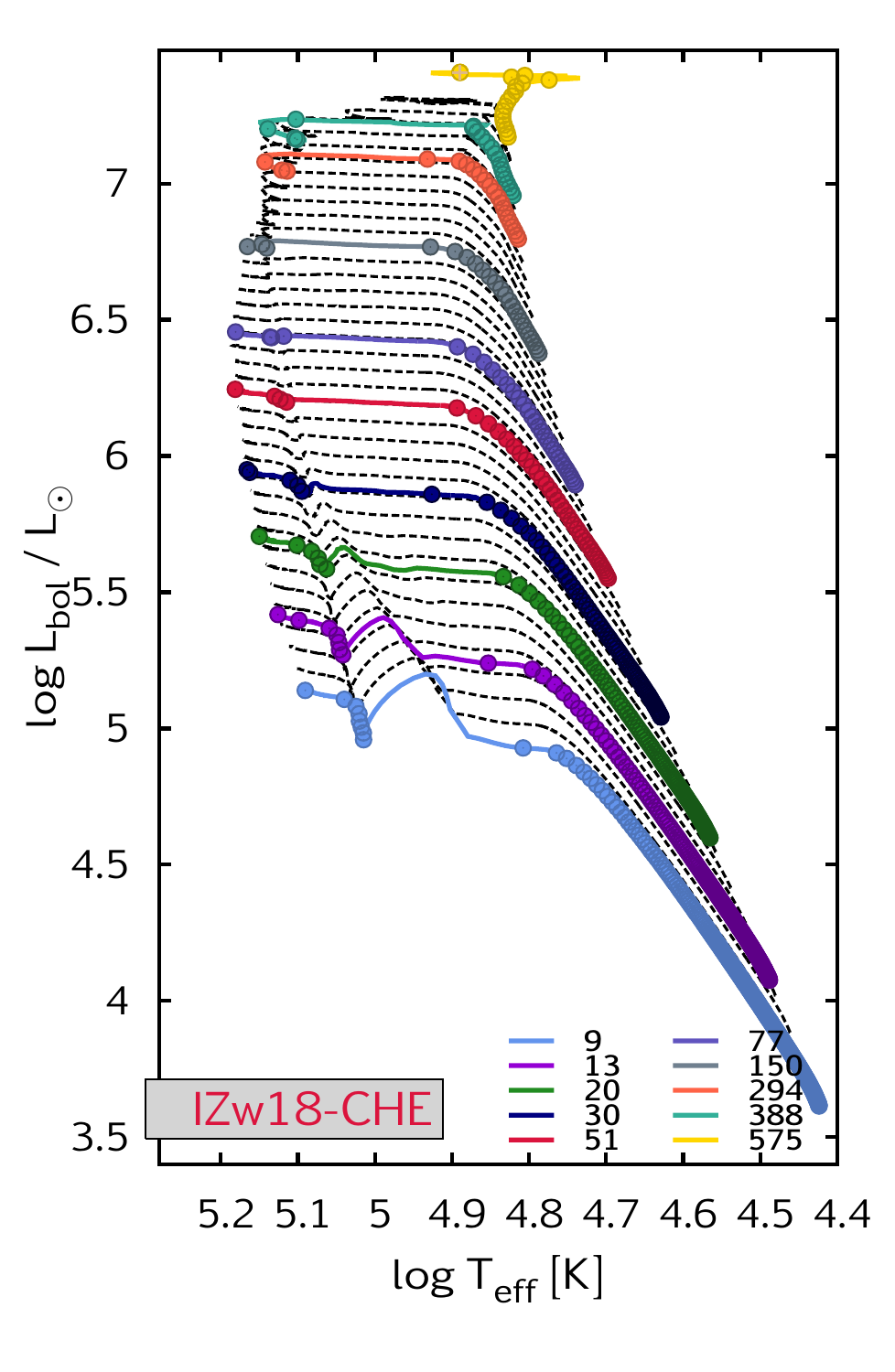}
        \caption{Hertzsprung--Russell diagrams. Initial mass is color-coded (between 9 and $\sim$500~M$_{\odot}$); dots mark every 10$^5$\,yr of evolution along the stellar models. Dashed black lines represent interpolated tracks (up to 500~M$_{\odot}$); brown crosses mark the phases in which the direct extension method (Sect.~\ref{sec:extrapol}) has been applied. For details about the models and their postprocessing, see Sects.~\ref{sec:stellarmodels} and \ref{sec:extrapol}, respectively, and for details about the interpolated tracks, see Sect.~\ref{sec:interpol}.}
        \label{fig:HRD}
\end{figure*}

\subsection{Design of the grids}\label{sec:design}

To be able to create synthetic populations out of these models in a consistent way, we extended the parameter space. In particular, we (i) computed new models up to 500~M$_{\odot}$ in all the grids, (ii) computed new grids for a good metallicity coverage with equal steps, and (iii) either simulated the post-main-sequence evolution or provided a reliable approximation for this phase. Thus we designed eight consistent grids with equal log-metallicity steps from Galactic down to the most metal-poor clusters observed. Additionally, (iv) we provide one grid with chemically homogeneous evolution. 

Every grid contains ten stellar models, starting with 9~M$_{\odot}$.
We did not fix any consistent value for the mass steps between the ten models in the grids. Instead, their initial masses were chosen in a way that facilitated the best interpolation between them. For example, we ensured that we properly covered the part of the HR diagram in which the models show blue loops or luminous blue variable-type features. \textbf{Indeed, because} these effects are highly metallicity dependent, the design of the grids was kept flexible in terms of the mass step, so that abrupt changes in the models were followed properly.  

\textsf{(i) New models up to 500~M$_{\odot}$. \ } We extended the published grids (MW, LMC, SMC, and IZw18) with very massive models (up to around 500~M$_{\odot}$).
For consistency, all these new models also had an initial rotational rate of 100~km~s$^{-1}$, and all their physical ingredients were the same as in the published models. The only two exceptions were the models with 250 and 500~M$_{\odot}$ at MW metallicity, for which the initial rotational rate was set to 0~km~s$^{-1}$ for convenience. We find that this does not make any practical difference in the evolution: at this mass and metallicity, mass loss is so strong already at the beginning of the main sequence that these stars would spin down very soon in any case. 

\textsf{(ii) New grids for a good metallicity coverage. \ }
We present here four new grids (cf. Table~\ref{tab:Z}). Two of them were designed so that the metallicity gap between SMC and IZw18 is filled. The other two belong to metallicities below IZw18, down to 1/250th~Z$_{\mathrm{MW}}$. Because these metallicities are typical of star-forming dwarf galaxies, we call these grids dwarfA, dwarfB, dwarfD, and dwarfE (dwarfC is consistent with the IZw18 grid). With these, our models allow studying the formation of even the lowest-metallicity globular clusters (with [Fe/H]~$\sim$~$-$2.3). 

\textsf{(iii) Late phases of evolution. \ } Most of the models in this work were computed with the Bonn code until core-helium exhaustion. In some cases, however, this was not possible due to numerical reasons. 
High-metallicity models above $\sim$40 or 60~M$_{\odot}$ and lower-metallicity models above $\sim$100~M$_{\odot}$ have inflated envelopes, and in this state, the computations become numerically challenging\footnote{The massive and very massive LMC and IZw18 models in \citet{Kohler:2015} and \citet{Szecsi:2015} were only computed and published up to the end of the main sequence mainly for this same reason.}.
It is difficult to ensure that all our models completely include the post-main-sequence phase (i.e., core-helium-burning).
Here we offer a new solution that is described in Sect.~\ref{sec:extrapol}. With this new method, we approximate for the remaining evolution, completing the last phases of our stars.

\textsf{(iv) Models with chemically homogeneous evolution.}
We provide one grid of fast-rotating, chemically homogeneously evolving models with [Fe/H]~$=$~$-$1.7 called IZw18-CHE. 
In this grid, the initial composition of the models is the same as in the IZw18 grid, but the initial rotational velocity is 500~km~s$^{-1}$ (except in the case of the 9~M$_{\odot}$ model, where it is somewhat lower, 450~km~s$^{-1}$, to avoid reaching critical rotation).
The main-sequence phase of these models (up to 300~M$_{\odot}$) was published and analyzed by \citet{Szecsi:2015}, and their atmospheres were studied by \citet{Kubatova:2019}. Here we complete this grid with two new very massive models (388~M$_{\odot}$ and 575~M$_{\odot}$) as well as the post-main-sequence phase of all of the models (computed properly for almost all of them, except for the highest mass, where we had to  include the remaining phases in an approximate way; cf.~Sect.~\ref{sec:extrapol}). 

We only included core hydrogen- and core helium-burning in our published stellar models. This is justified by the fact that core carbon-burning and subsequent burning phases constitute only~$\lesssim$\,1\% of a massive star's life, during which no significant contribution to stellar feedback is expected (except for the supernova explosion; cf. Sect.~\ref{sec:core}). For example, by simulating the core carbon-burning phase of the 26~M$_{\odot}$ model in our IZw18 grid, we find that this lasts for $\sim$7300~years, which is a mere 0.12\% of the total 6.21~Myr lifetime of the model. The mass that is lost during this time is about 0.02~M$_{\odot}$ , which is an order of magnitude less than what is lost during the whole evolution. Because some of our models experience numerical difficulties in their late phases in any case (we approximate for them in Sect.~\ref{sec:extrapol}), omitting carbon-burning and beyond from the present version of BoOST is not expected to cause additional discrepancies in our simulated populations from the point of view of stellar feedback and wind properties.

We publish the stellar model grids as simple tables. They include the following quantities as functions of time (cf. also the Readme file attached to the table, as well as Appendix \ref{sec:units}): the stellar mass, $M$ as a function of time, the effective temperature of the surface, $T_\mathrm{eff}$, the bolometric luminosity, $L_\mathrm{bol}$, the stellar radius, $R$, the mass loss, $\dot{M}$, the logarithm of the surface gravity, $\log(g)$, the rotation velocity at the surface at the equator, $v_\mathrm{surf}$, the critical rotation velocity, $v_\mathrm{crit}$, the Eddington factor, $\Gamma$, and abundances of 34 isotopes at both the stellar surface and in the center of the star, as listed below. Additionally, the mass of the final He core and CO core are included as a proxy for the mass of the compact object remnant (cf. Sect.~\ref{sec:core}). 

The Bonn code simulates nuclear reaction networks for the following 34 isotopes: $^{1}$H, $^{2}$H, $^{3}$He, $^{4}$He, $^{6}$Li, $^{7}$Li, $^{7}$Be, $^{9}$Be, $^{8}$B, $^{10}$B, $^{11}$B, $^{11}$C, $^{12}$C, $^{13}$C, $^{12}$N, $^{14}$N, $^{15}$N, $^{16}$O, $^{17}$O, $^{18}$O, $^{19}$F, $^{20}$Ne, $^{21}$Ne, $^{22}$Ne, $^{23}$Na, $^{24}$Mg, $^{25}$Mg, $^{26}$Mg, $^{26}$Al, $^{27}$Al, $^{28}$Si, $^{29}$Si, $^{30}$Si, and $^{56}$Fe. They are all included in the published tables.

\subsection{HR diagrams}

\begin{figure*}\centering
        \includegraphics[width=\ratioo\columnwidth,page=2]{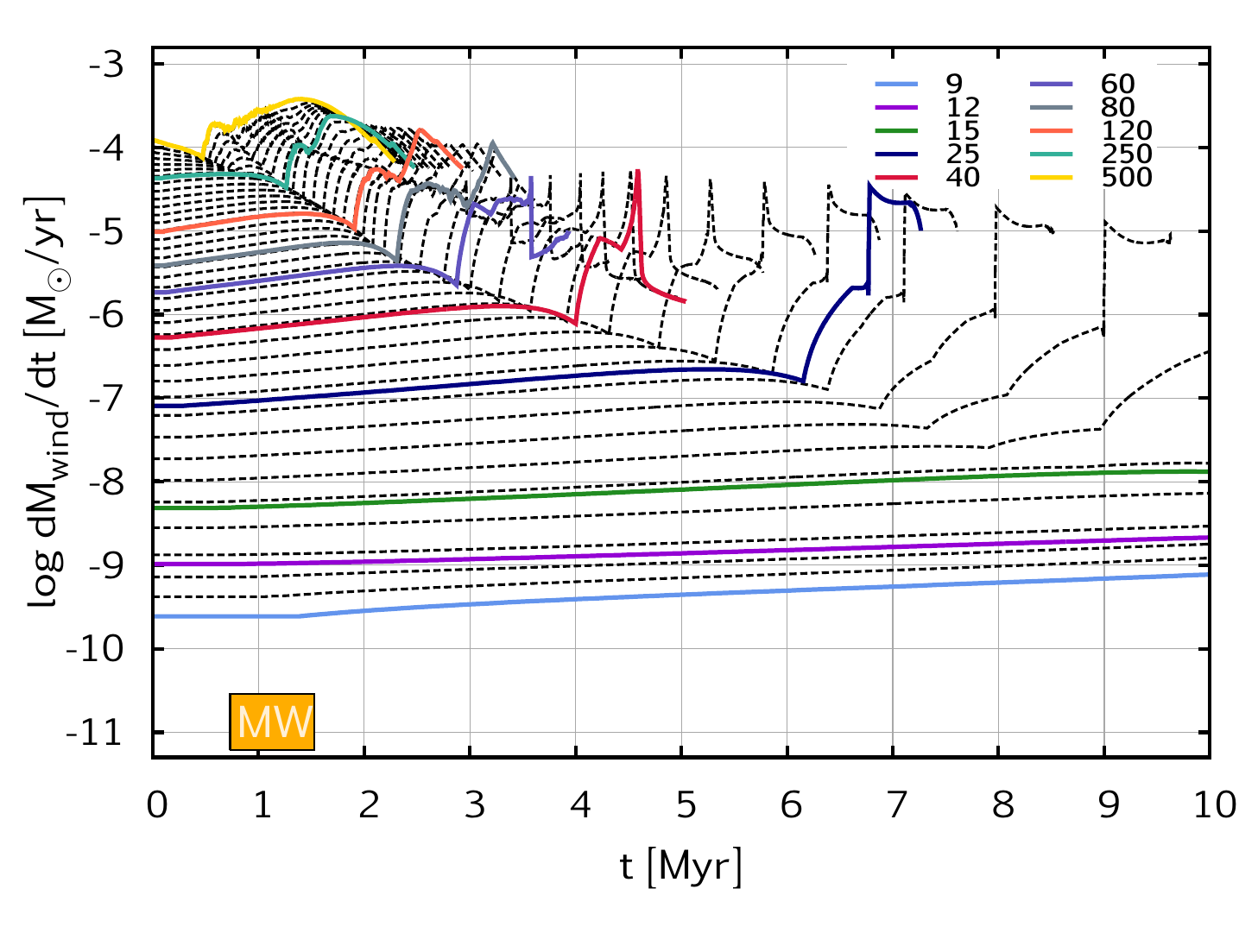}\hspace{2pt}
        \includegraphics[width=\ratioo\columnwidth,page=2]{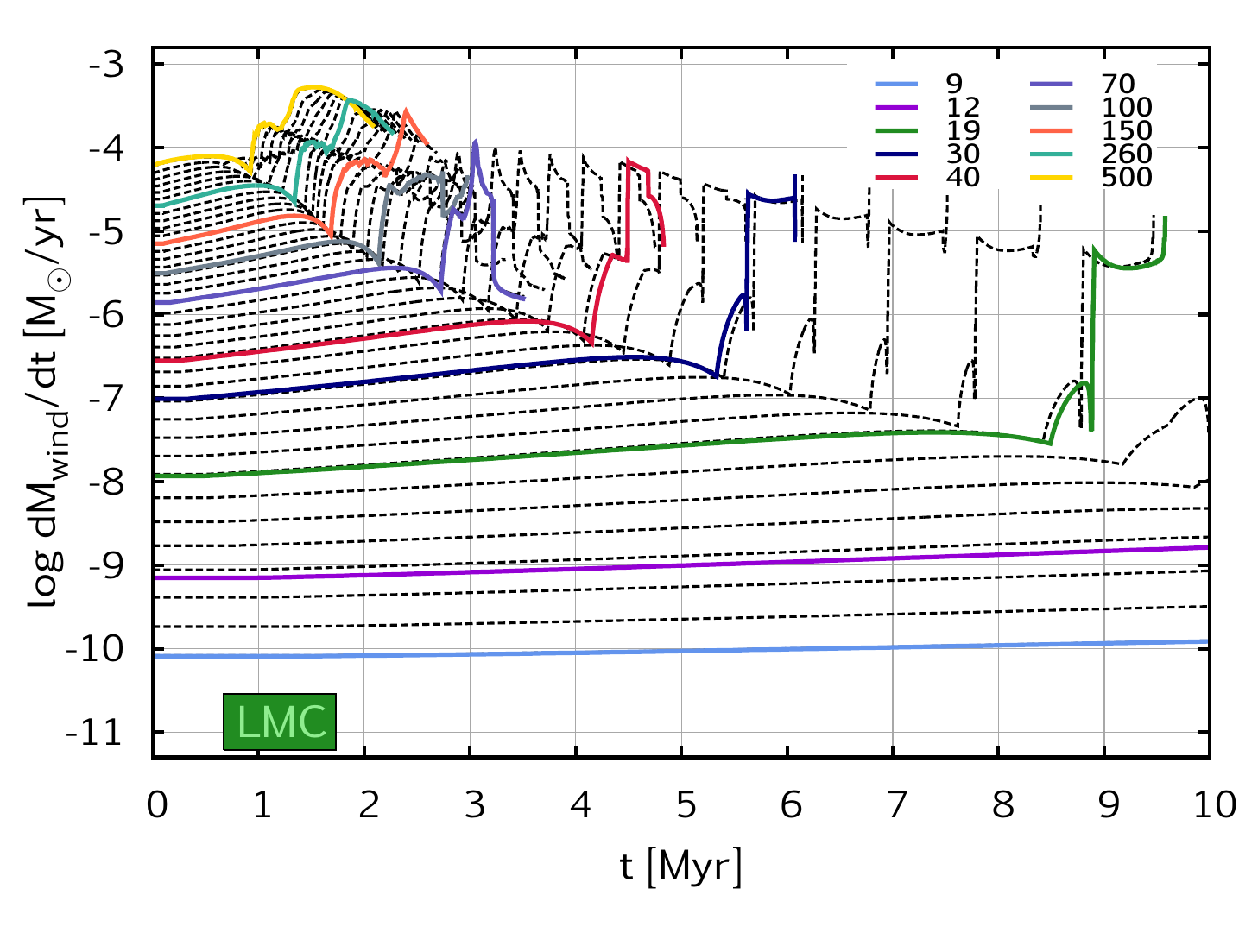}\hspace{2pt}
        \includegraphics[width=\ratioo\columnwidth,page=2]{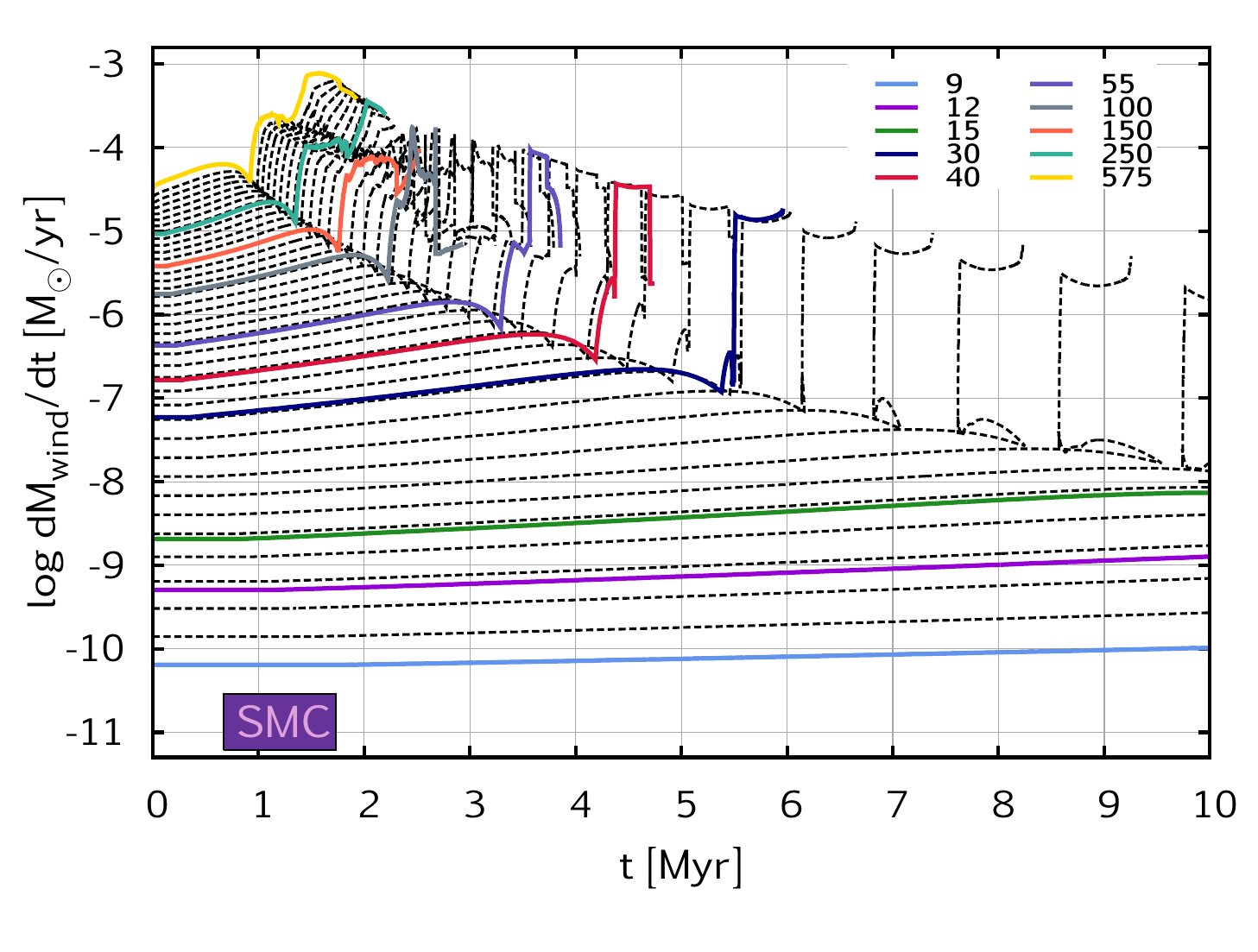}\vspace{2pt} \\
        \includegraphics[width=\ratioo\columnwidth,page=2]{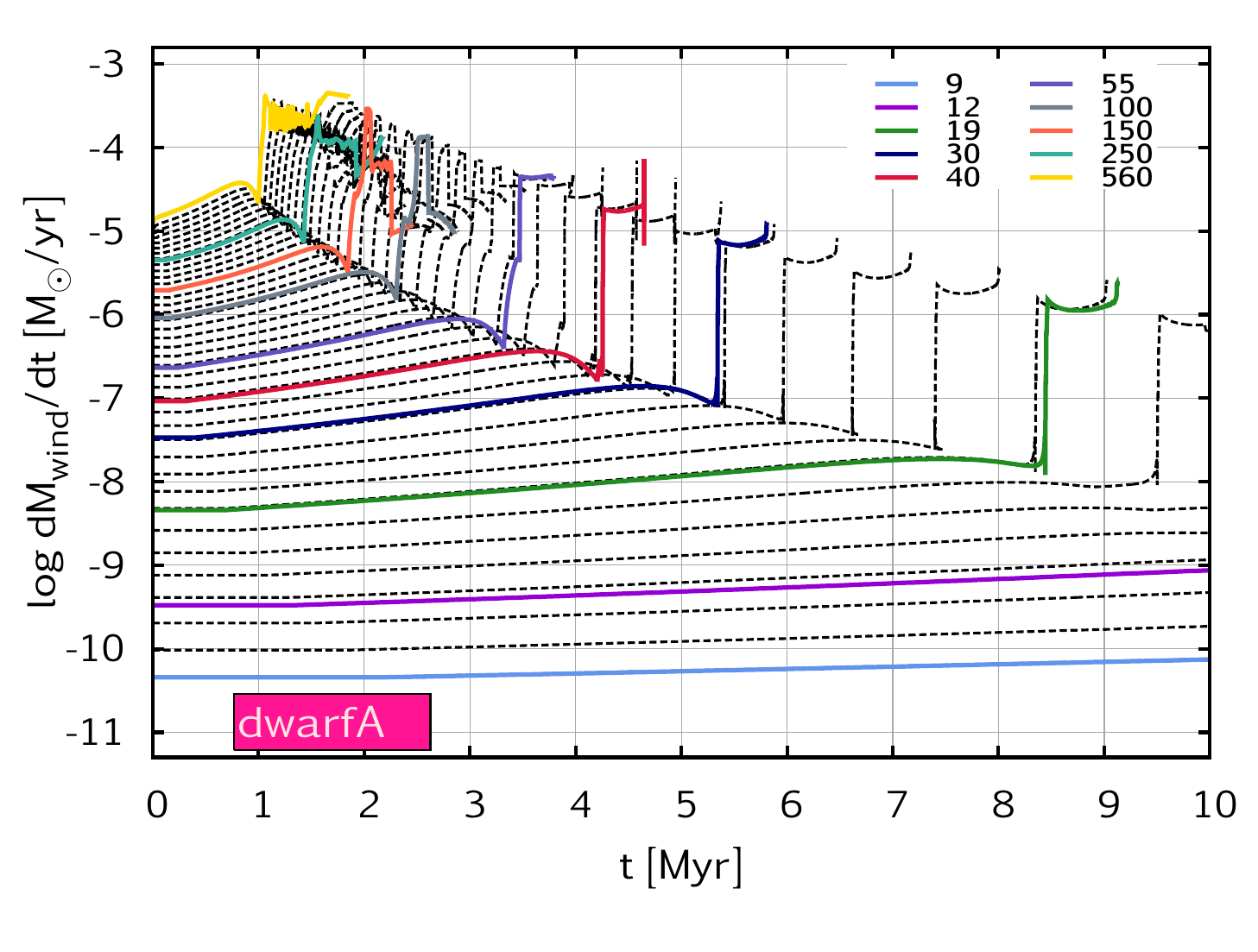}\hspace{2pt}
        \includegraphics[width=\ratioo\columnwidth,page=2]{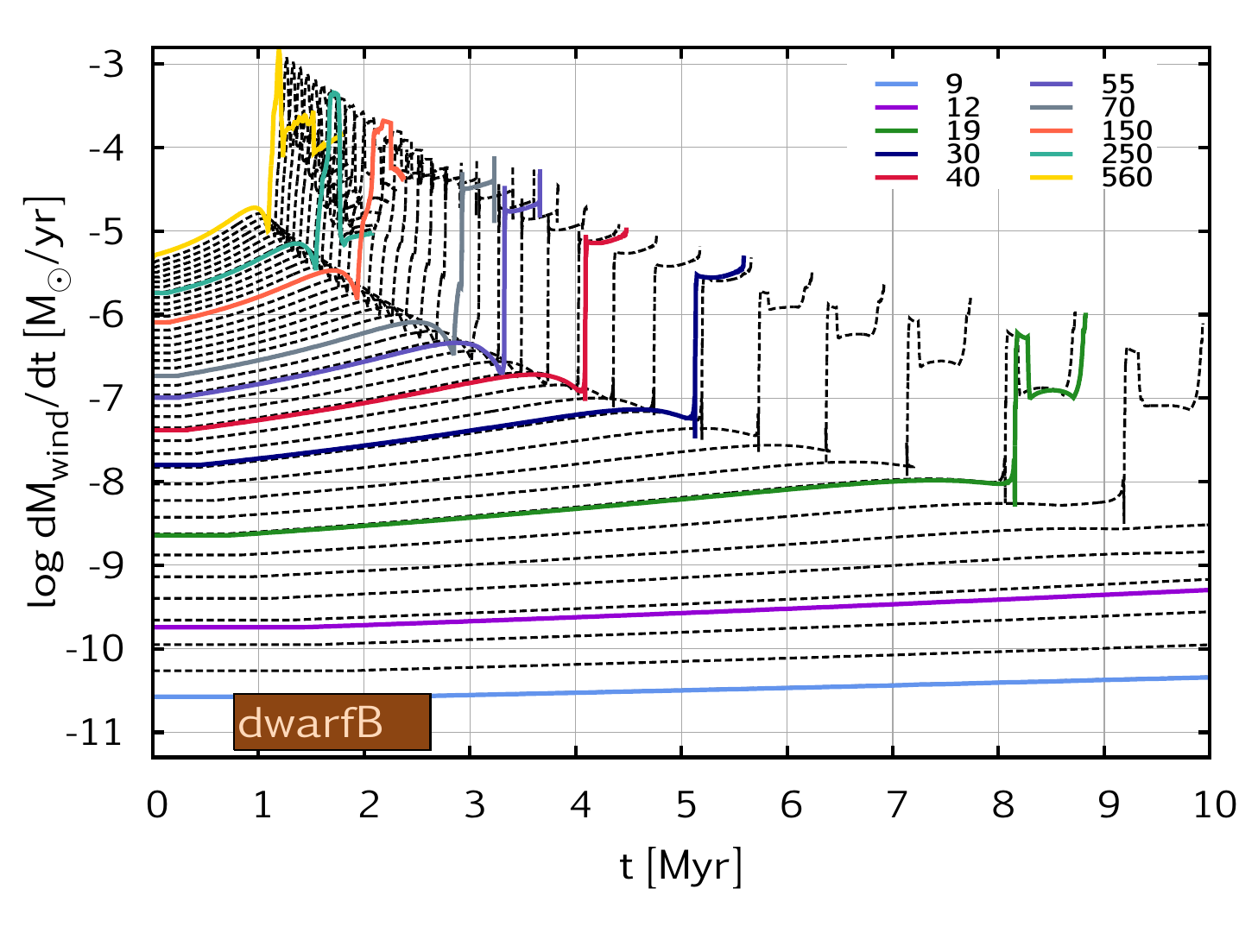}\hspace{2pt}
        \includegraphics[width=\ratioo\columnwidth,page=2]{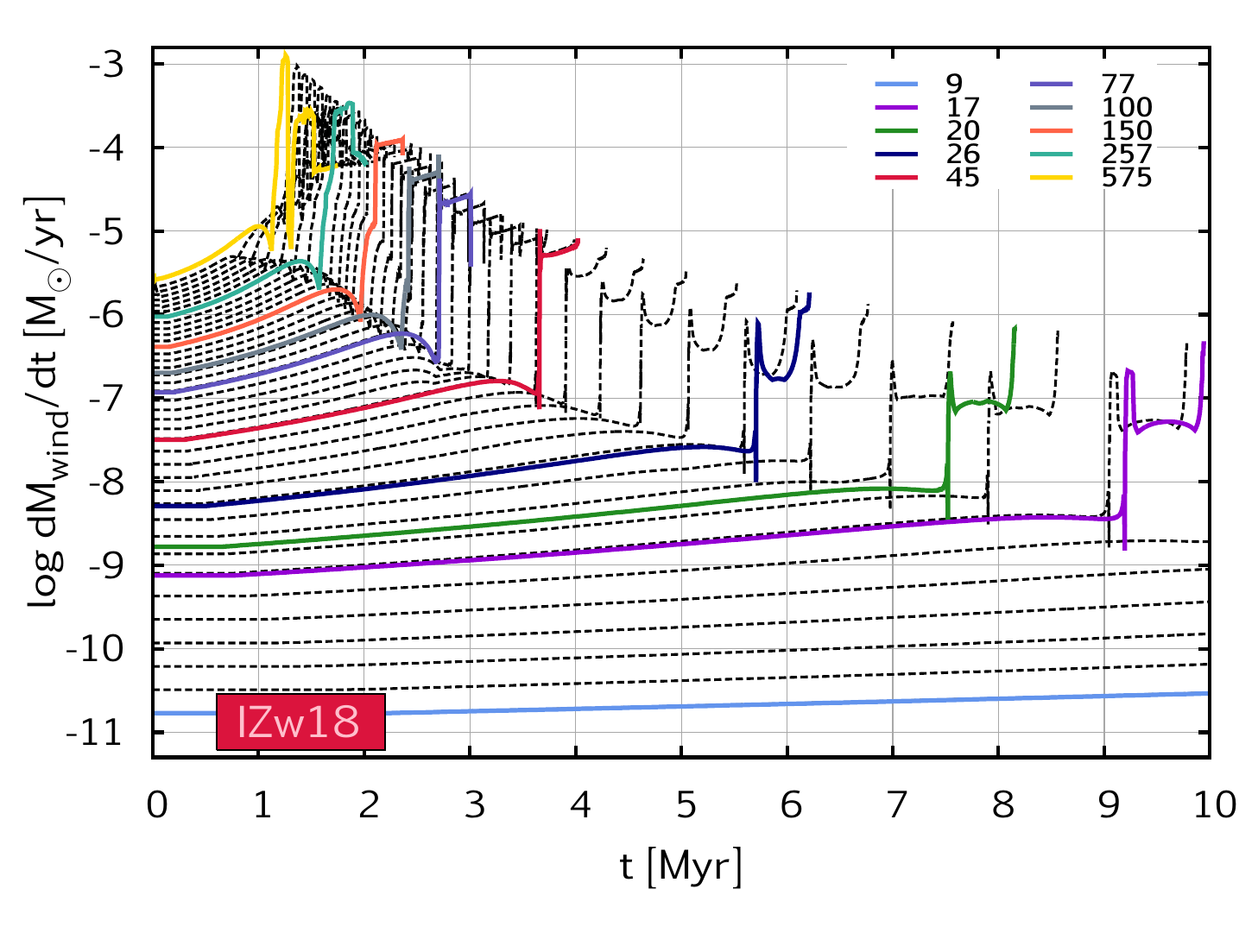}\vspace{2pt} \\
        \includegraphics[width=\ratioo\columnwidth,page=2]{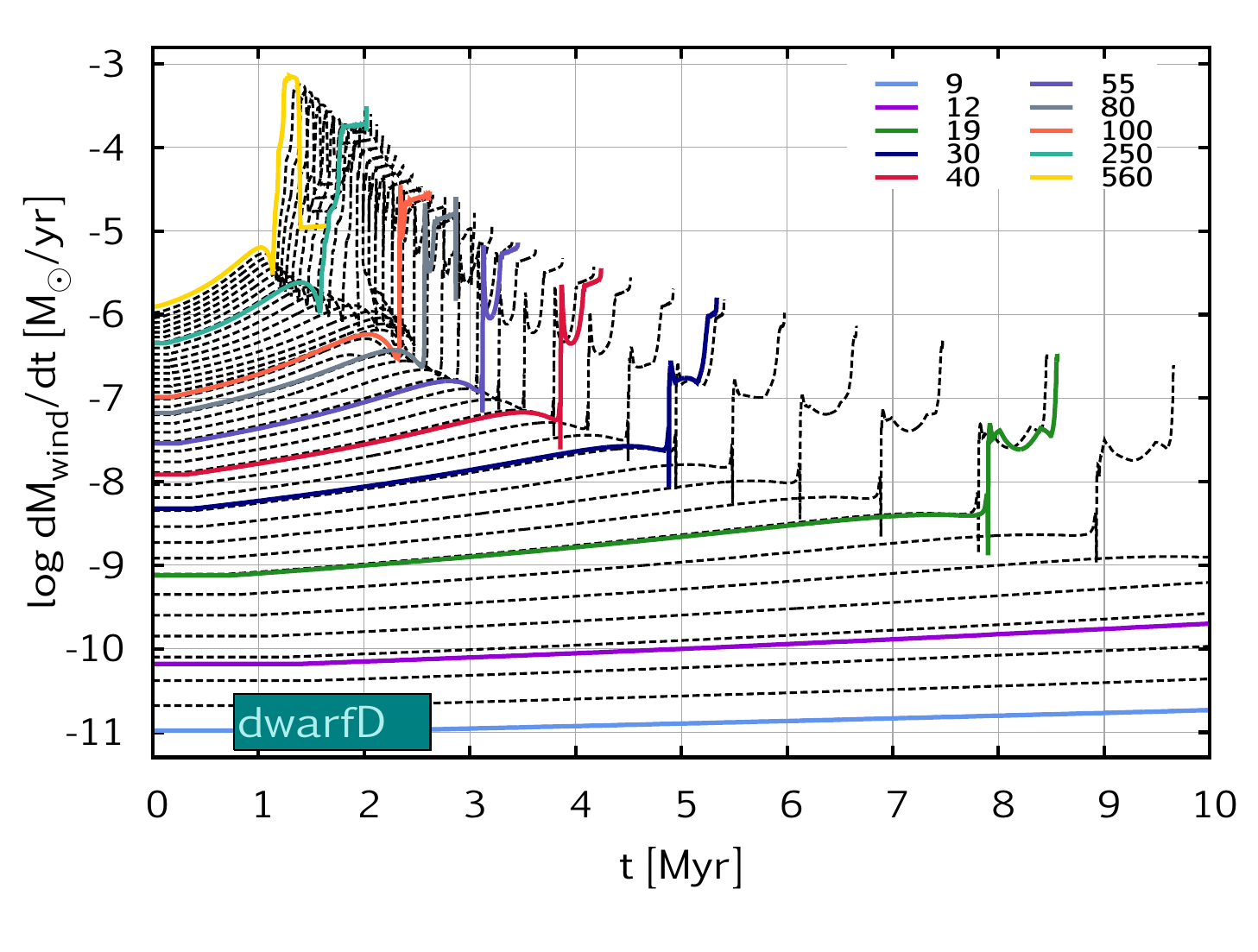}\hspace{2pt}
        \includegraphics[width=\ratioo\columnwidth,page=2]{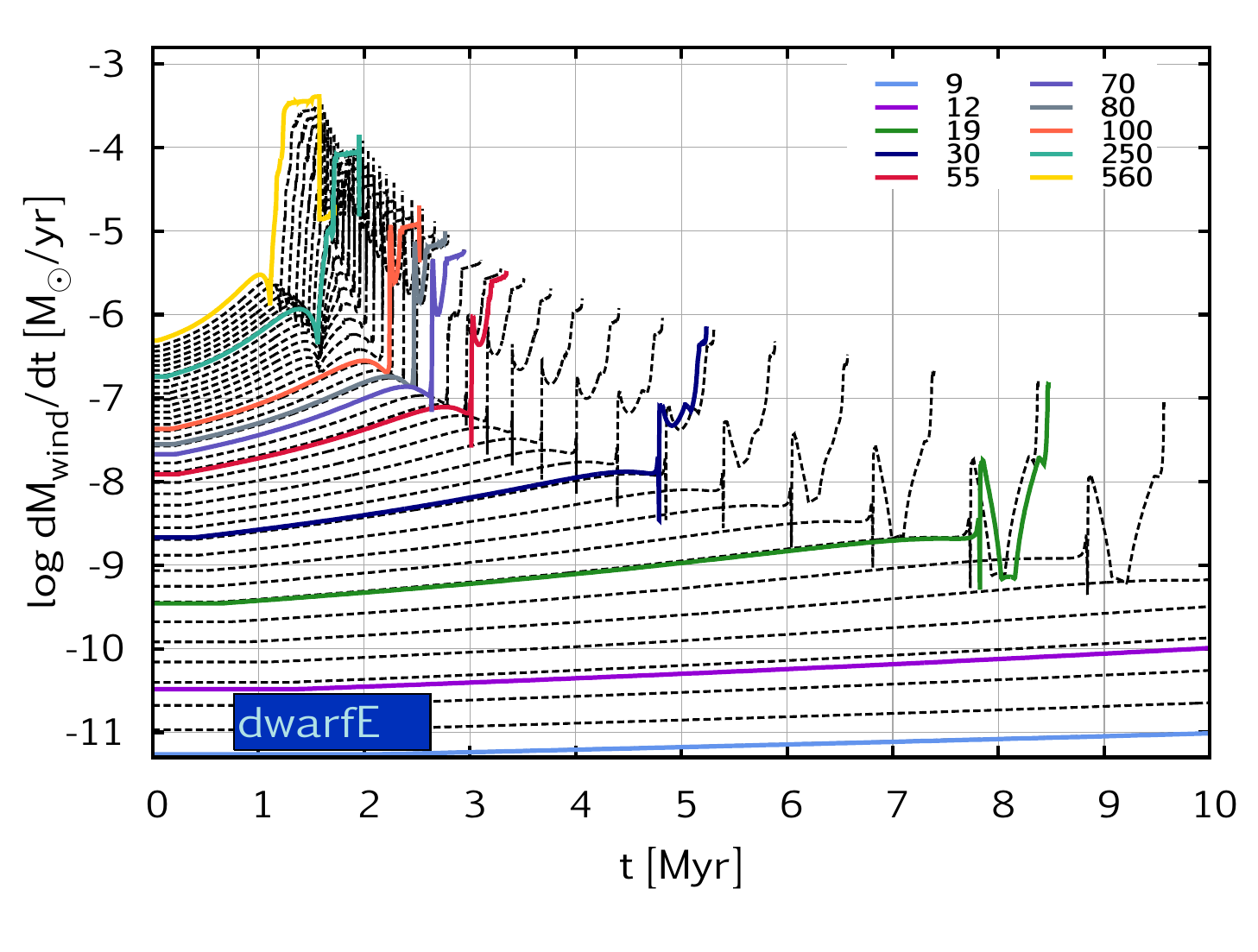}\hspace{2pt}
        \includegraphics[width=\ratioo\columnwidth,page=2]{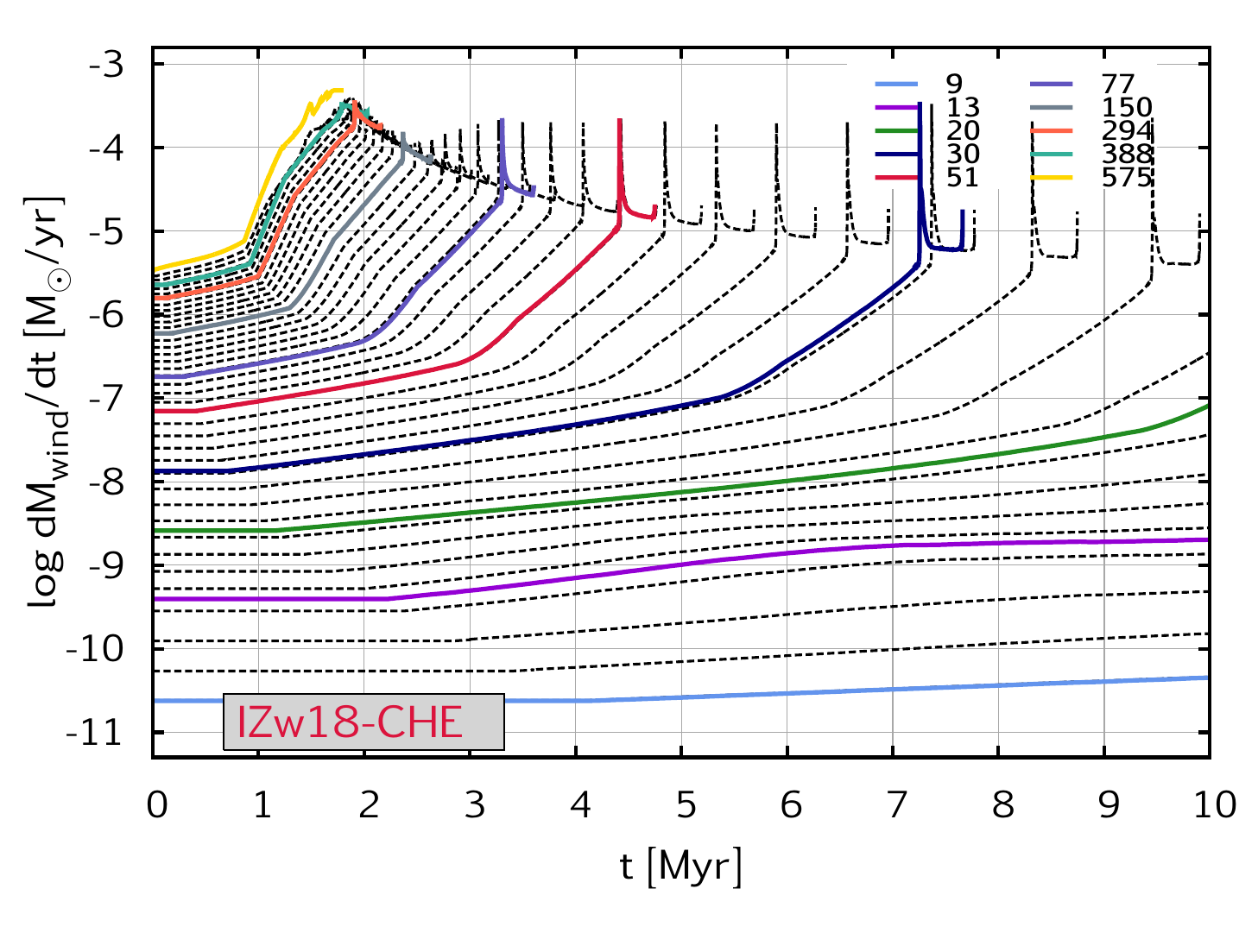}
        \caption{Time evolution of the mass-loss rate. The initial mass is color-coded (between 9 and $\sim$500~M$_{\odot}$). The dashed black lines represent interpolated tracks (up to 500~M$_{\odot}$); cf. Fig.~\ref{fig:HRD}.. While here only shown until 10~Myr, the files published in the BoOST project contain data until the end of the lifetimes of the longest-living model in the population ($\sim$30~Myr).}\label{fig:Mdot}
\end{figure*}

\begin{figure*}\centering
        \includegraphics[width=\ratioo\columnwidth,page=3]{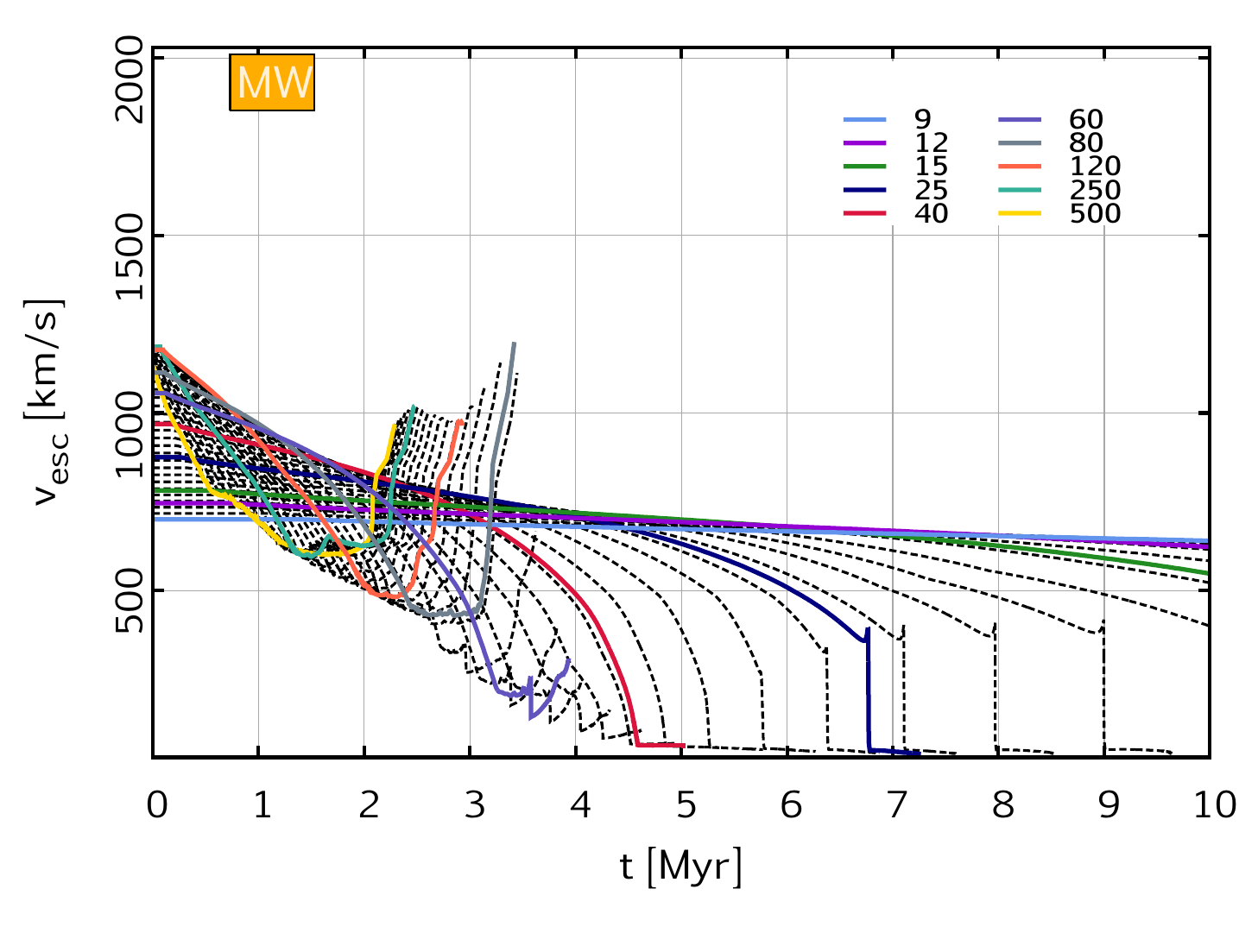}\hspace{2pt}
        \includegraphics[width=\ratioo\columnwidth,page=3]{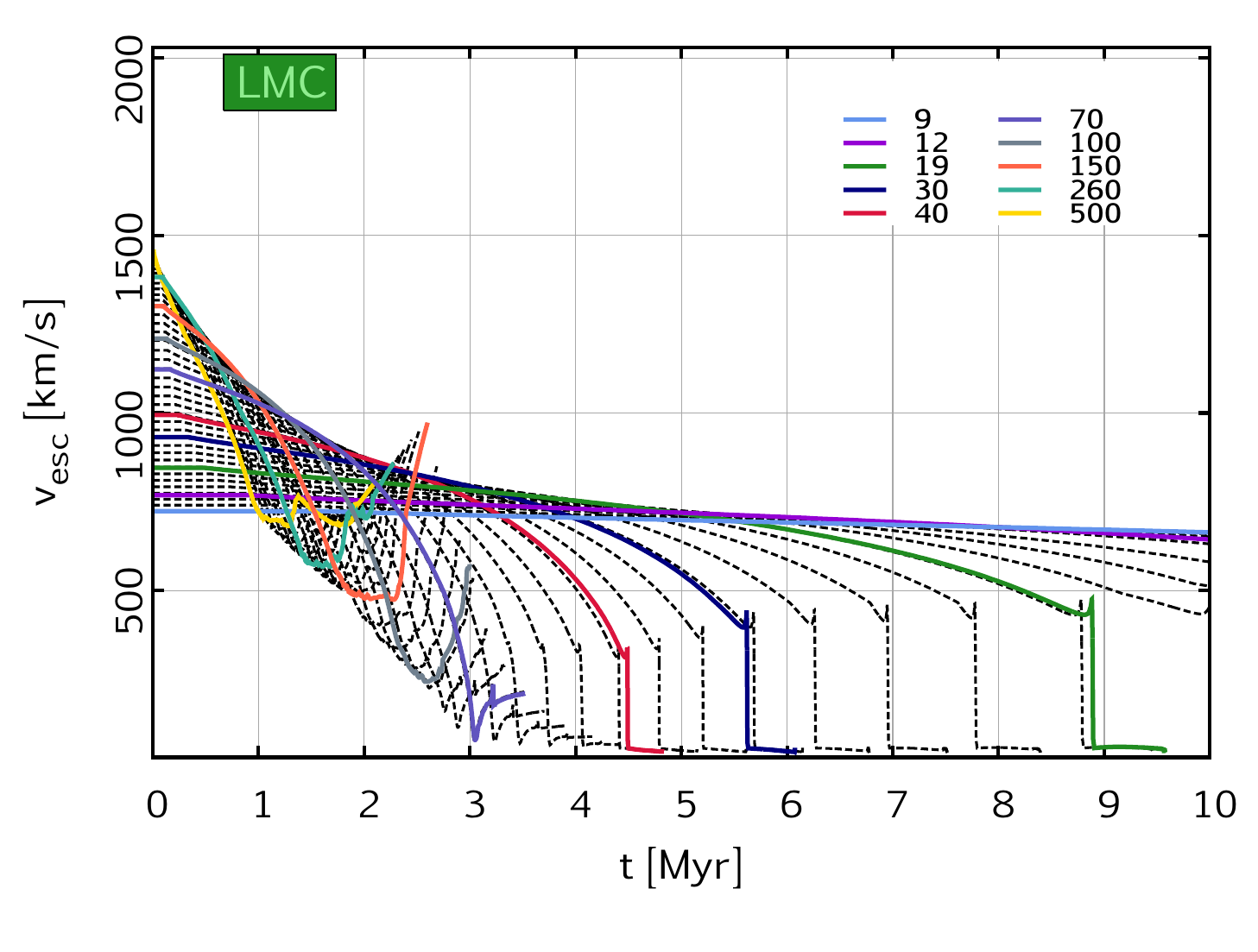}\hspace{2pt}
        \includegraphics[width=\ratioo\columnwidth,page=3]{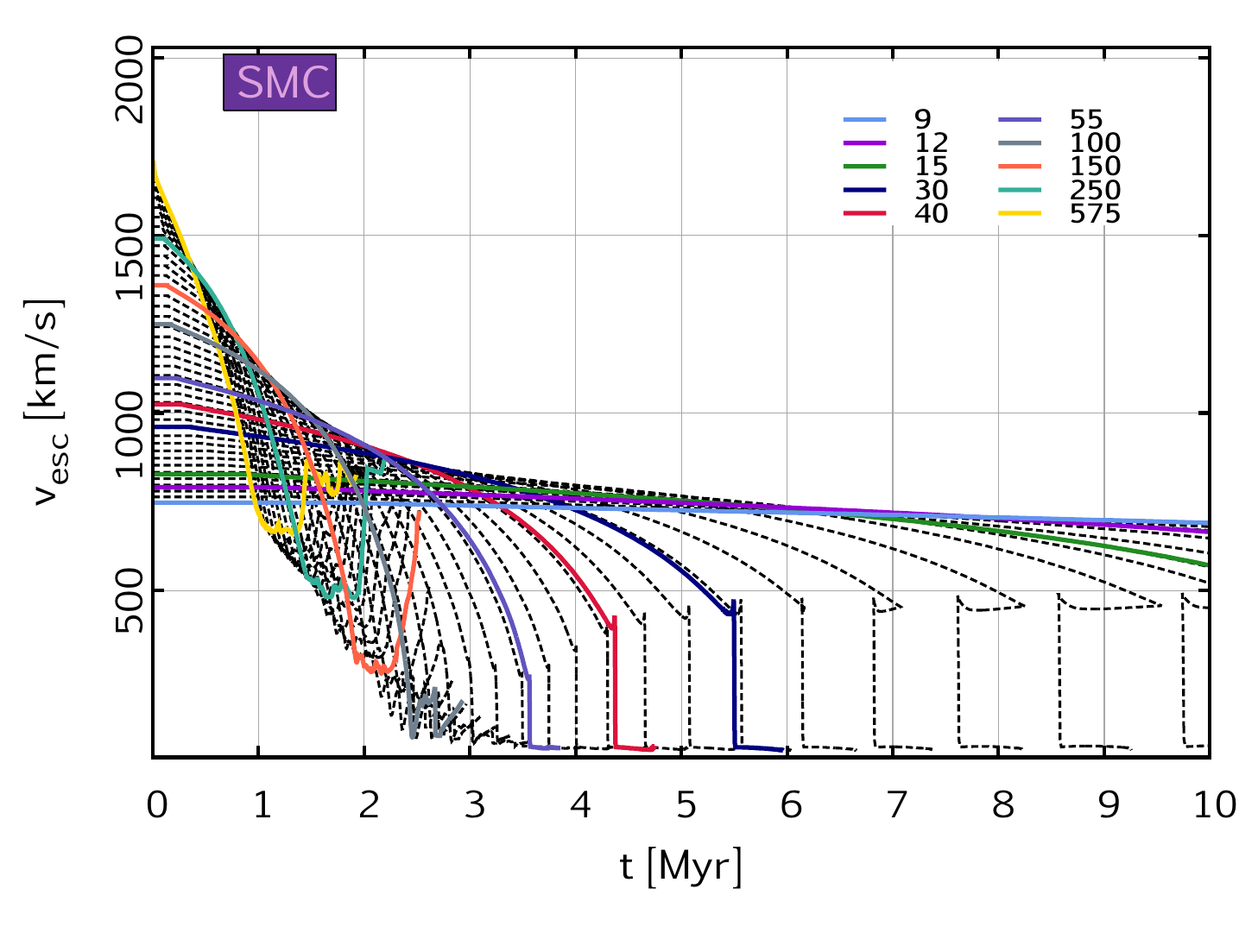}\vspace{2pt} \\
        \includegraphics[width=\ratioo\columnwidth,page=3]{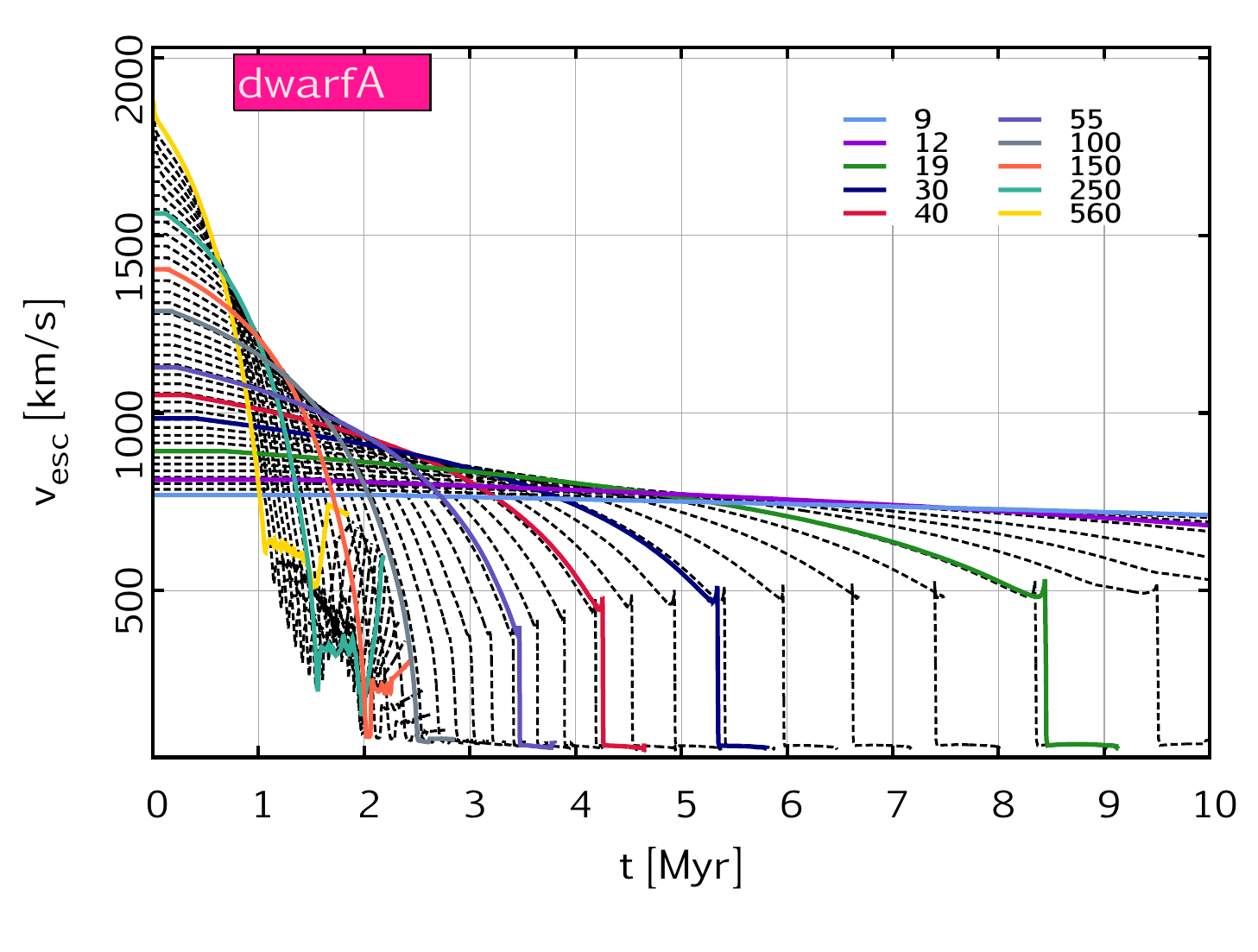}\hspace{2pt}
        \includegraphics[width=\ratioo\columnwidth,page=3]{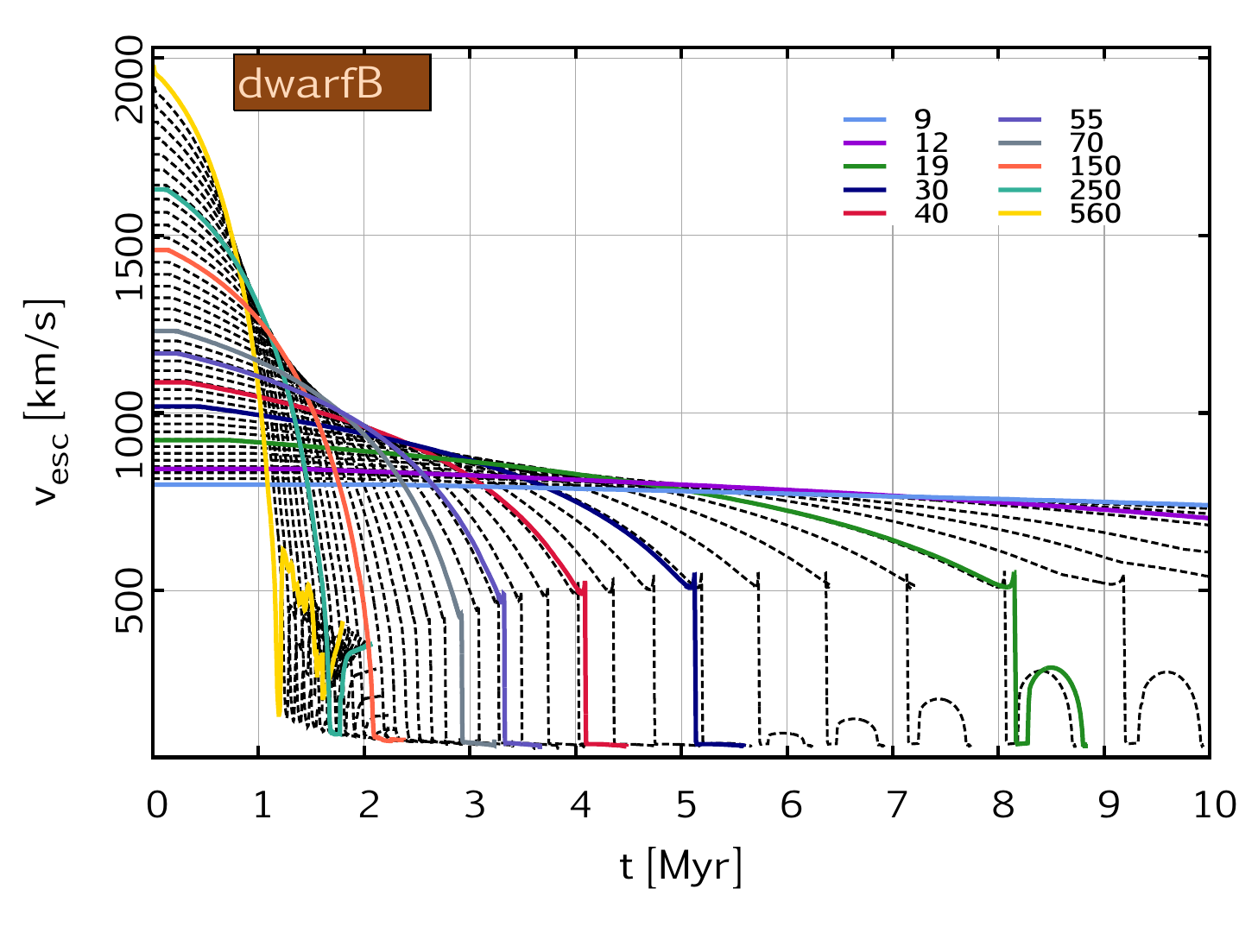}\hspace{2pt}
        \includegraphics[width=\ratioo\columnwidth,page=3]{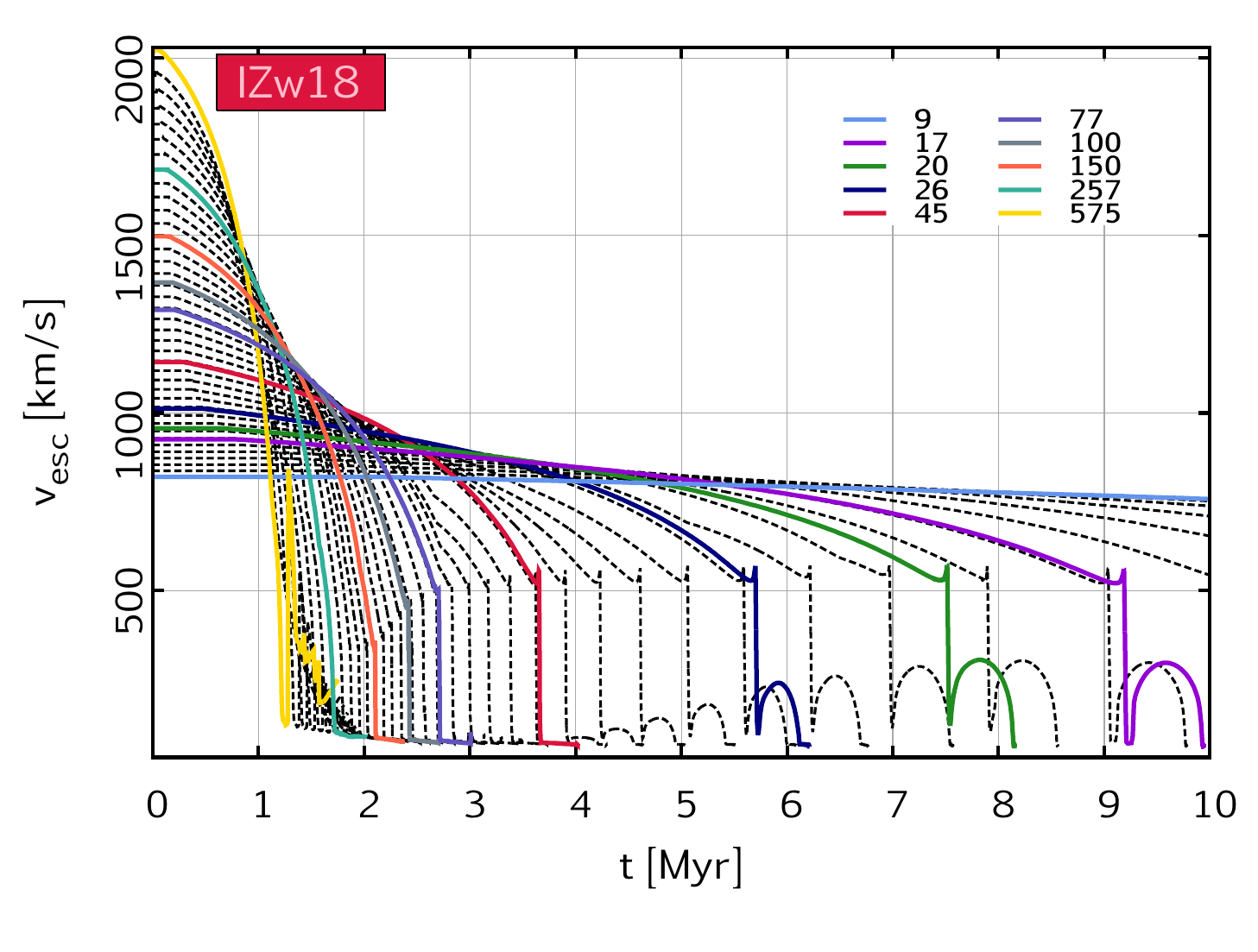}\vspace{2pt} \\
        \includegraphics[width=\ratioo\columnwidth,page=3]{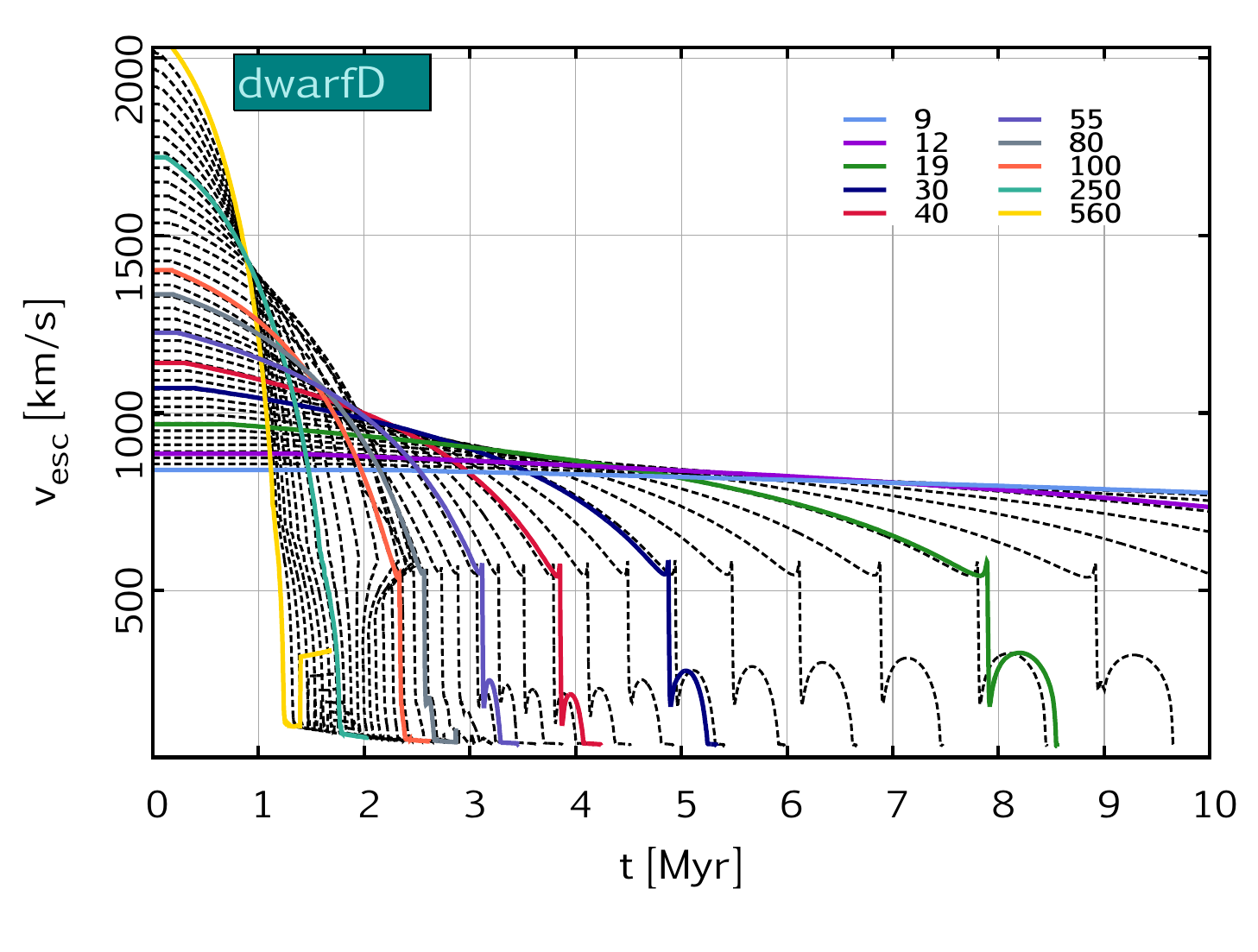}\hspace{2pt}
        \includegraphics[width=\ratioo\columnwidth,page=3]{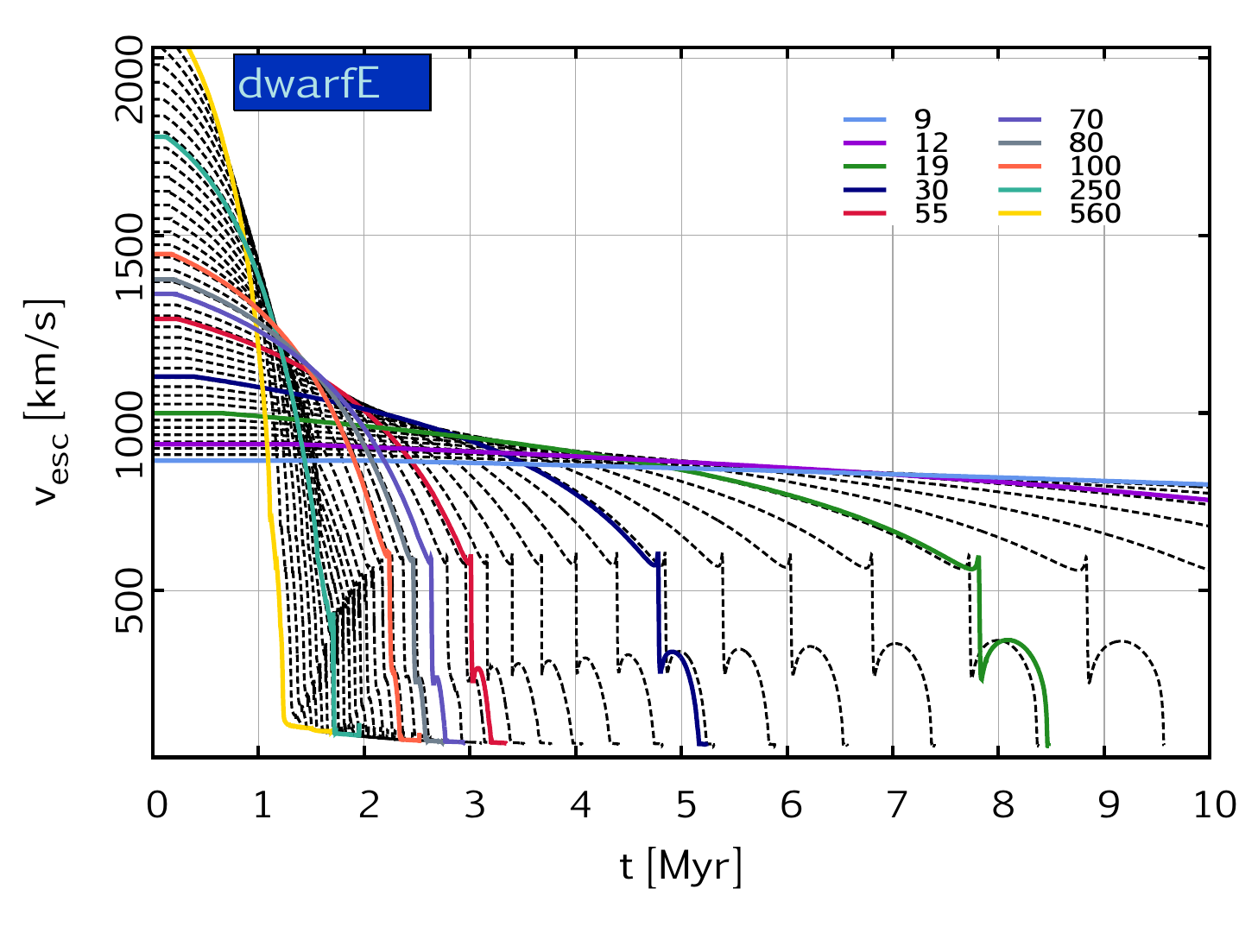}\hspace{2pt}
        \includegraphics[width=\ratioo\columnwidth,page=3]{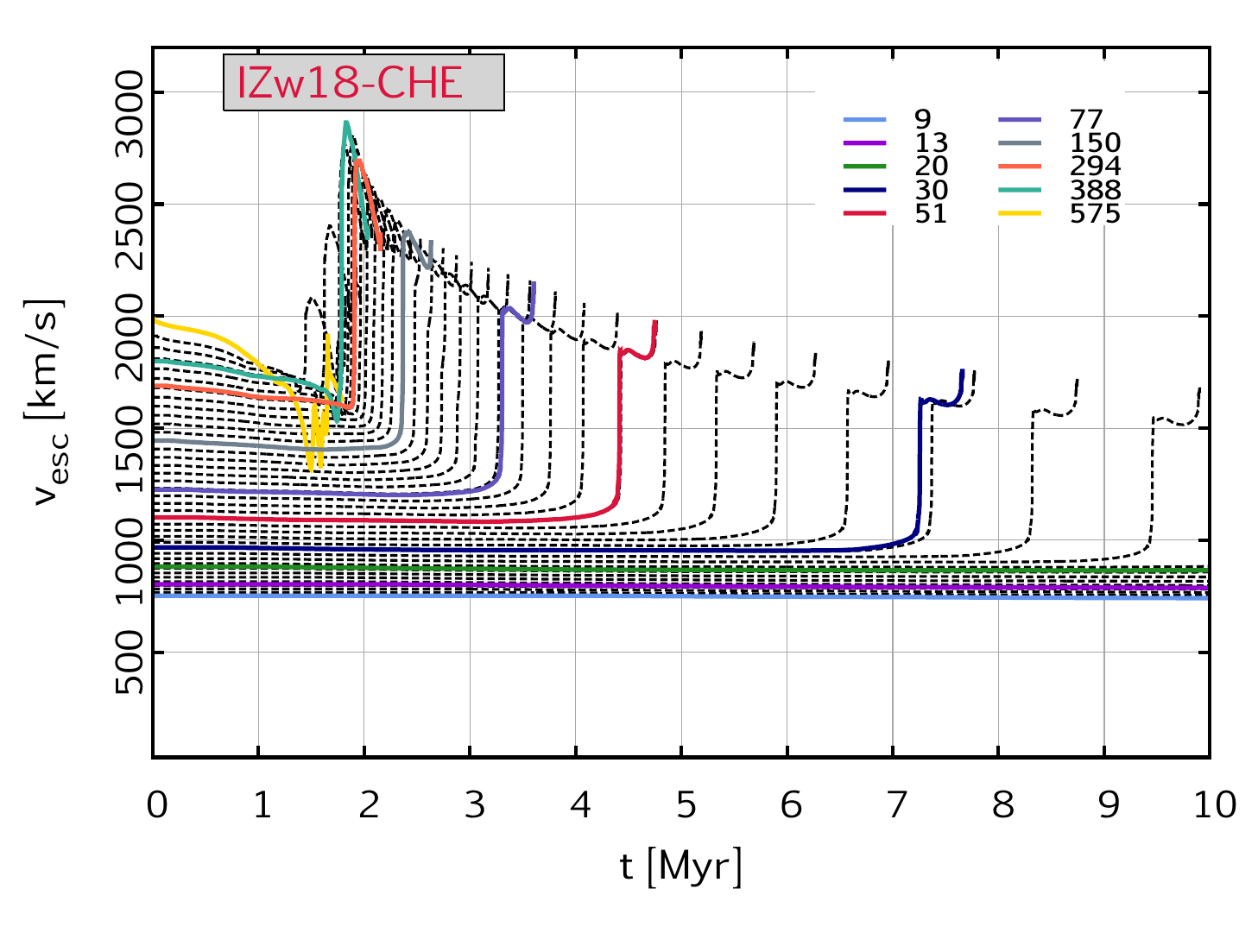}
        \caption{Time evolution of the escape velocity; cf. Eq.~(\ref{eq:wind}). The initial mass is color-coded (between 9 and $\sim$500~M$_{\odot}$). The dashed black lines represent interpolated tracks (up to 500~M$_{\odot}$); cf. Figs.~\ref{fig:HRD} and~\ref{fig:Mdot}. While here only shown until 10~Myr, the files published in the BoOST project contain data until the end of the lifetimes of the longest-living model in the population ($\sim$30~Myr).}\label{fig:vwind}
\end{figure*}

\begin{figure*}\centering
        \includegraphics[width=\ratioo\columnwidth,page=4]{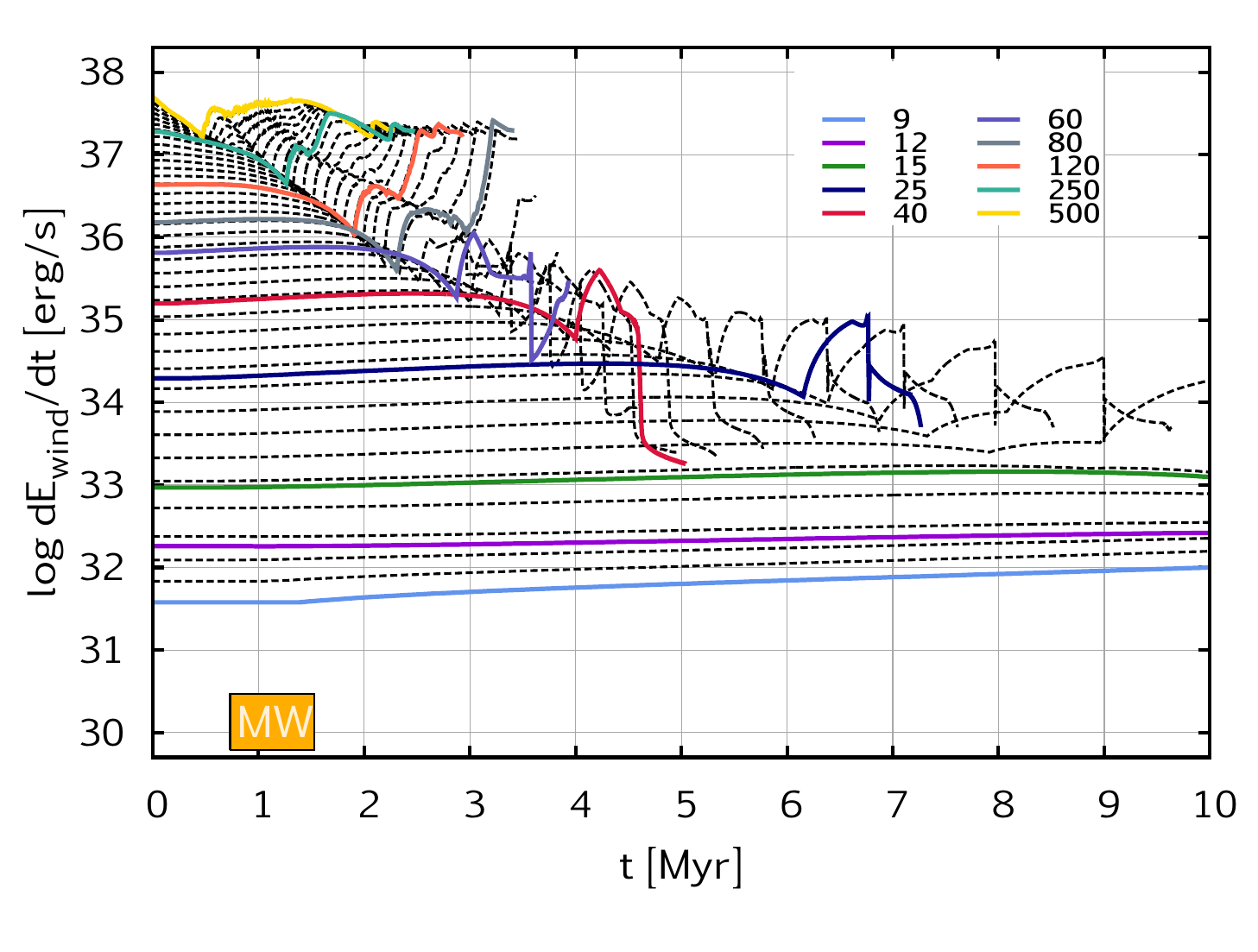}\hspace{2pt}
        \includegraphics[width=\ratioo\columnwidth,page=4]{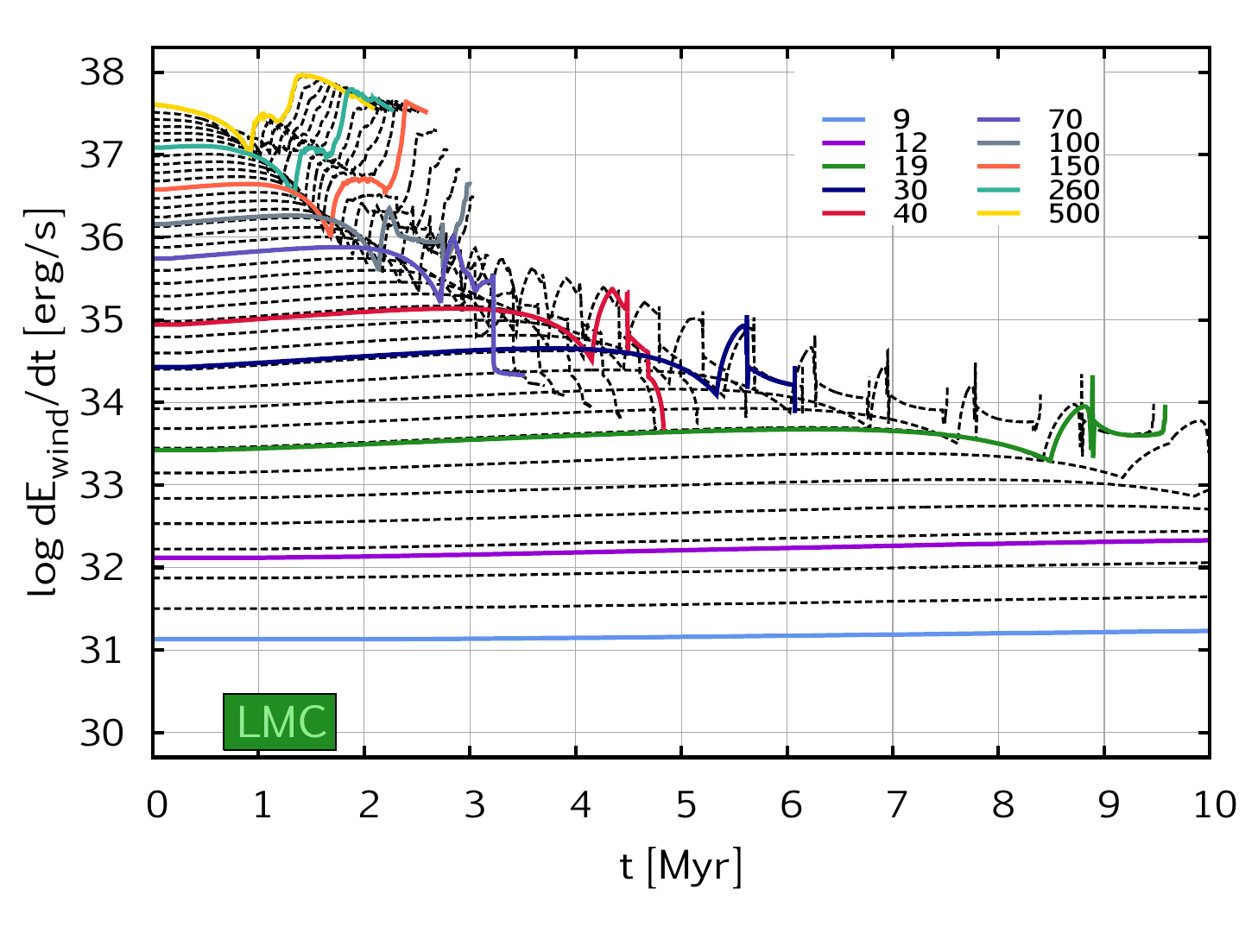}\hspace{2pt}
        \includegraphics[width=\ratioo\columnwidth,page=4]{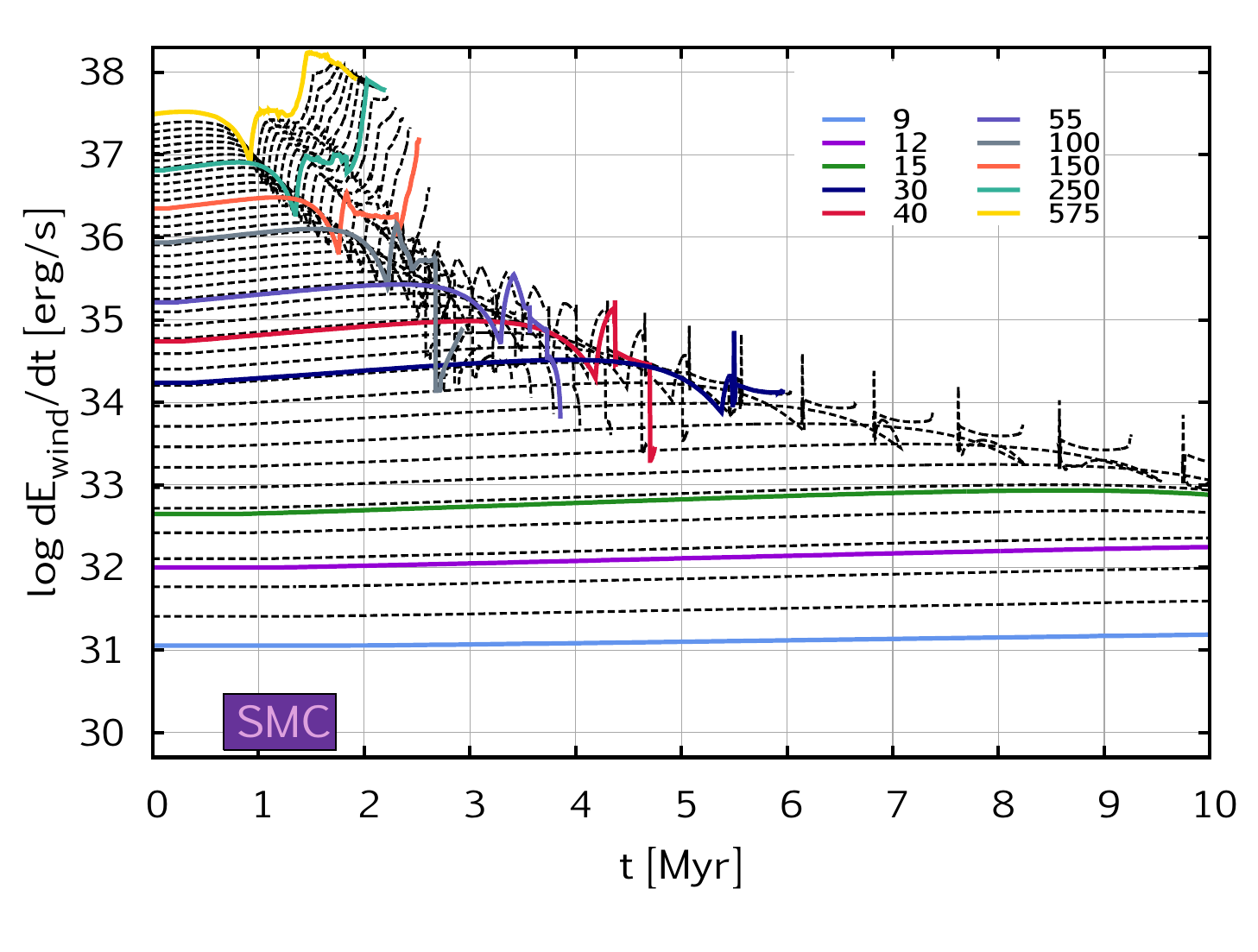}\vspace{2pt} \\
        \includegraphics[width=\ratioo\columnwidth,page=4]{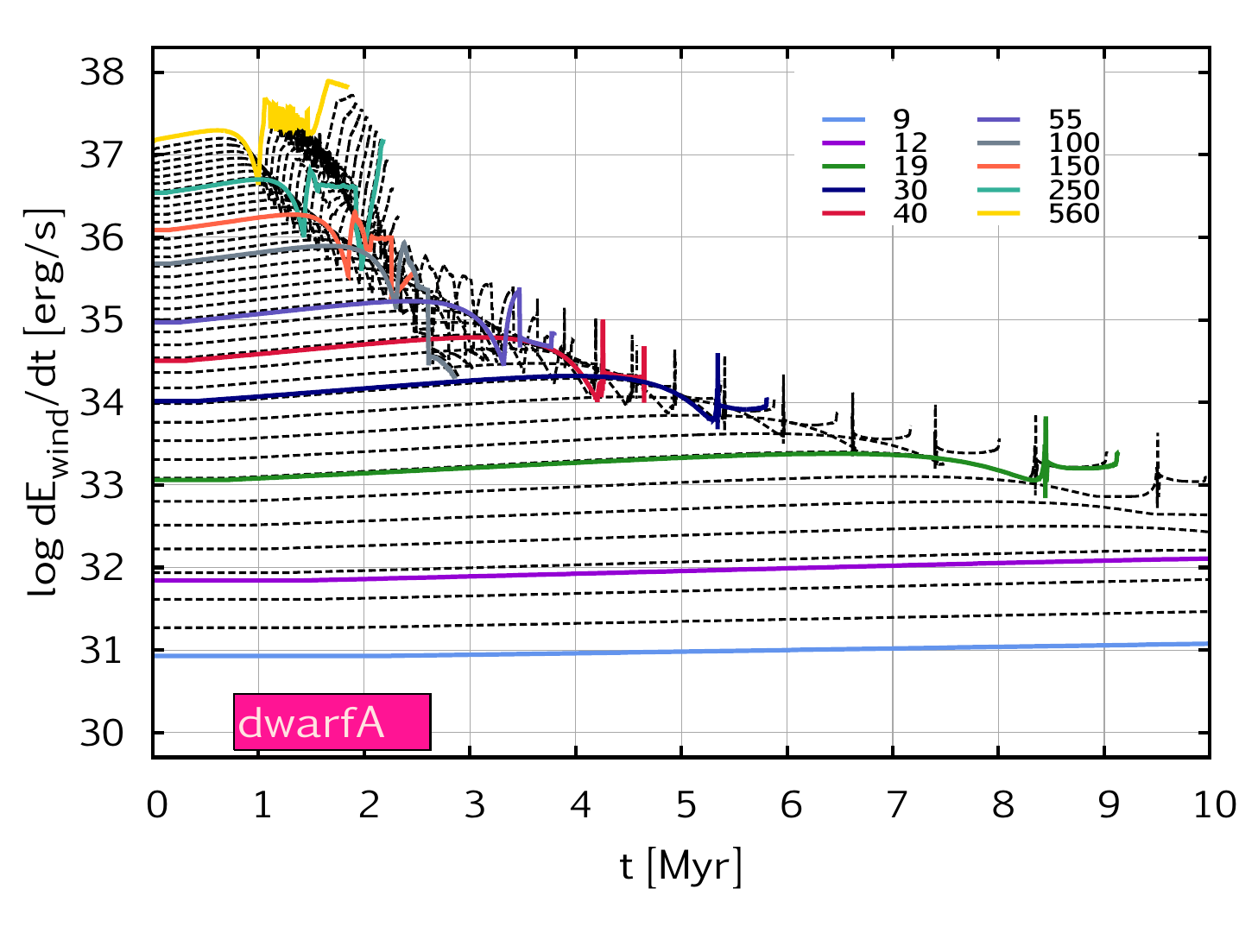}\hspace{2pt}
        \includegraphics[width=\ratioo\columnwidth,page=4]{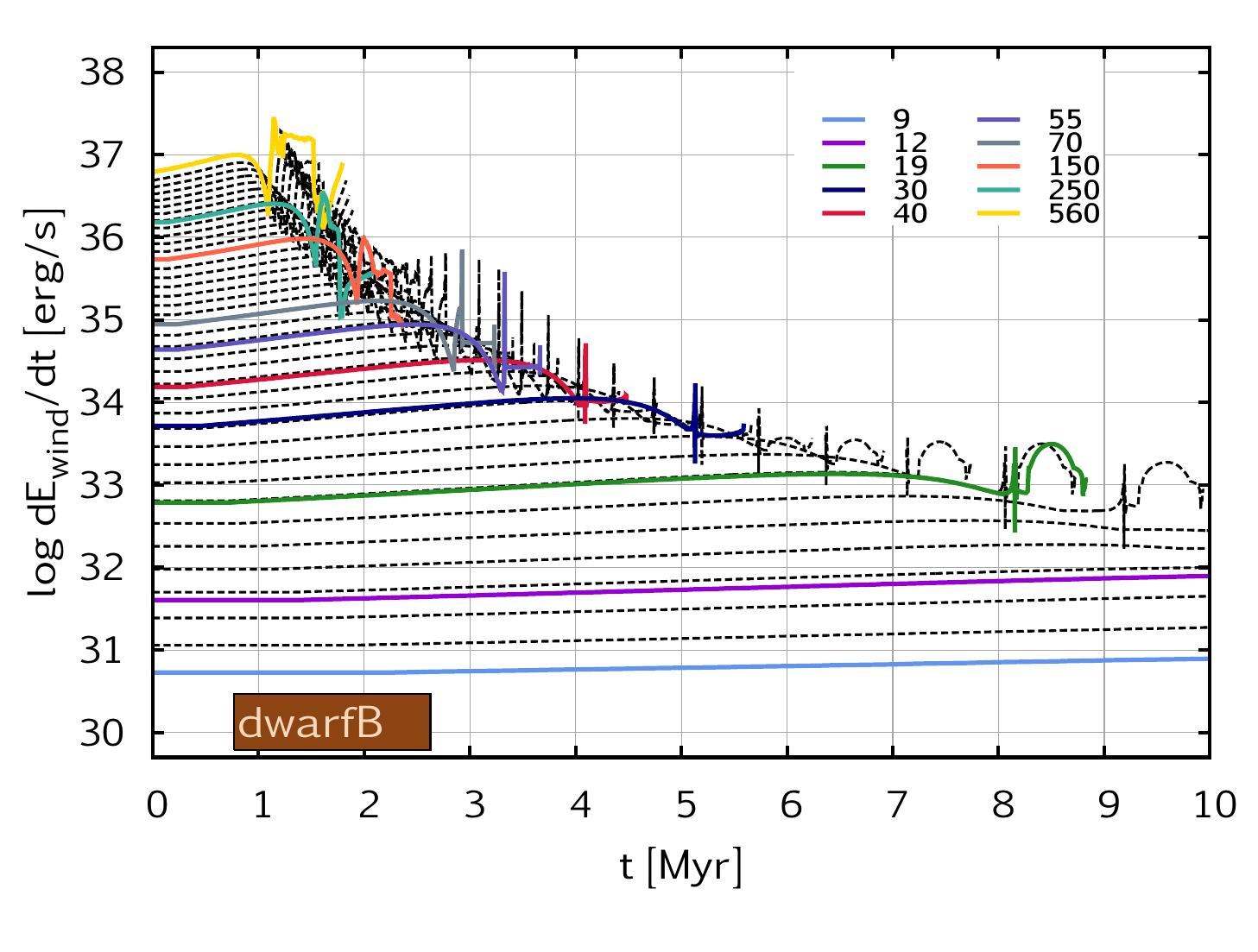}\hspace{2pt}
        \includegraphics[width=\ratioo\columnwidth,page=4]{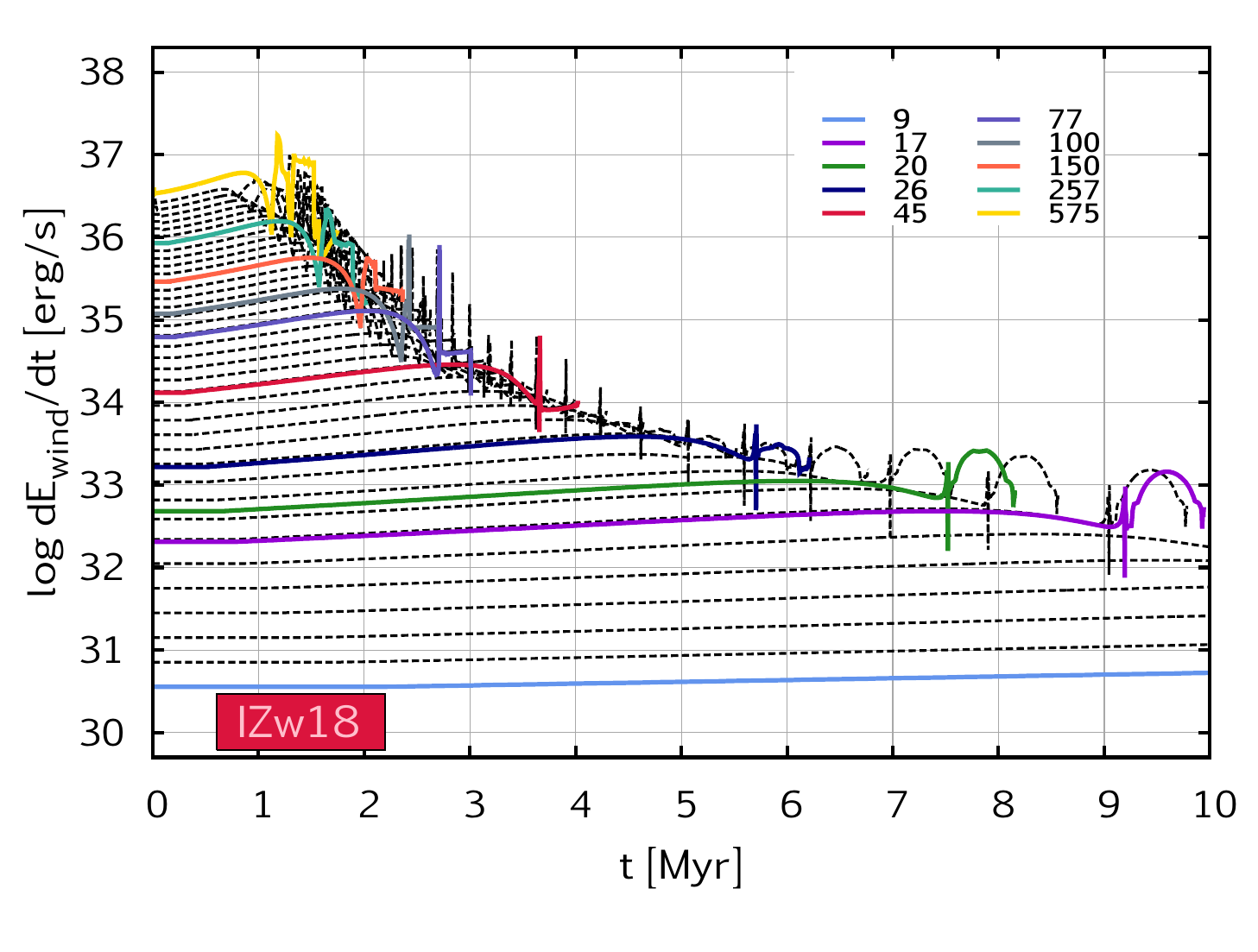}\vspace{2pt} \\
        \includegraphics[width=\ratioo\columnwidth,page=4]{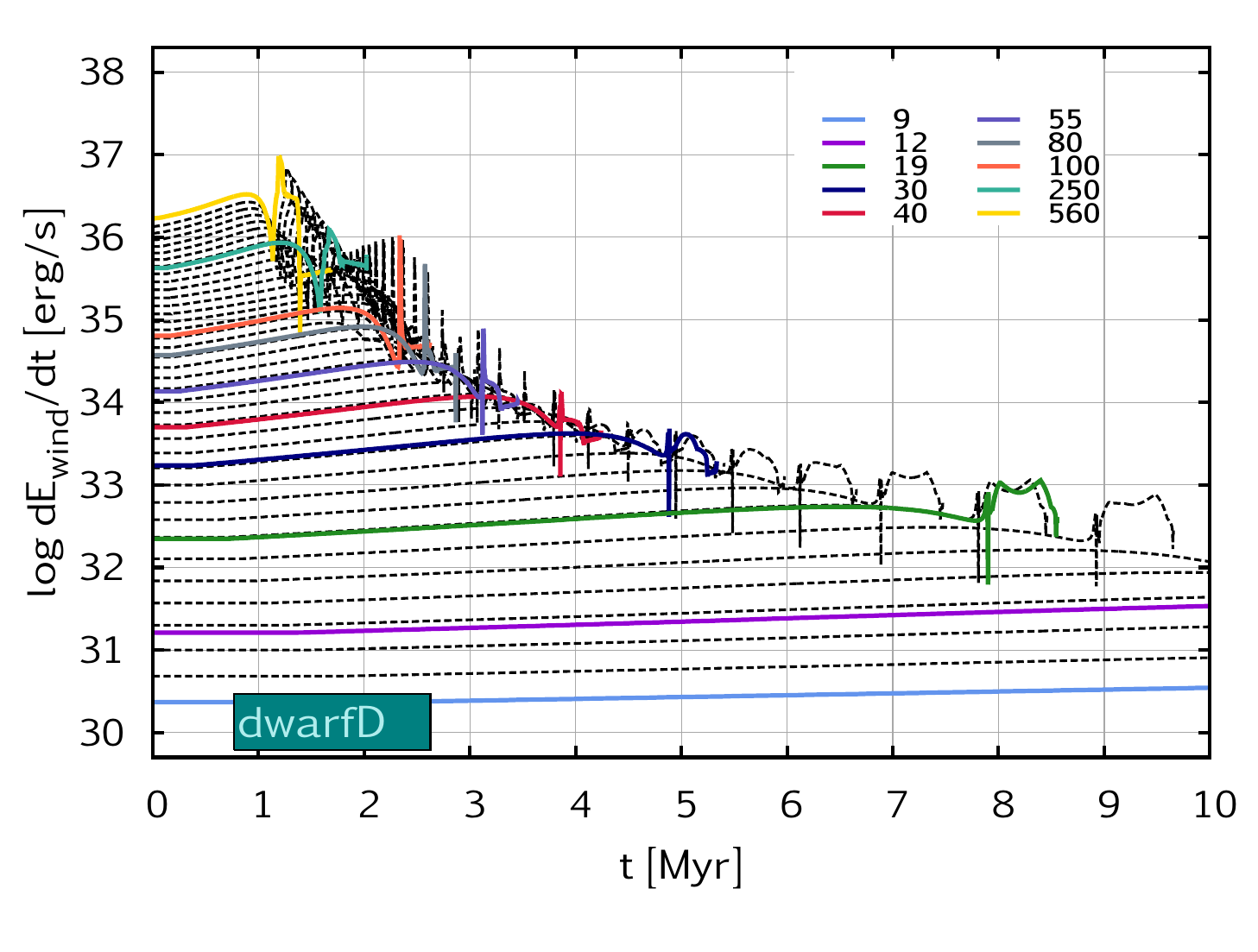}\hspace{2pt}
        \includegraphics[width=\ratioo\columnwidth,page=4]{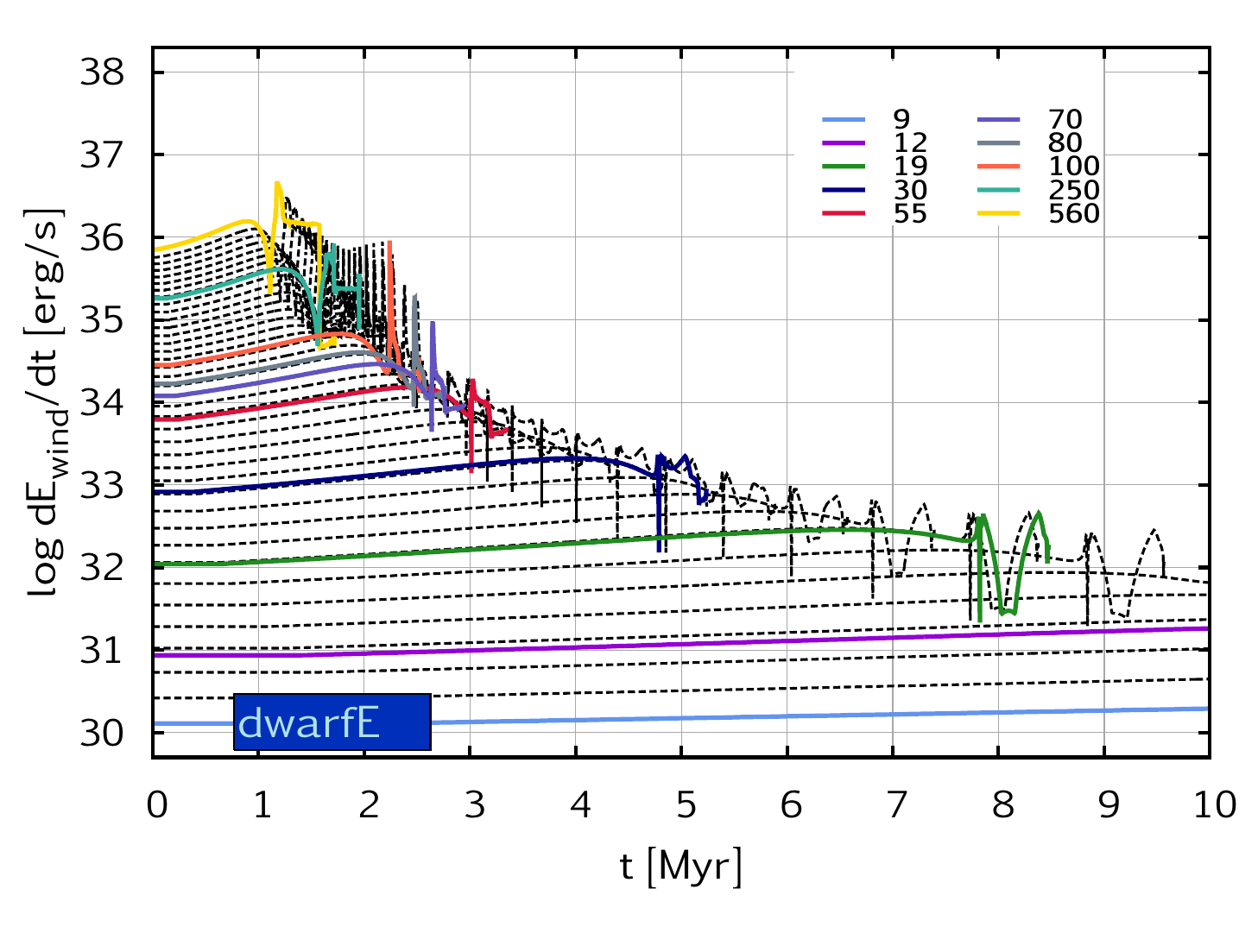}\hspace{2pt}
        \includegraphics[width=\ratioo\columnwidth,page=4]{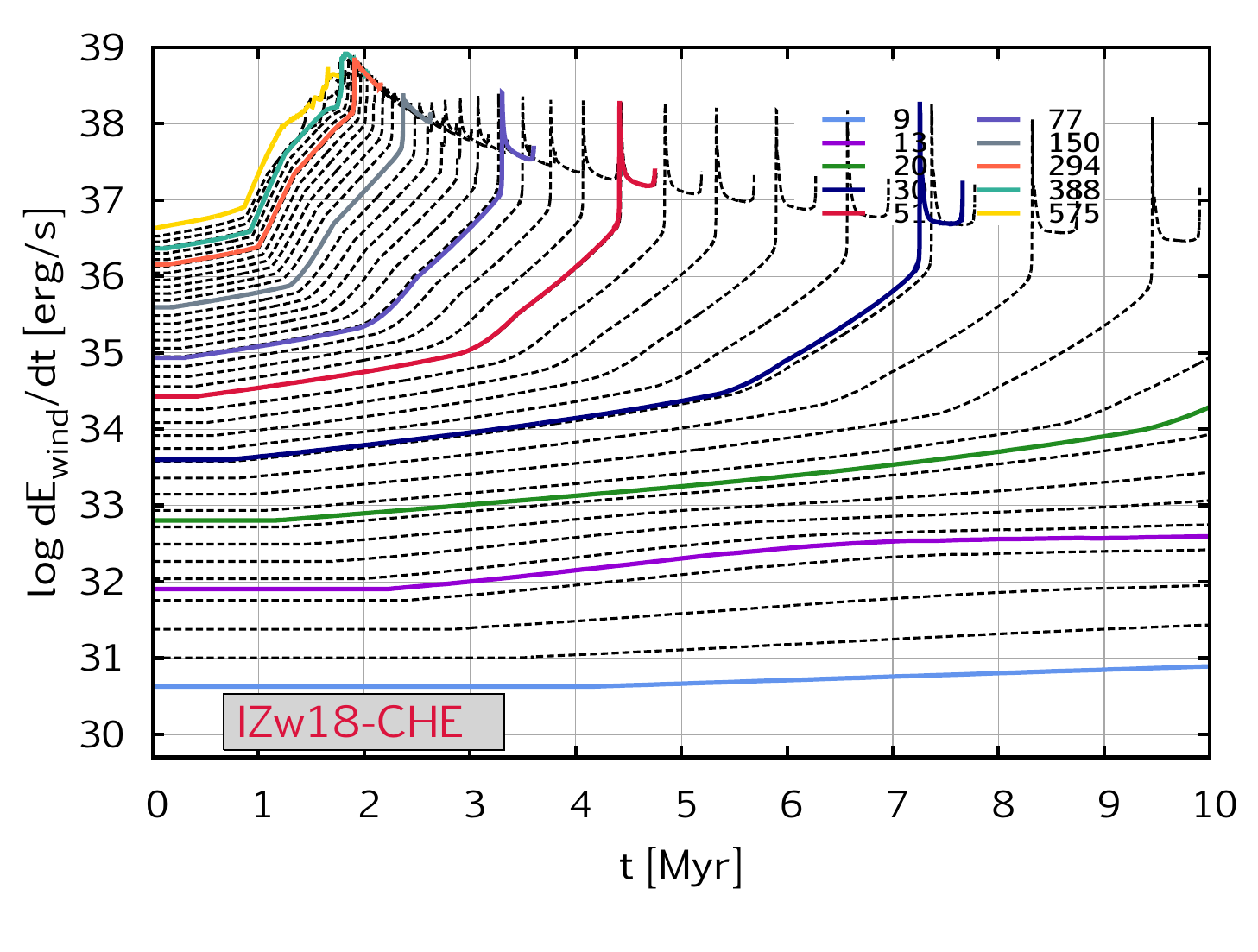}
        \caption{Time evolution of the wind kinetic energy rate. The initial mass is color-coded (between 9 and $\sim$500~M$_{\odot}$). The dashed black lines represent interpolated tracks (up to 500~M$_{\odot}$); cf. Figs.~\ref{fig:HRD}\,--\,\ref{fig:vwind}. While here only shown until 10~Myr, the files published in the BoOST project contain data until the end of the lifetimes of the longest-living model in the population ($\sim$30~Myr).}\label{fig:Edot}
\end{figure*}

Hertzsprung--Russell diagrams of all the BoOST stellar models are shown in Fig.~\ref{fig:HRD}. Additionally, Figures~\ref{fig:Mdot}--\ref{fig:Edot} present various diagnostic diagrams showing important properties of stellar feedback such as mass-loss rate, wind velocity, and kinetic energy of the winds. 
As expected, both the zero-age main sequence and the supergiant branch shift to higher effective temperatures when the metallicity is lower due to the lower opacities. Furthermore, there are two features that gradually change from high to low metallicity: very luminous supergiants at high masses, 
and blue supergiants at lower masses (in the blue loop of the evolutionary models). 

{The luminous supergiants are the natural result of envelope inflation in these massive stars. As explained by \citet{Sanyal:2015}, for example, in stellar models close to the Eddington limit \citep{Langer:1997}, density and pressure-inversion regions can develop in the outer regions. In the absence of any user intervention (cf. the discussion in Sect.~\ref{sec:eddington}), the code deals with this by increasing the physical extent of the star, that is, by inflating the envelope and thus producing luminous supergiants (cf. also Sect.~5 of \citealt{Szecsi:2015} as well as \citealt{Sanyal:2017}). These special supergiants have been shown to possibly contribute to the formation of globular clusters (in two different scenarios, the first presented by \citealt{Szecsi:2018} and the second in \citealt{Szecsi:2019}). Thus, with the metallicity coverage the BoOST grids provide, the door opens to studying the role of these supergiants in cluster and star formation research and beyond.
}

{The blue loop is known to be sensitive to any change in physical parameters during the evolution. For example, \citet{Schootemeijer:2019} presented stellar models with an SMC composition between 9$-$100~M$_{\odot}$ using various semiconvective and overshooting parameters. They showed that the presence or absence of blue loops depends on these parameters, as well as on the applied rotational velocity. In short, the phenomenon was found to be tightly linked to internal mixing. They did not study the effect of metallicity, but our models show that this influences blue loops as well. \textbf{With both semiconvection and overshooting fixed} ($\alpha_{\rm sc}$~$=$~1.0 and $\alpha_{\rm over}$~$=$~0.335; cf. Sect.~\ref{sec:ingredients}), and with the same initial rotational velocity chosen for all the models (100~km~s$^{-1}$), we find no blue loops at metallicities above that of the SMC (consistent with \citealt{Schootemeijer:2019}), while with decreasing metallicity, the feature becomes increasingly prominent. This may provide a way to improve our models in the future: because blue supergiants are observed in the SMC \citep{Humphreys:1991,Kalari:2018}, a next version of the BoOST grids, for example, may be computed with testing a higher semiconvective parameter.}

\section{Interpolation and population synthesis: Presenting \textsc{synStars}}\label{sec:interpol}      

{
        We have developed the simple stellar population synthesis code {\sc synStars} written in Python with libraries {\sc NumPy} \citep{numpy} and {\sc SciPy} \mbox{\citep{2020SciPy}}. In addition to the actual population synthesis (to be discussed in Sect.~\ref{sec:popsyn}), {\sc synStars} is also able to interpolate between our precomputed stellar models (those presented in Sect.~\ref{sec:stellarmodels}). Below we describe how the interpolation of the tracks is implemented (including preprocessing in Sect.~\ref{sec:EEPs}) and what the published tables contain in Sect.~\ref{sec:interpolation}.
        
        \subsection{Finding equivalent evolutionary phases} 
        \label{sec:EEPs}
        
A stellar population consists of stars of varying initial masses. \textbf{To construct them, we interpolate} between the precomputed stellar models from Sect.~\ref{sec:stellarmodels}. During the process, it is important to ensure that the resulting set of interpolated tracks shows only gradual changes.
However, the evolutionary models of stars vary significantly between non-neighboring masses, especially within the wide mass range covered in this project. For example, a star with 9~M$_{\odot}$ has a different evolutionary path in the HR diagram from a star with 60~M$_{\odot}$ or from a star with 500~M$_{\odot}$. To correctly interpolate the evolutionary sequence of a star using sequences of neighboring masses, it is thus common to determine EEPs \citep[ ][]{Prather:1976,Bergbusch:1992,Bergbusch:2001,Pietrinferni:2004} between stellar models. EEPs are identified by using evolutionary features that occur across the range of stellar tracks, such as the amount of hydrogen burned in the core, and they represent different phases during the evolution of a star.
        
For our models we identified seven EEPs (labeled A to G). Each EEP was further subdivided into an equal number of points, so that each phase of stellar evolution was represented by a fixed number of points (i.e., lines in the data file) to ensure that the $n$th point in one model has a comparable interpretation in another model. The seven EEPs for different stellar models are shown in Figs.~\ref{fig:fixpoints1}--\ref{fig:fixpoints3}. The method of identifying these EEPs is explained below. 
        
        \begin{figure*}\centering
                \includegraphics[width=\ratiooo\columnwidth,page=2,angle=270]{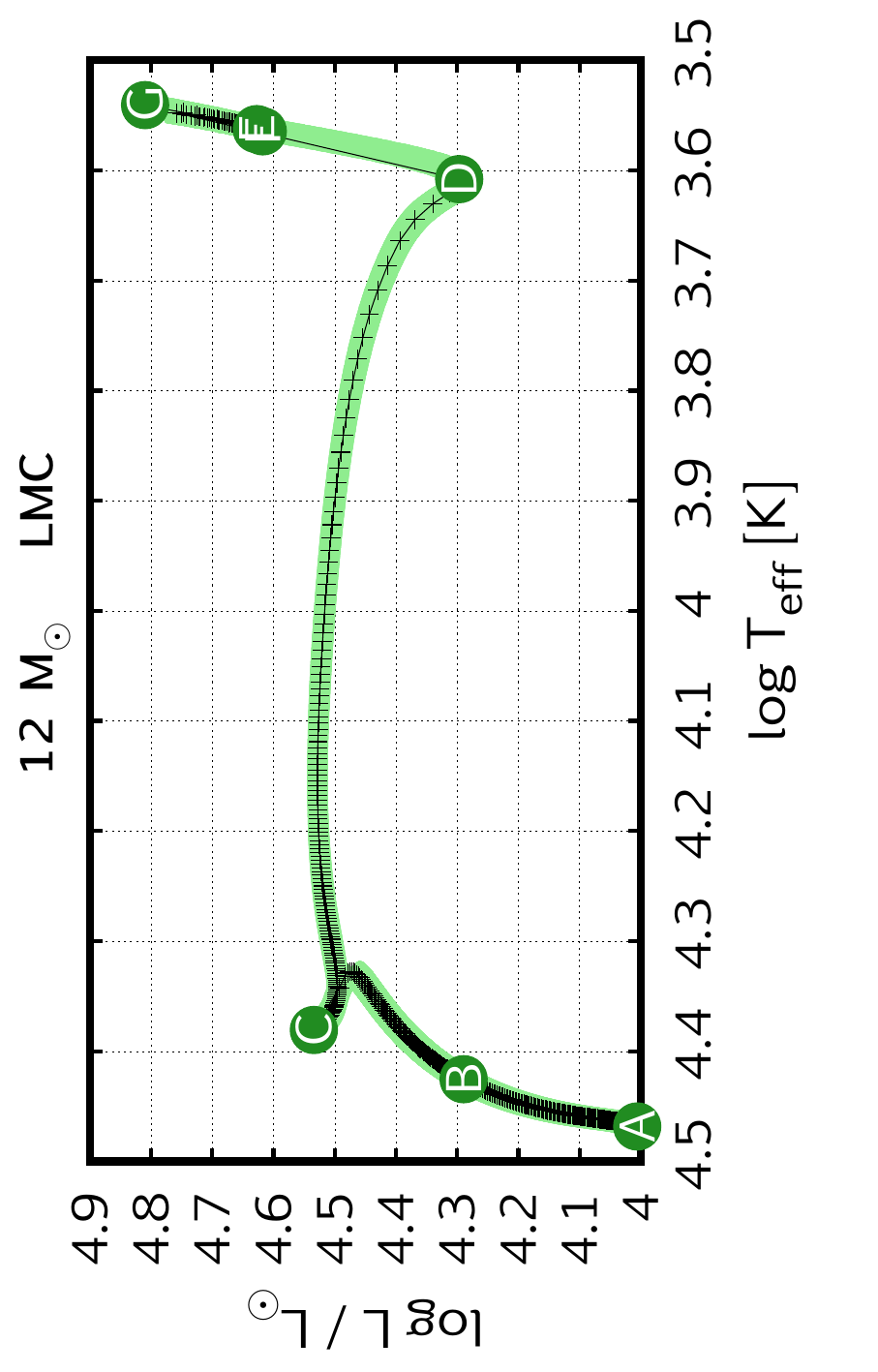}\hspace{2pt}
                \includegraphics[width=\ratiooo\columnwidth,page=2,angle=270]{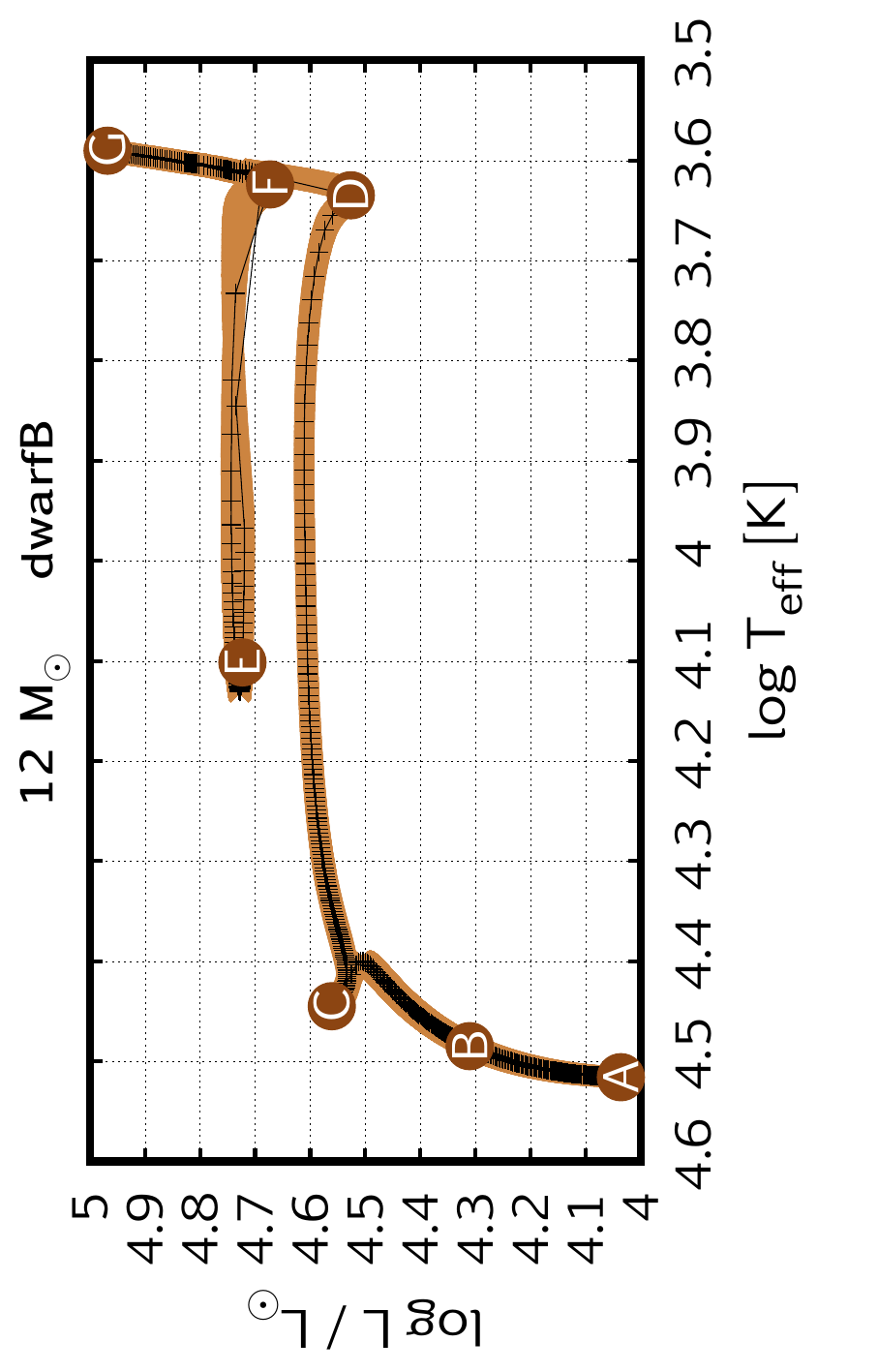}\vspace{-5pt}   \\
                \includegraphics[width=\ratiooo\columnwidth,page=2,angle=270]{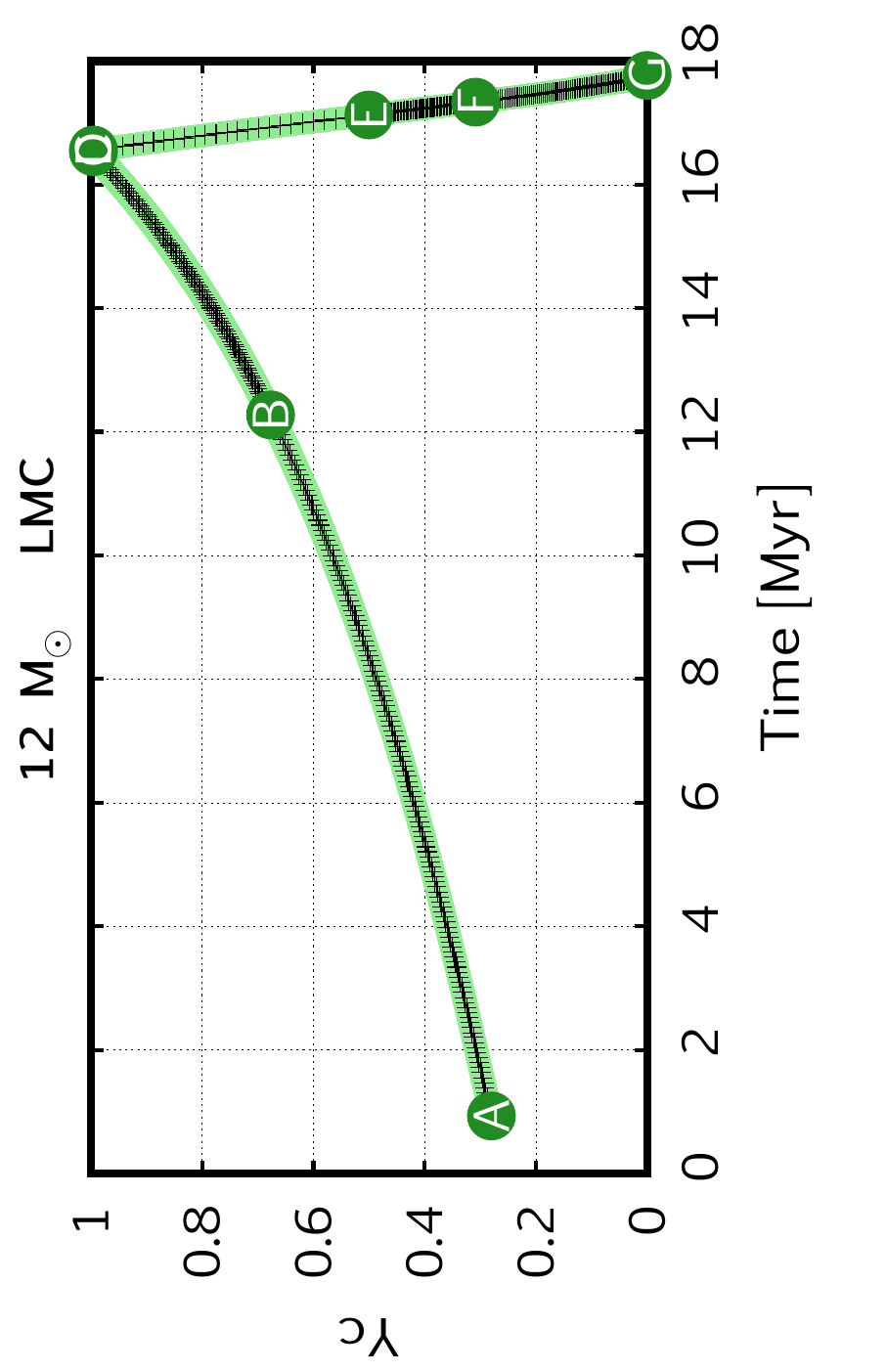}   \hspace{2pt}
                \includegraphics[width=\ratiooo\columnwidth,page=2,angle=270]{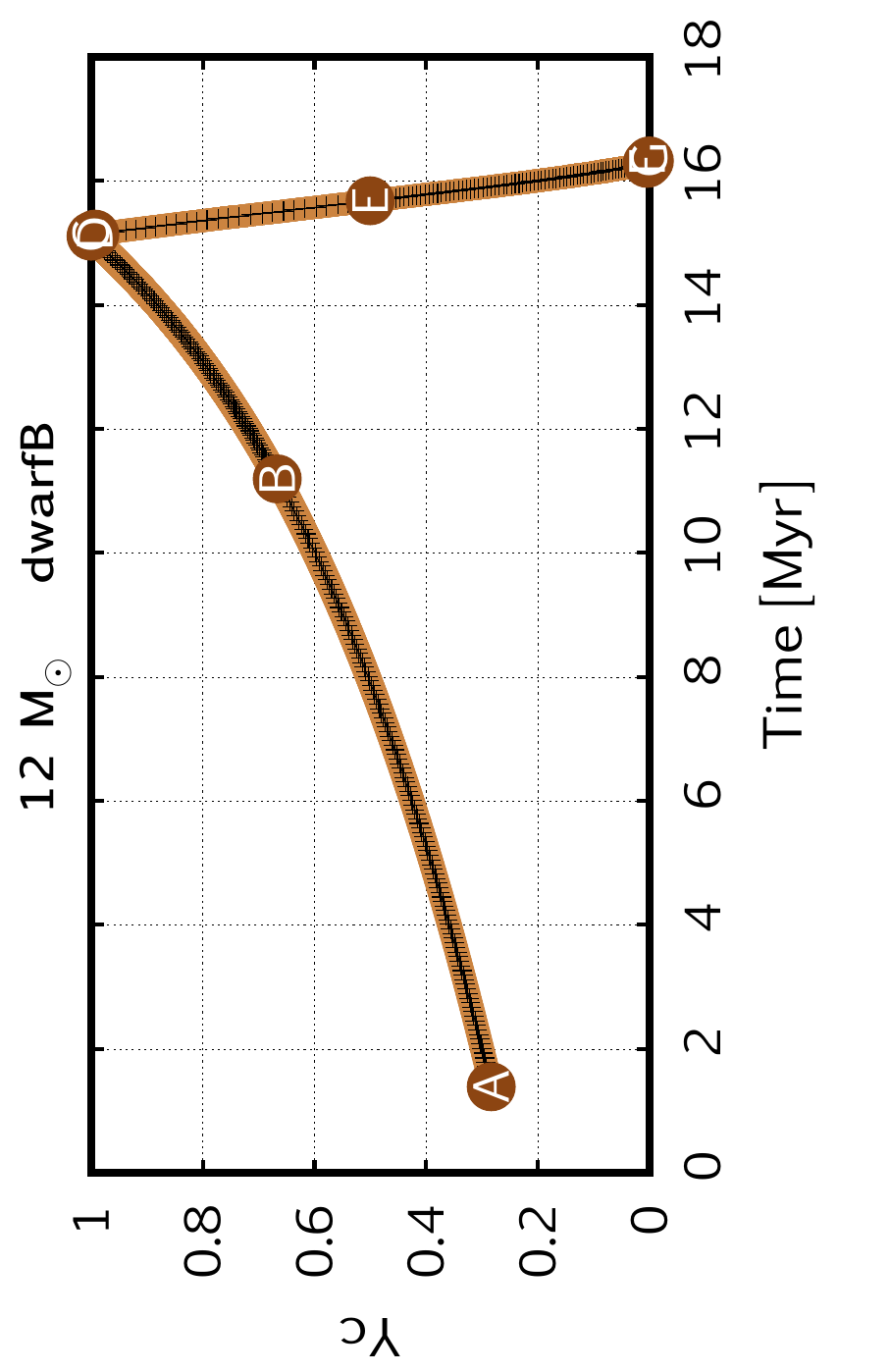}\vspace{-5pt}\\
                \includegraphics[width=\ratiooo\columnwidth,page=2,angle=270]{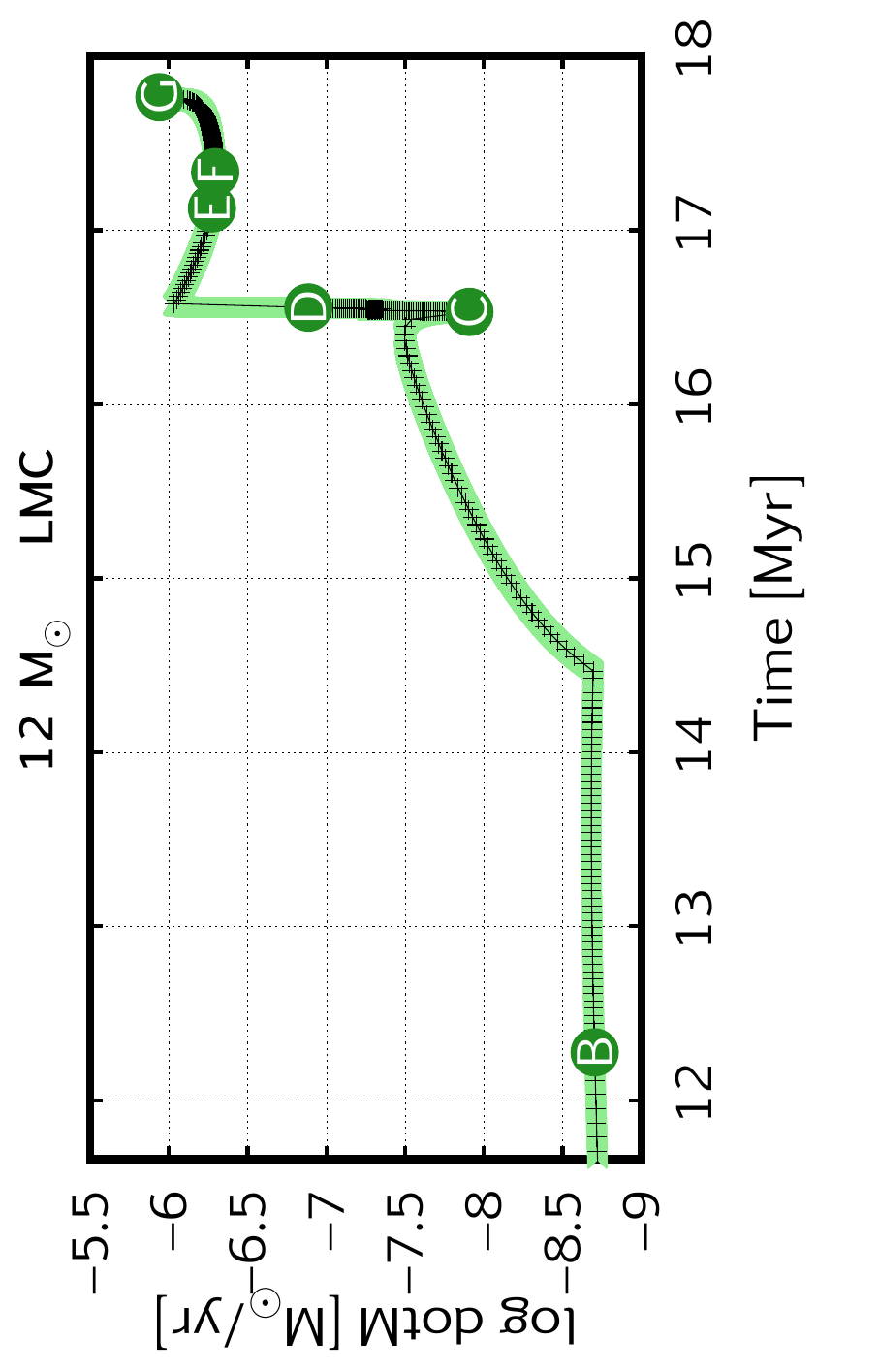}  \hspace{2pt}
                \includegraphics[width=\ratiooo\columnwidth,page=2,angle=270]{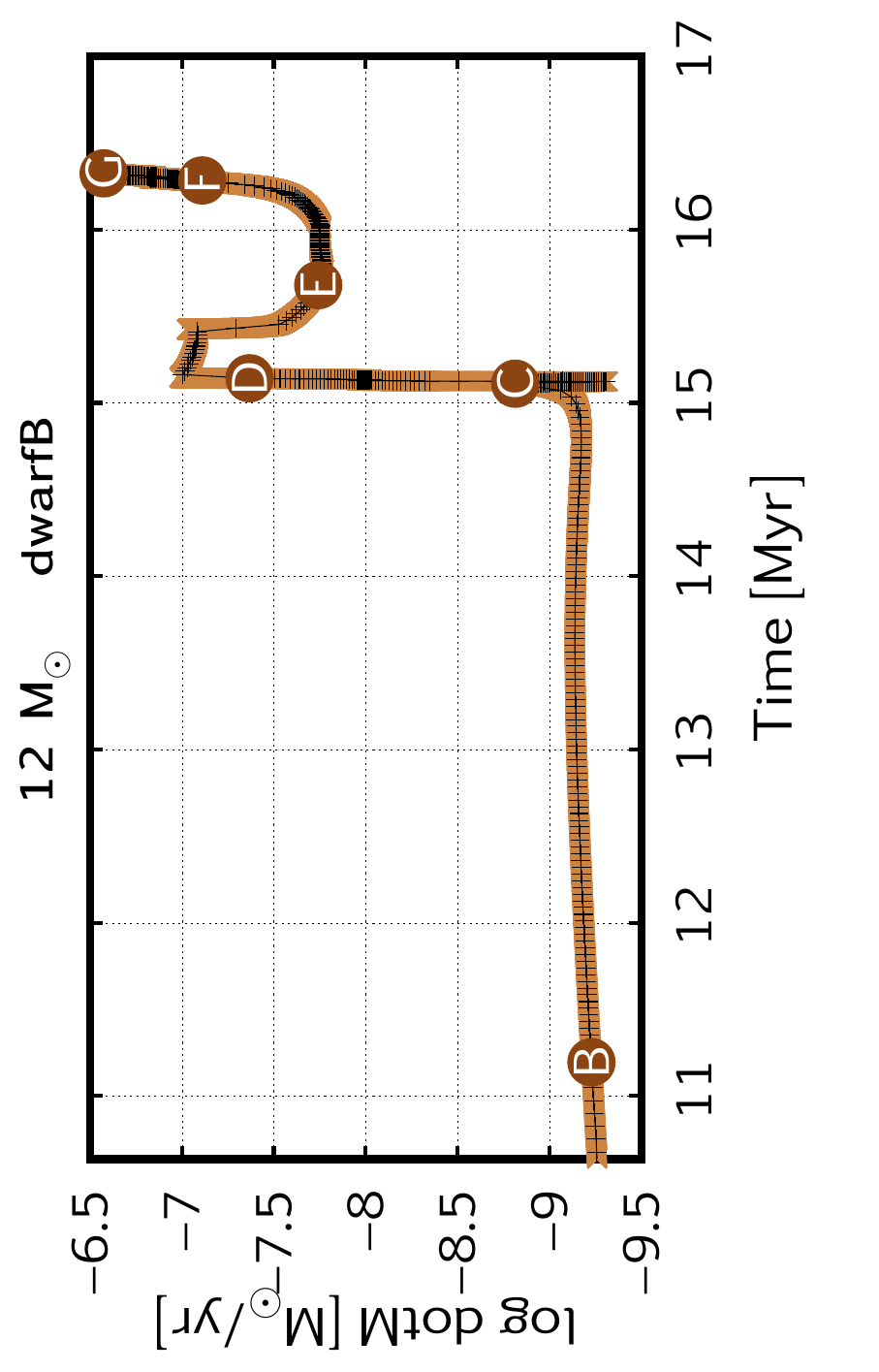}\vspace{-5pt}\\        
                \caption{Position of EEPs (i.e., fixed points in evolution) during the lifetime of some typical models. Colored lines represent the original output of our computations with the Bonn code, and black lines and crosses mark the filtered version (consistently containing the same number of dots between EEPs). The seven EEPs are labeled A--G; our method of choosing them is explained in the text.}
                \label{fig:fixpoints1}
        \end{figure*}
        
        \begin{figure*}\centering
                \includegraphics[width=\ratiooo\columnwidth,page=2,angle=270]{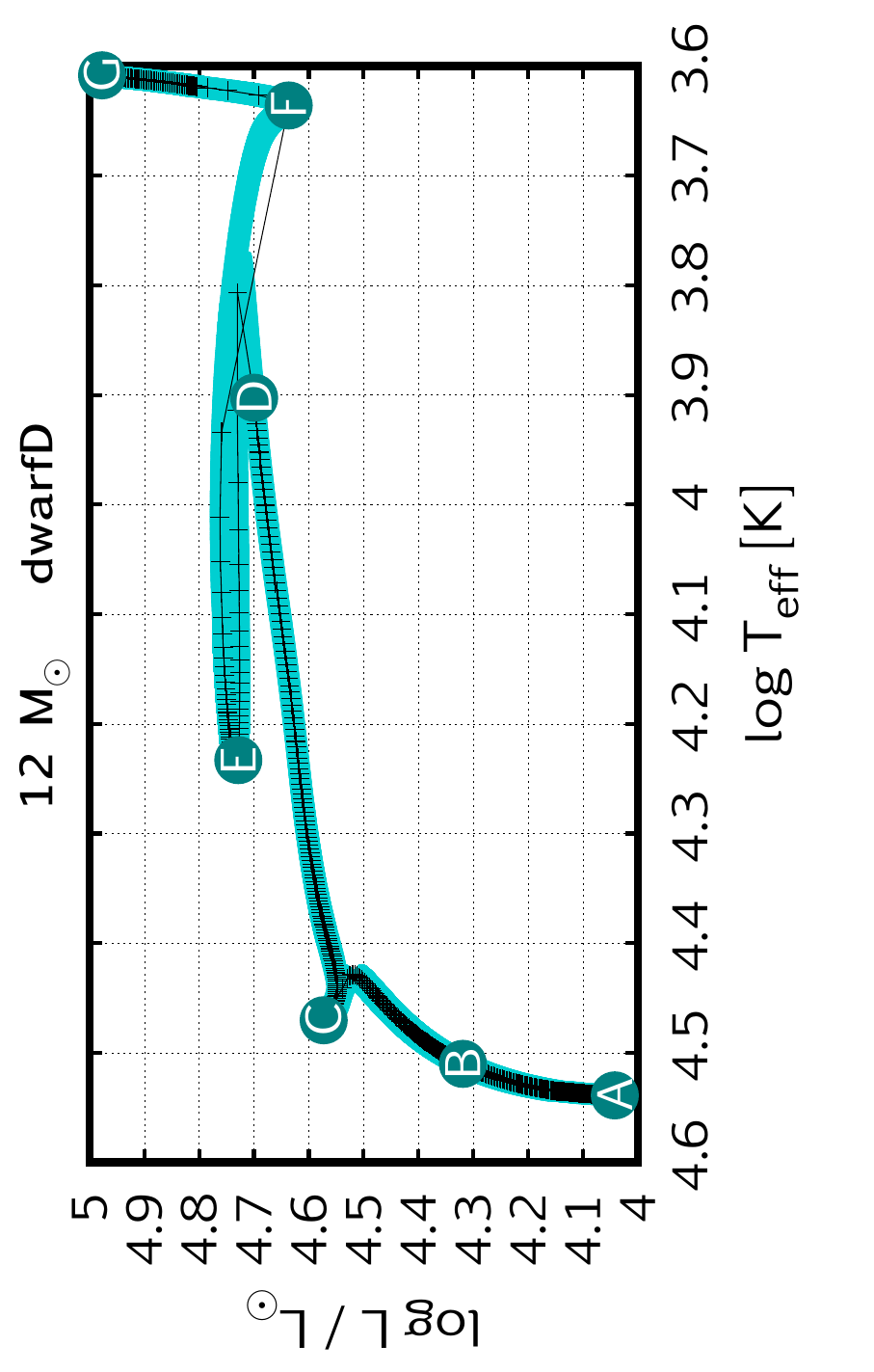}\hspace{2pt}
                \includegraphics[width=\ratiooo\columnwidth,page=8,angle=270]{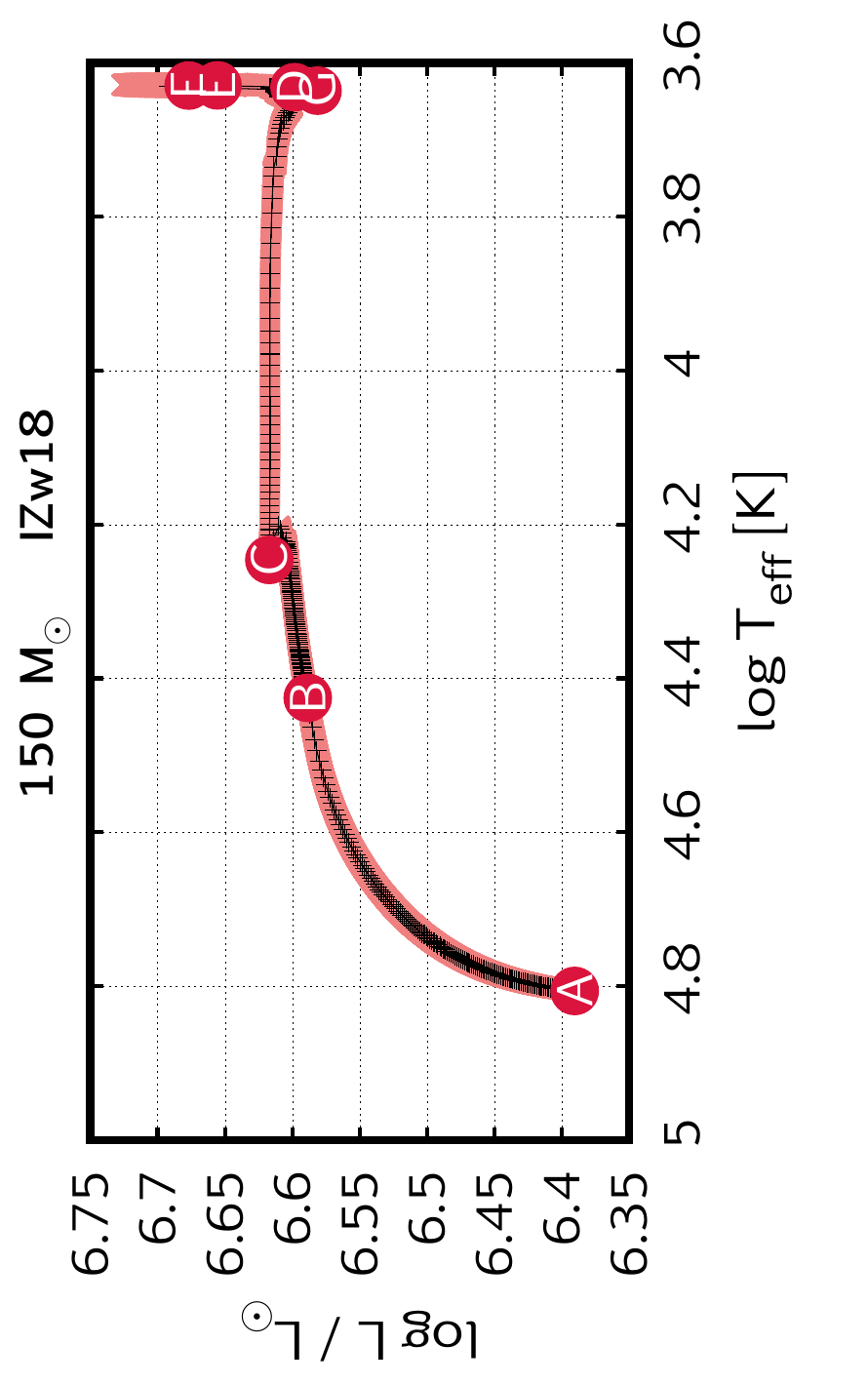}\vspace{-5pt}    \\
                \includegraphics[width=\ratiooo\columnwidth,page=2,angle=270]{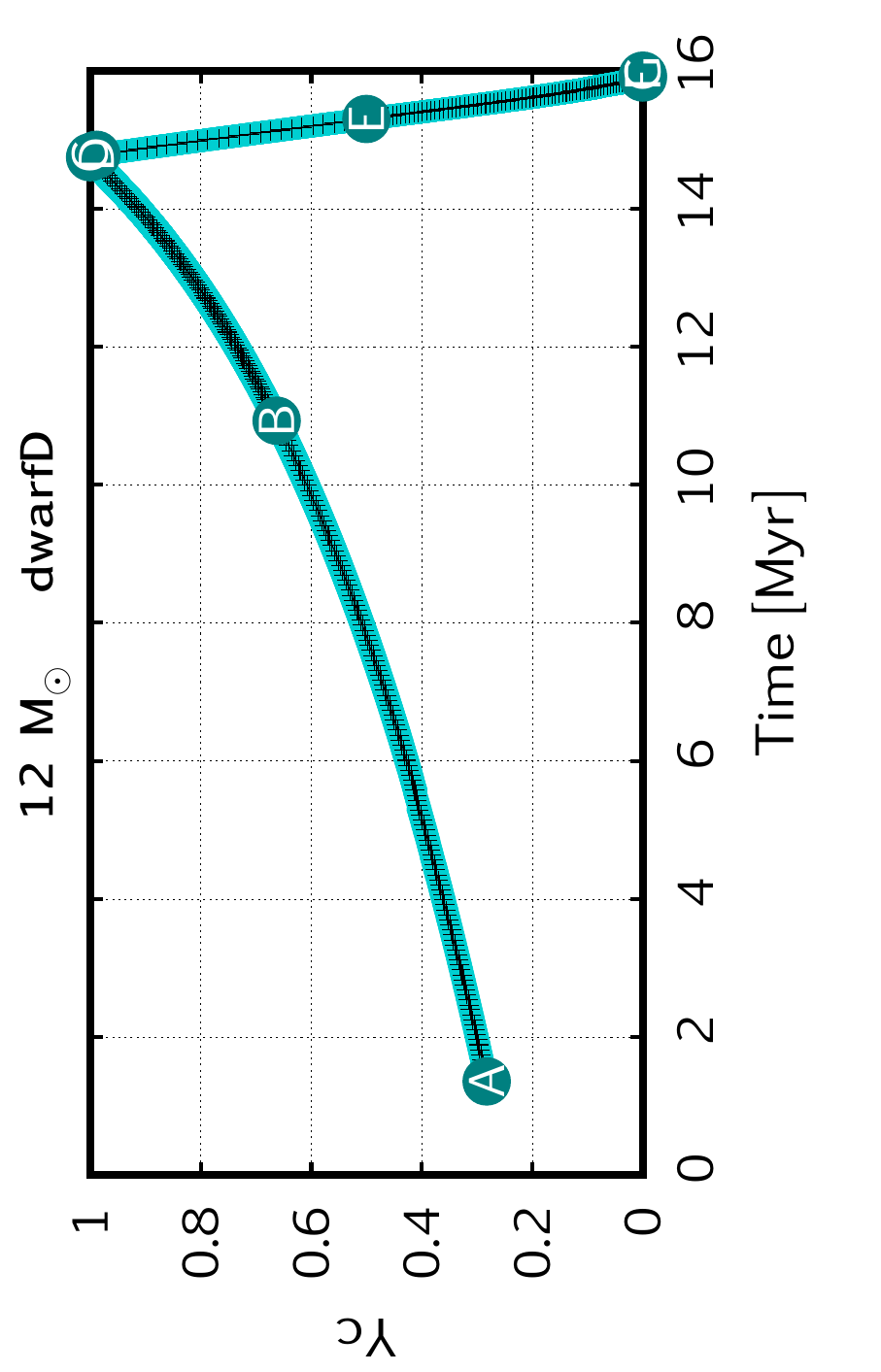}        \hspace{2pt}
                \includegraphics[width=\ratiooo\columnwidth,page=8,angle=270]{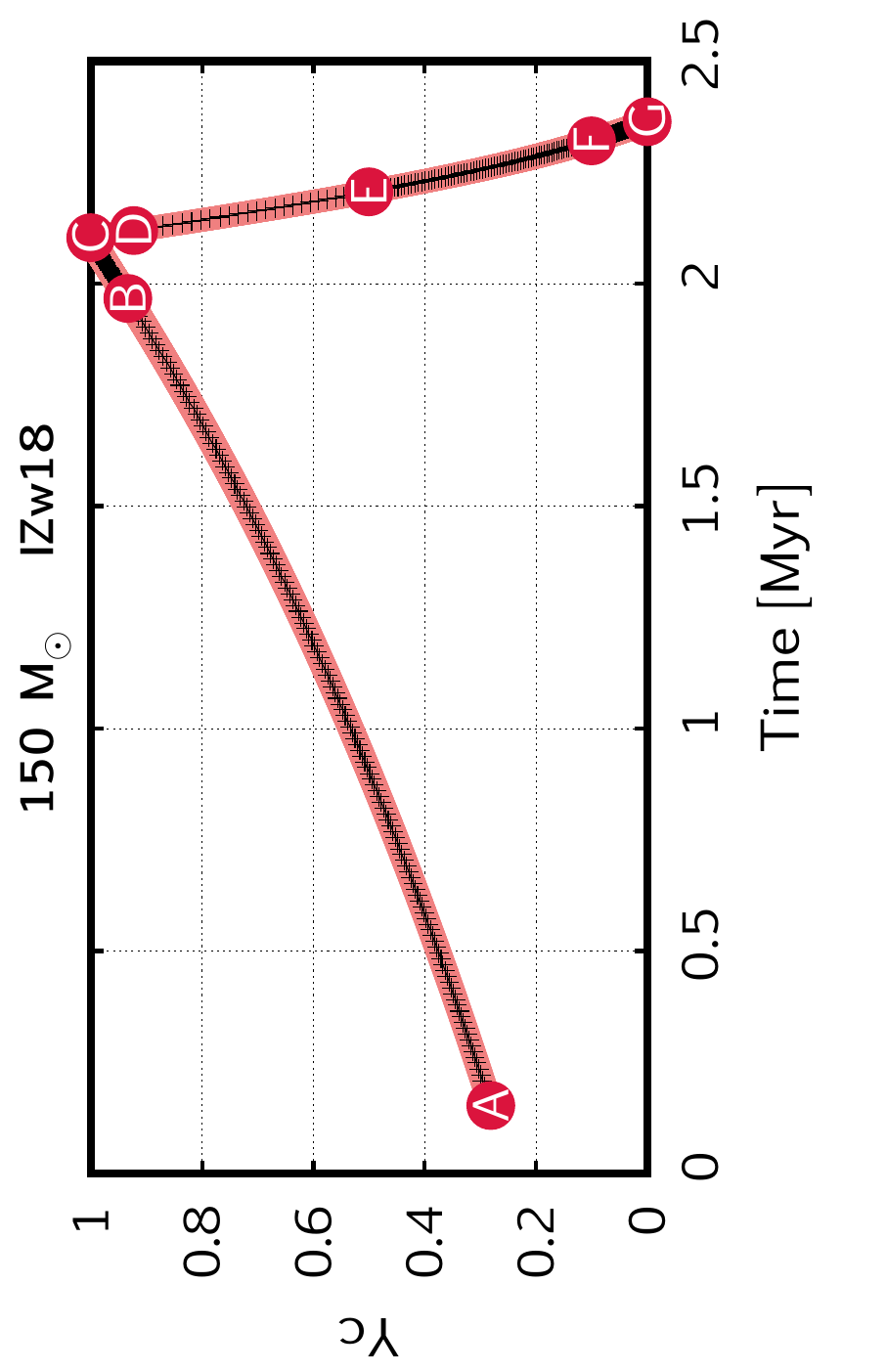}\vspace{-5pt}\\
                \includegraphics[width=\ratiooo\columnwidth,page=2,angle=270]{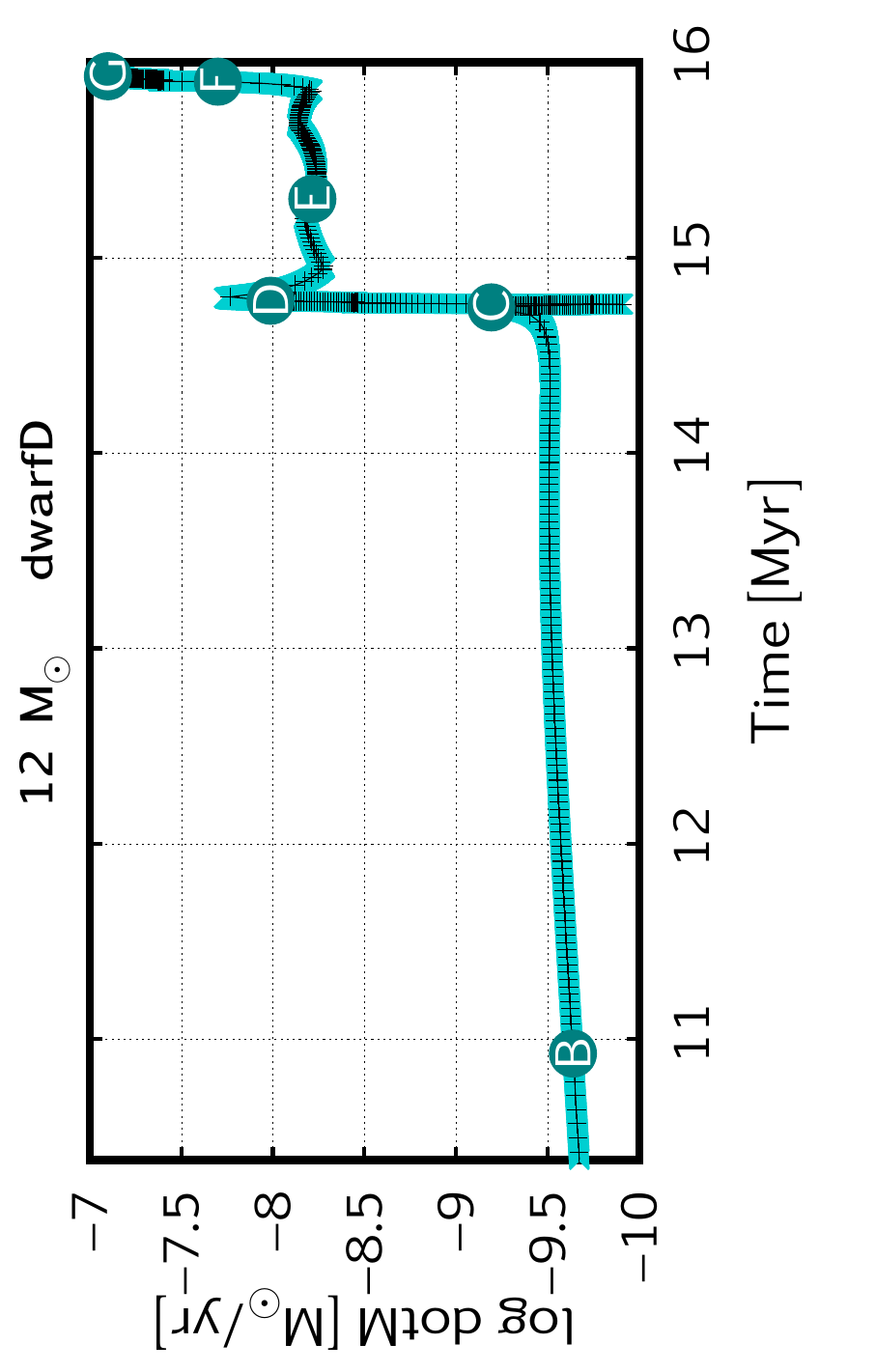}       \hspace{2pt}
                \includegraphics[width=\ratiooo\columnwidth,page=8,angle=270]{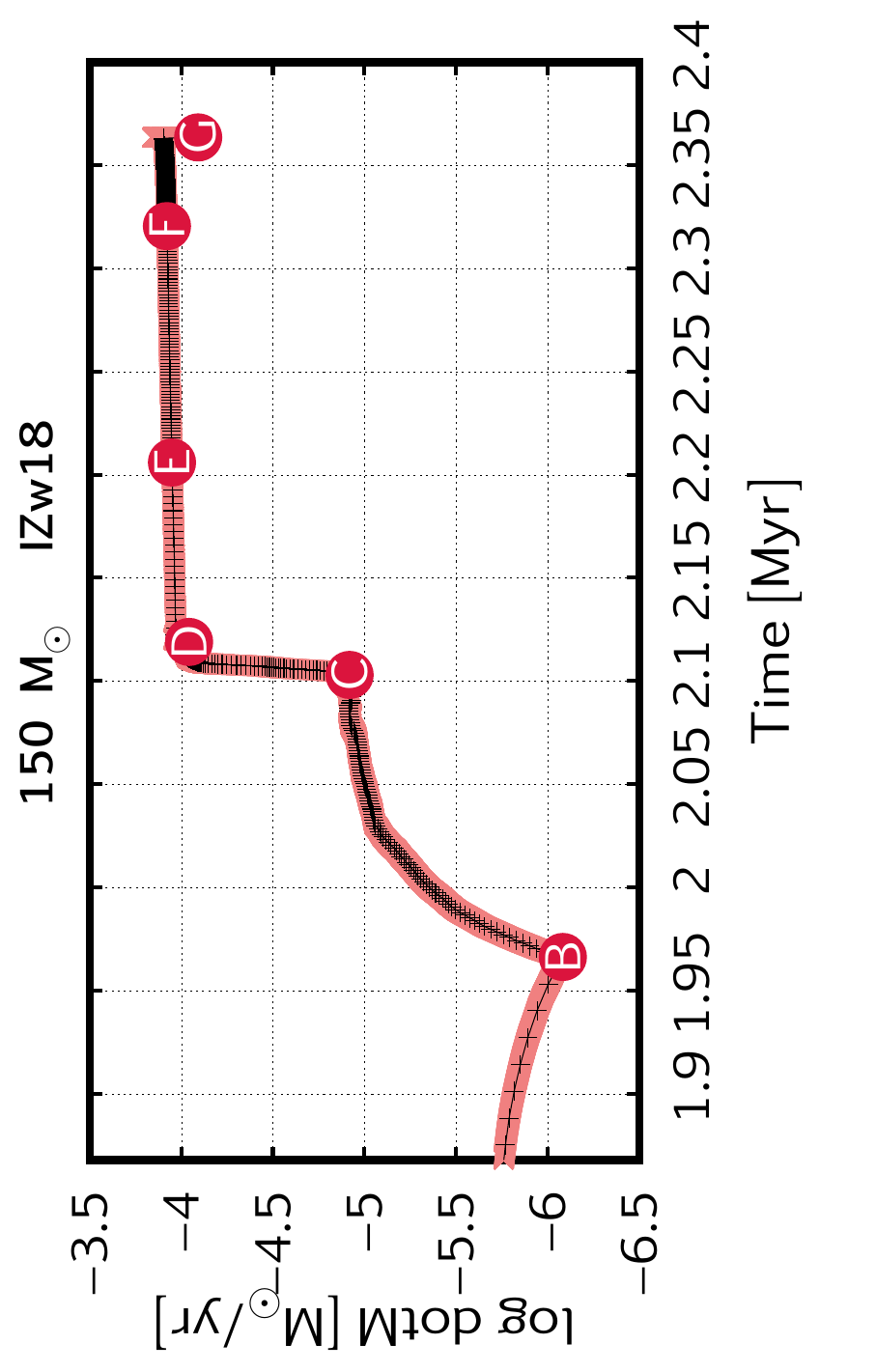}\vspace{-5pt}\\ 
                \caption{Same as Fig.~\ref{fig:fixpoints1} but for a model with prominent blue loop (without a starting point in the red supergiant branch, \textit{left}) and one with inflated envelope (\textit{rigth}). For more details on the latter, which is a core-hydrogen-burning supergiant, we refer to e.g. \citet{Sanyal:2015,Szecsi:2015,Sanyal:2017,Szecsi:2018,Szecsi:2019}}\label{fig:fixpoints2}
        \end{figure*}
        
        \begin{figure*}\centering
                \includegraphics[width=\ratiooo\columnwidth,page=6,angle=270]{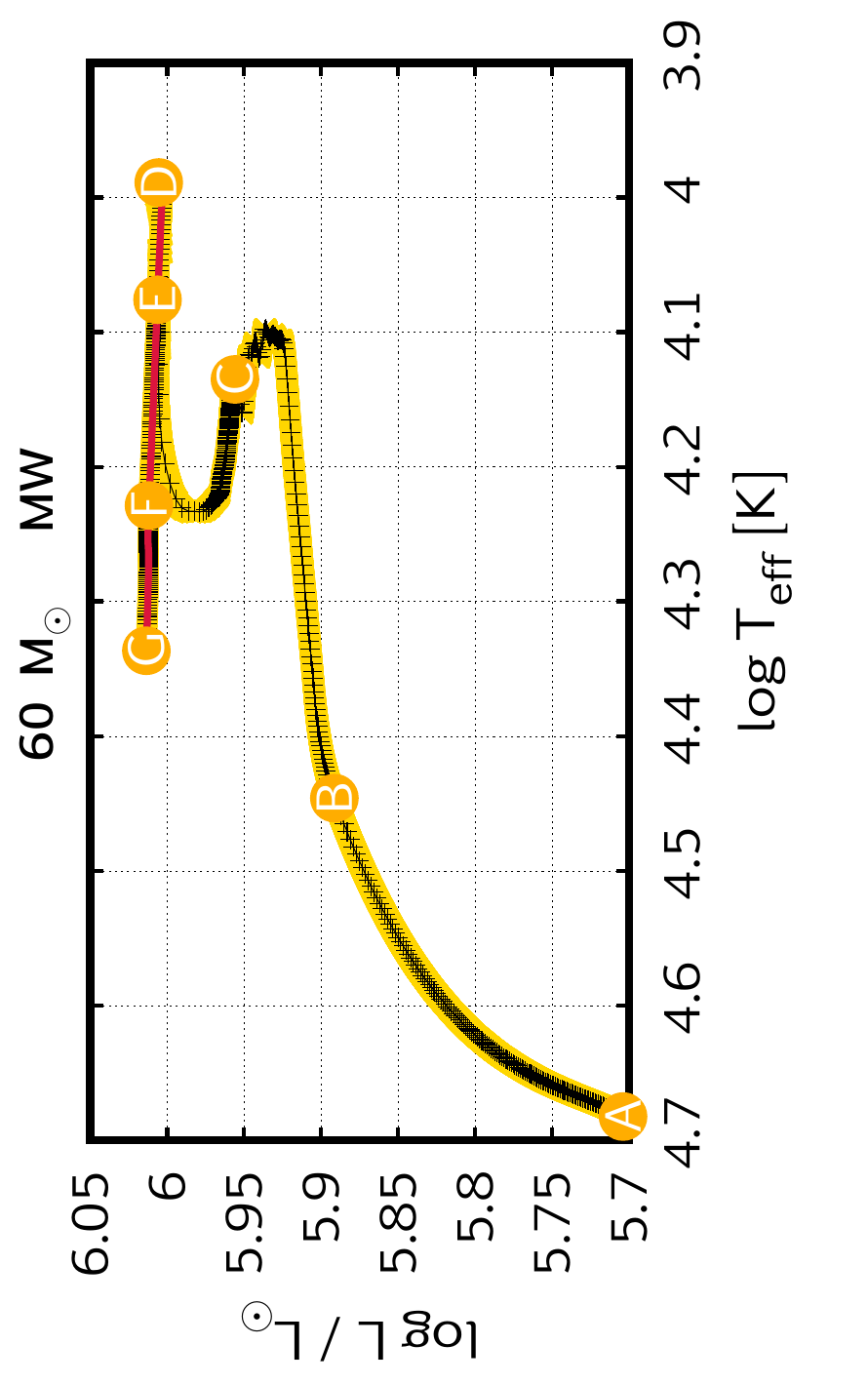}\hspace{2pt}
                \includegraphics[width=\ratiooo\columnwidth,page=10,angle=270]{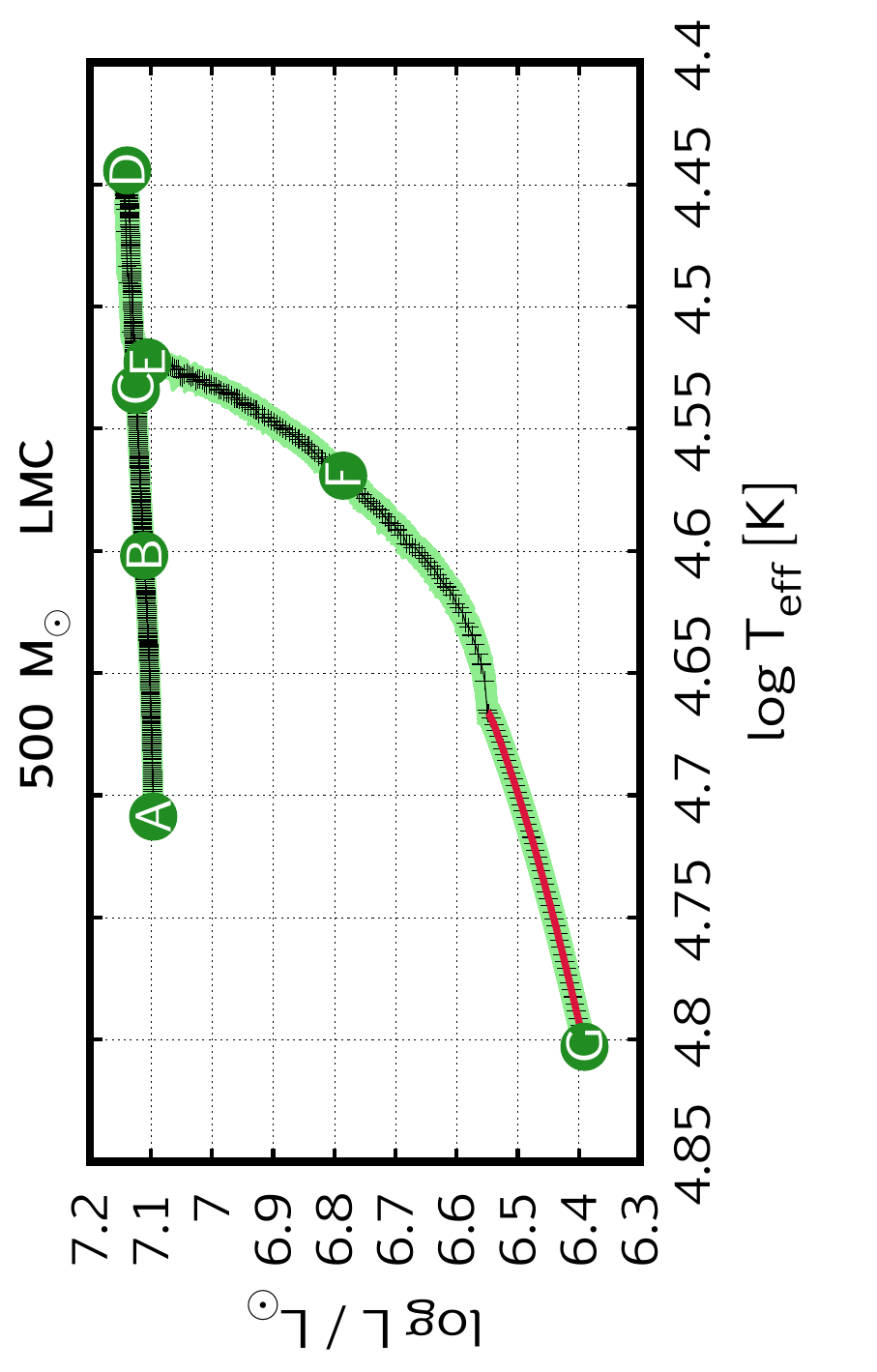}\vspace{-5pt}     \\
                \includegraphics[width=\ratiooo\columnwidth,page=6,angle=270]{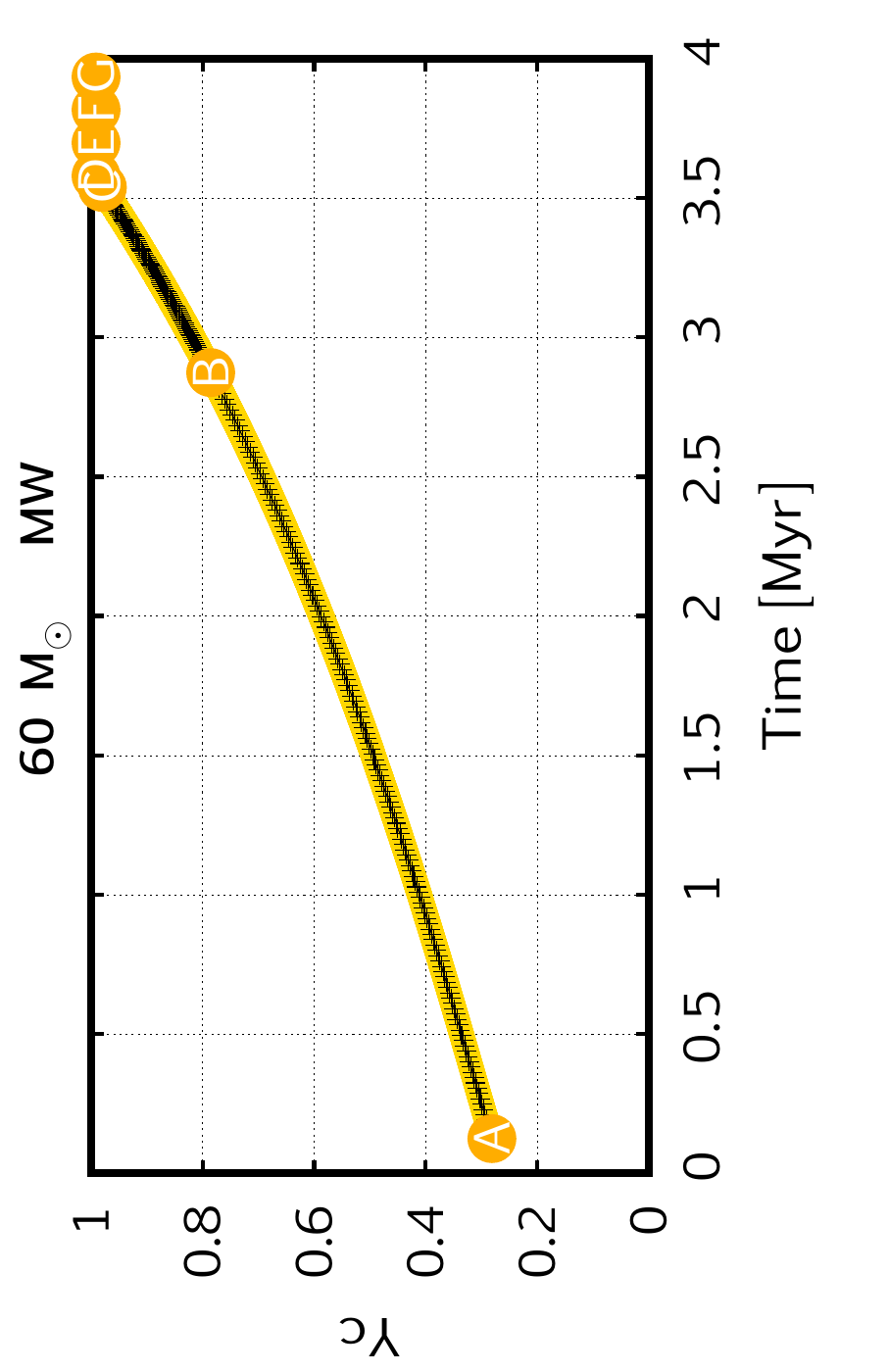}    \hspace{2pt}
                \includegraphics[width=\ratiooo\columnwidth,page=10,angle=270]{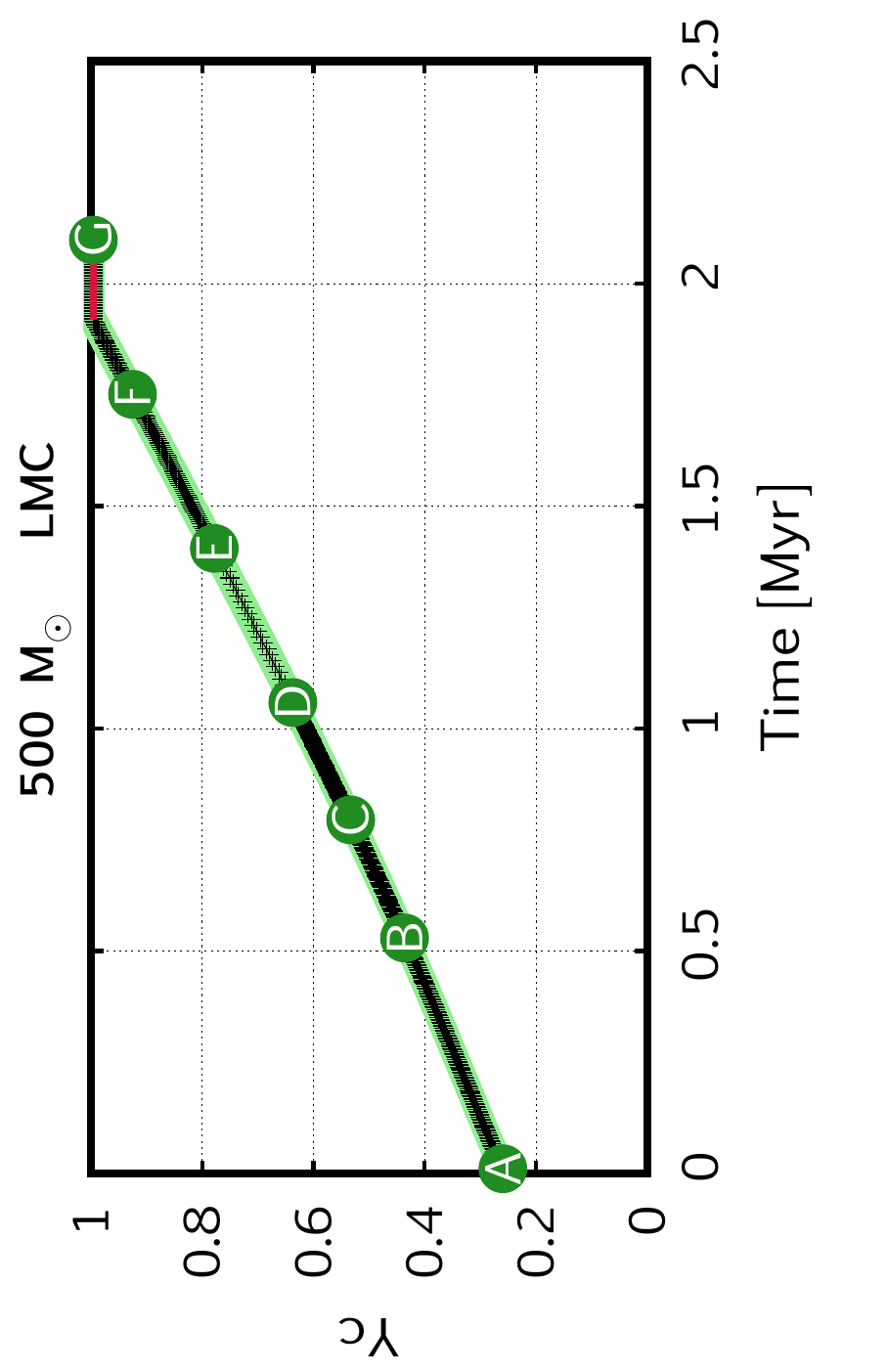}\vspace{-5pt}\\
                \includegraphics[width=\ratiooo\columnwidth,page=6,angle=270]{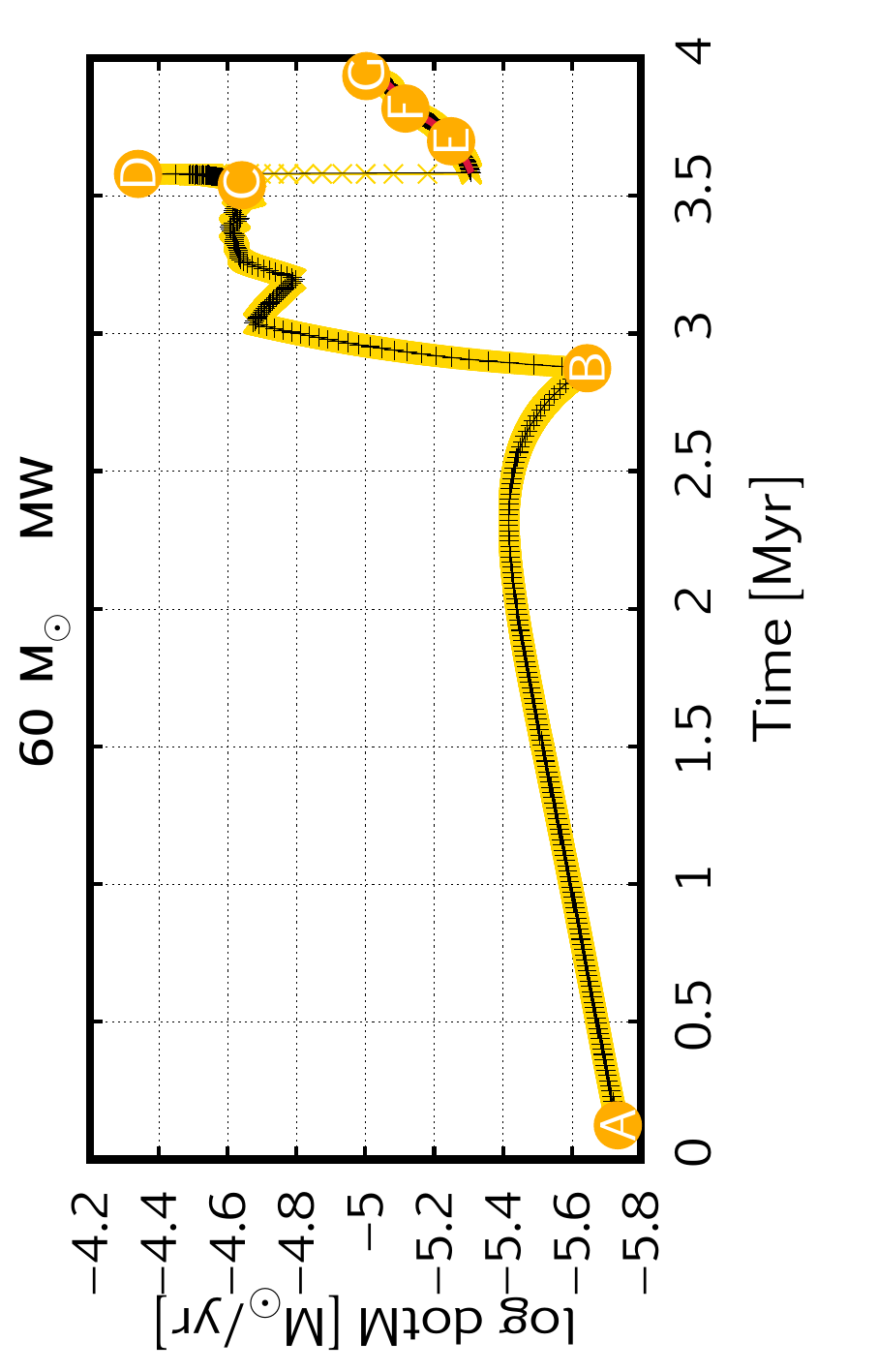}   \hspace{2pt}
                \includegraphics[width=\ratiooo\columnwidth,page=10,angle=270]{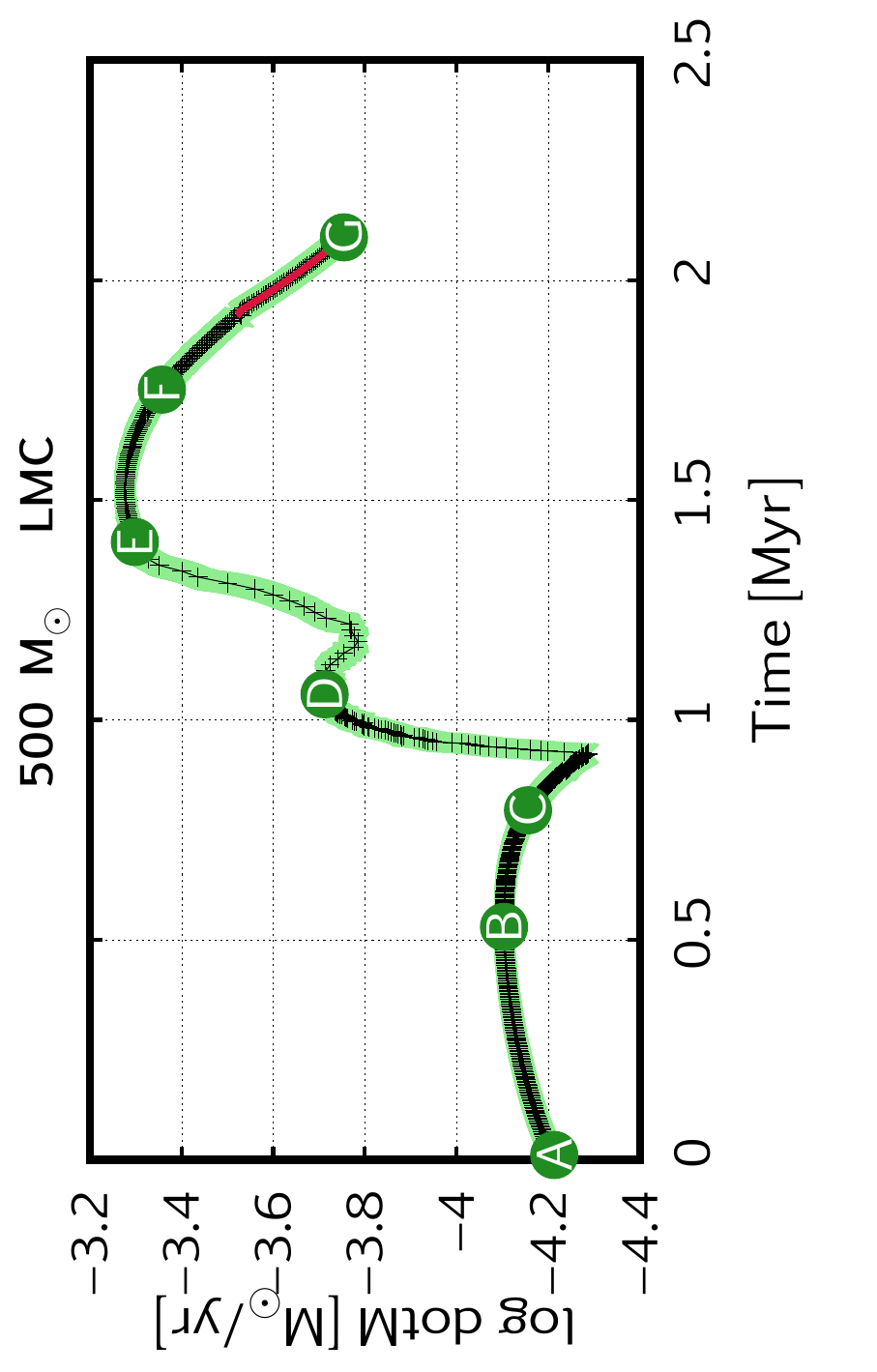}\vspace{-6pt}\\  
                \caption{Same as Fig.~\ref{fig:fixpoints1} but for models that show features associated with a luminous blue variable phase before becoming Wolf-Rayet stars. In both cases, we use our newly developed method for a direct extension in the post-main-sequence phase (that is, removing mass layers one by one from the last computed model's envelope and correcting for the values of the surface properties, as explained in Sect.~\ref{sec:extrapol}). Red line marks this phase of stellar life.}\label{fig:fixpoints3}
        \end{figure*}
        
        The first EEP (A) is the zero-age main sequence (excluding the initial hook phase caused by hydrogen ignition). The second EEP (B) is the local minimum of the mass-loss rate corresponding to the bistability jump,  that is, when it occurs during the main sequence. If the local minimum of the mass-loss rate occurs after the main sequence ends, we chose the second EEP simply to be at $\text{about}$~three-fourths of the main-sequence lifetime. Assigning our second EEP to the local minimum of the mass-loss rate ensures that the interpolation behaves nicely at quick changes in mass loss, which is an important requirement when these models are to be applied to studying their feedback on star formation.
        
        The third EEP (C) is the tip of the hook at the end of the main sequence. However, for very massive stars, this hook is not visible; in this case, we simply chose a point close to core hydrogen-exhaustion. 
        The fourth EEP (D) is the bottom of the red supergiant branch where the luminosity has a local minimum. If this was not visible, for instane, in the case of an extreme blue loop without a base at the red supergiant branch, a point in the middle of the loop blueward progression was chosen. Some examples are shown in Fig.~\ref{fig:fixpoints2}.
        
        The fifth EEP (E) is the middle of the helium-burning phase. It corresponds to the hottest point of the blue loop if there is one, otherwise, Y$_{\rm C}$~$\sim$~0.5. The sixth EEP (F) is chosen near the end of core helium-burning when the model has a small dip in luminosity. 
        If this was not visible, we chose a point at around Y$_{\rm C}$~$\sim$~0.1. The seventh EEP (G) is the end of the helium-burning phase where Y$_{\rm C}$~$=$~0.0, and we did not include carbon burning in our models.

        For the most massive models at high metallicities, we chose the EEPs to be equally distributed in time during the luminous blue variable phase because these stars have no systematically recognizable surface features except for their extremely strong winds. Some examples are shown in Fig.~\ref{fig:fixpoints3}. 
        
        The precomputed stellar evolutionary models were filtered so that the published files are in EEP format.
        To choose the number of points between each EEP, we followed the convention established by \citet{Choi:2016}. Our EEPs occur at lines 1, 151, 252, 403, 429, 505, and 608 for each stellar track. We did not include the pre-main-sequence phase of stars in our models: all models start their evolution at the zero-age main sequence. Consequently, BoOST models have 200 fewer lines than those in \citet{Choi:2016}. The simulation of the pre-main-sequence phase for these very massive stars is hardly physical because a clear picture of how they form in reality is still lacking. Thus, following a pre-main-sequence path that imitates the behavior of low-mass stars on the pre-main-sequence is not more realistic than starting out with a homogeneous zero-age main-sequence model and evolving the star from there on, as we did.

        \subsection{Interpolation with \textsc{synStars}}\label{sec:interpolation}
        
        The precomputed and EEP-filtered stellar models were read in by \textsc{synStars} for each model characterized by its initial mass, $M_0$. Several additional quantities were calculated by \textsc{synStars}, namely, the velocity of the stellar wind, $v_\mathrm{wind}$, was determined with the procedure suggested by \citet{Lamers:1999} and \citet{Vink:2001} following from the theory of line-driven winds,
        \begin{equation}
                v_\mathrm{wind} = \left\{
                \begin{array}{lll}
                        1.3v_\mathrm{esc} & \mathrm{for} & T_\mathrm{eff} < 21\;\mathrm{kK}\\ 
                        2.6v_\mathrm{esc} & \mathrm{for} & T_\mathrm{eff} > 21\;\mathrm{kK}
                \end{array}
                \right.\label{eq:wind}
        ,\end{equation}
        where $v_\mathrm{esc} = (2GM/R)^{(1/2)}$ is the escape velocity from the stellar surface and $G$ is the constant of gravity. Additionally, following \citet{Leitherer:1992}, for instance, the wind velocity was corrected for the metallicity of the wind material, $Z$, by multiplying it by a factor $(Z/Z_\odot)^{0.13}$. Furthermore, the wind power was given by $L_\mathrm{mech} = \dot{M} v_\mathrm{wind}^2/2$.
        
        Interpolated tracks with $M_0$ between the EEP-filtered models were then calculated with {\sc synStars}. The interpolation was performed separately for the stellar age, $t$, and for all the other quantities, using the {\sc SciPy} function {\sc InterpolatedUnivariateSpline} implementing the spline interpolation method of a given order. The stellar age, $t(M_0)$, was interpolated in the $log(t) - log(M_0)$ space using splines of the second order by default. This default can be changed by the user; however, we found while testing various choices that the first-order interpolation can lead to unphysical discontinuities in quantities integrated over the stellar population  \citep[cf.][who documented a similar effect]{Cervino:2001}. All the other quantities, $Q(M_0)$, were also interpolated in the $log(Q) - log(M_0)$ space, but the default order for them is $1$, that is, the interpolation is linear, to avoid errors due to overshooting for quantities that change abruptly (e.g., abundances).
}

The interpolated tracks computed with \textsc{synStars} are presented in Fig.~\ref{fig:HRD} in the HR-diagram, and in Figs.~\ref{fig:Mdot}--\ref{fig:Edot} in terms of mass-loss rate, wind velocity, and kinetic energy deposition rate of the wind. As these quantities are typically needed to study stellar feedback in star formation, we especially ensured that they did not include any numerical artifacts. 

We used only the initial mass as the basis for the interpolation.
In some of the grids we used stellar models with 560 or 575~M$_{\odot}$ as the highest mass (cf.~Sect.~\ref{sec:design}), but still took the upper limit M$_{\mathrm{top}}$~$=$~500~M$_{\odot}$ for the interpolated set of tracks to be consistent.

\begin{figure}[!t]\centering
        \includegraphics[width=\ratioA\columnwidth,angle=0]{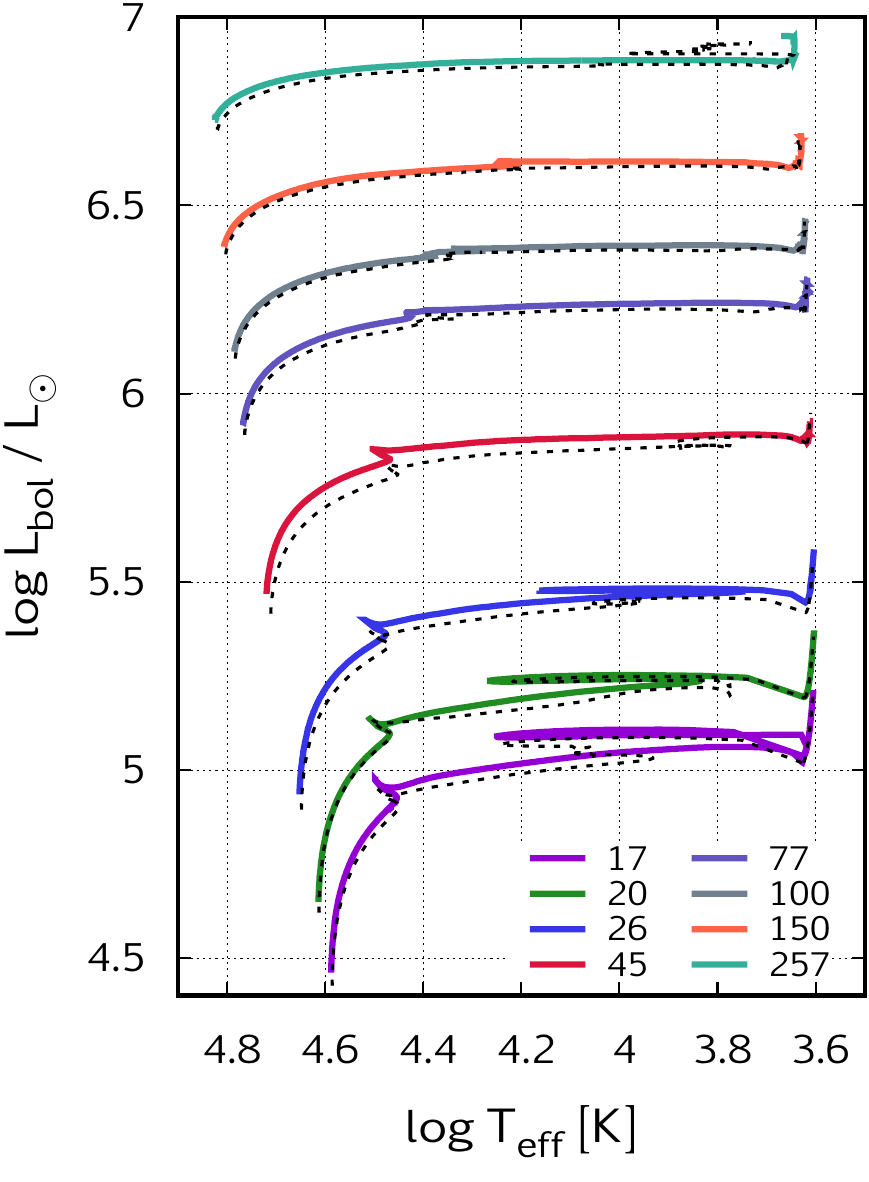}
        \includegraphics[width=\ratioB\columnwidth,angle=0]{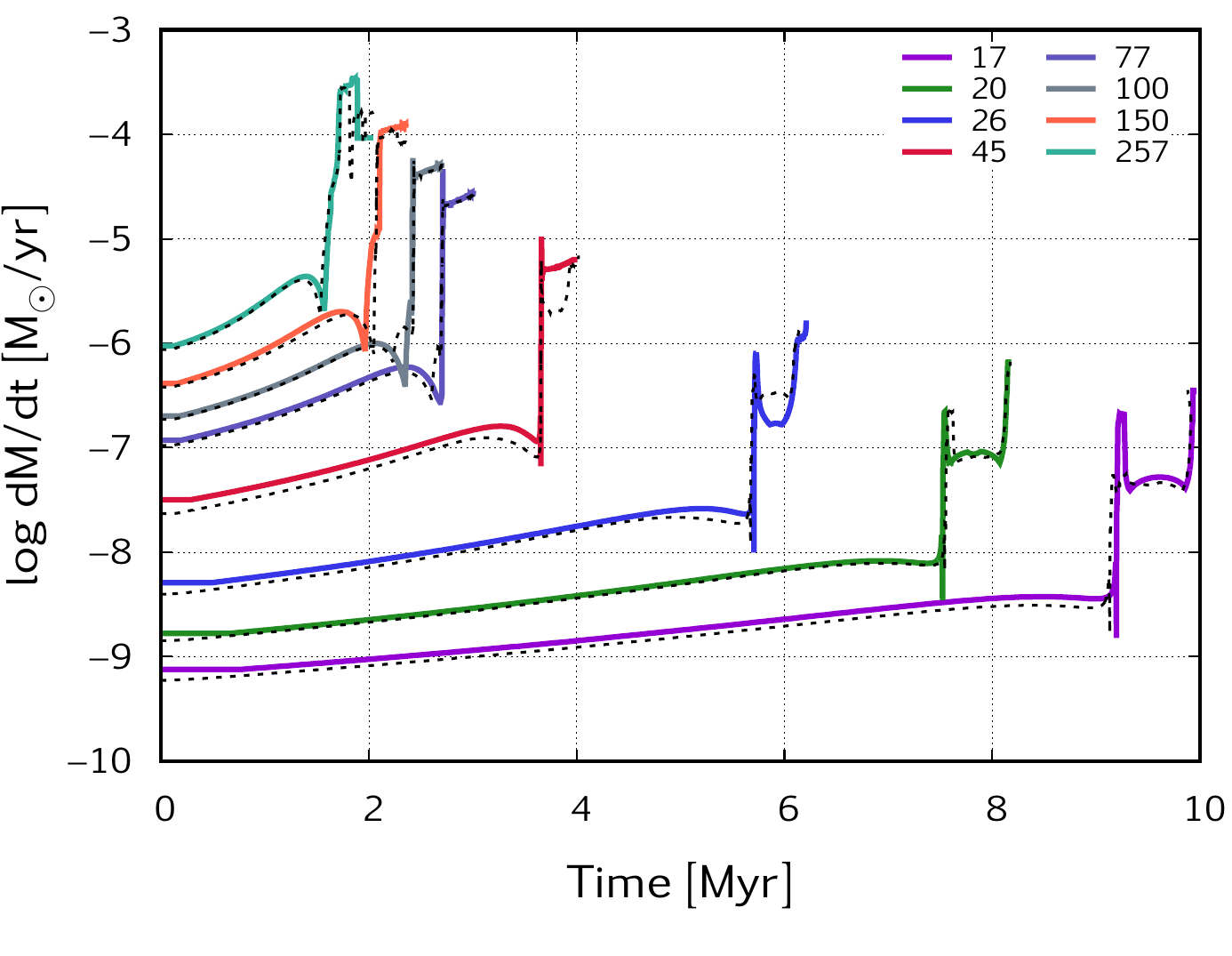}
        \caption{
                Validating the interpolation. Colored lines represent the original stellar models (IZw18 grid), dotted black lines represent the interpolated tracks created by removing the corresponding original while performing these test interpolations. See also Sect.~\ref{sec:interpolation}.
        }\label{fig:interp}
\end{figure}

We tested the validity of our interpolation method by comparing interpolated tracks and their corresponding original models in Fig.~\ref{fig:interp}. The largest difference in terms of surface temperature is 0.47\,dex, and in terms of luminosity, the difference is 0.08\,dex, meaning that the interpolated tracks match the original models well within the error margins of massive star evolution.

We publish the interpolated (synthetic) tracks as one large data table per grid. These files contain 1856 records between 9~M$_{\odot}$ and 498.4~M$_{\odot}$, equally distributed in $log(M_0)$, all having 608 lines. (Figs.~\ref{fig:HRD} and ~\ref{fig:Mdot}--\ref{fig:Edot} only show every 50th interpolated track for clarity.) Thus, the size of this data file is about 800\,MB. Tracks are marked by their initial mass values before their record starts (in M$_{\odot}$ and in cgs units). The following columns are provided (cf. the Readme file next to the published tables): initial mass, time, actual mass, wind velocity, kinetic energy generation rate of the wind, luminosity, radius, effective temperature, mask, type of interpolation, surface rotational velocity, critical rotational velocity, Eddington factor, flag marking whether the phase includes the direct extension method of Sect.~\ref{sec:extrapol} or not, helium-core mass, carbon-oxygen core mass, and surface mass fraction of the same 34 isotopes as in the original stellar models.

\subsection{Synthetic populations}\label{sec:popsyn}

Synthetic populations of single massive and very massive stars are also computed with {\sc synStars} based on the interpolated tracks. The population data we publish here represent massive star clusters with a total mass of 10$^7$~M$_{\odot}$ that were formed as the result of a single, burst-like star formation event in which the initial mass function follows a classical Salpeter distribution (with an upper mass limit of 500~M$_{\odot}$). The evolution of these populations is presented in the data files with equal time-steps up to $\sim$25~Myr, that is, when the life of the longer-living star of our models (the star with 9~M$_{\odot}$) ends. After this, such a cluster will only contain stars below 9~M$_{\odot}$, the contribution of which to feedback processes can typically be neglected. In addition, the compact object remnants of the dead massive stars are also expected to still be within the cluster (see Sect.~\ref{sec:core}, and we note that remnant types are not specifically listed in the published tables, only final core masses are).

The current version of {\sc synStars} is attached to the published data. Thus if needed, the user can feed the precomputed stellar grids to it and create their own interpolated tracks or their own synthetic populations. For example, the interpolated tracks and populations computed here have an upper mass of 500~M$_{\odot}$, but the user may need populations with an upper mass limit that is lower than this. Thus they can create their own synthetic population with 150 or 250~M$_{\odot}$ as an upper mass, for example, and even change the index of the mass function and the total mass of the stellar cluster. (For the highest mass achievable with \textsc{synStars}, see the precomputed models with the highest mass in the grids, Sect.~\ref{sec:design}.) A typical run of \textsc{synStars} creating thousands of tracks takes a few minutes on a normal workstation.

Alternatively, the user may wish to use their own population synthesis tool. This is also a possibility because one of the outputs of {\sc synStars} is the set of interpolated stellar tracks, and their resolution (bin size) can be defined simply as a command line input. These interpolated tracks can then be fed into any population synthesis code, thus providing great flexibility for the user.

In the synthetic population data files created by {\sc synStars} (not the same as the interpolated track data tables), the following quantities are given as a function of time (also see the Readme file attached to the tables): mass lost from stars in the form of fast stellar wind (i.e., faster than 100~km~s$^{-1}$), mass lost from stars in the form of slow stellar wind (i.e., slower than 100~km~s$^{-1}$), kinetic energy in the fast winds, integrated bolometric luminosity, integrated UV flux \citep[including corrections for optically thick stellar winds, following the method presented in][\textbf{cf., Chapt.\,4.5.1}]{Szecsi:2016}, time-integrated values of $\dot{\rm M}$ and $\dot{\rm E}$, that is,~the total mass and mechanical energy produced (in the form of winds and their power) by the all stars up to a given time, as well as the mass fraction of all 34 isotopes in the wind (i.e., integrated over the population). 

\subsection{{Red supergiant luminosity histogram}}\label{sec:lumhisto}

\begin{figure}[!b]
        \includegraphics[width=\ratioA\columnwidth,angle=0]{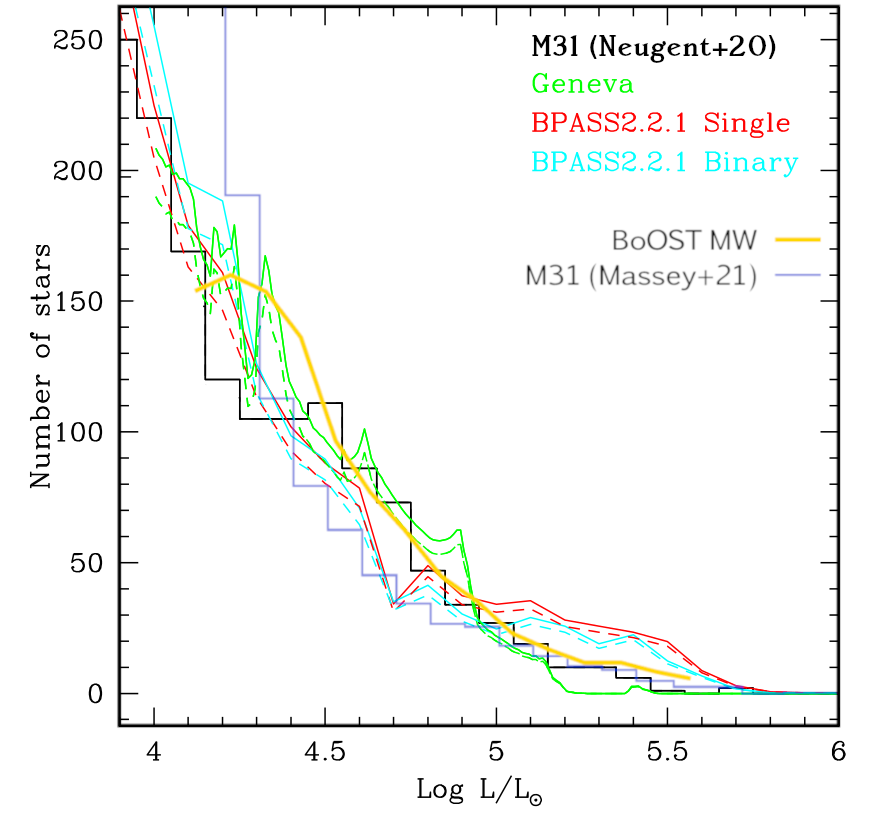}
        \caption{
                {Luminosity histogram of our BoOST models with an MW composition (yellow). The original is taken from \citet{Neugent:2020}, including M31 observations (black) and theoretical predictions from the Geneva (green) and BPASS stellar models (red and blue) for solar metallicity. 
                An even larger and more recent dataset of M31 observations from \citet{Massey:2021} is overplotted (violet). Our MW population follows the observed luminosity histograms as closely as the Geneva and BPASS stellar models. See Sect.~\ref{sec:lumhisto} for further details.}
        }\label{fig:lumhisto}
\end{figure}

{To test our stellar populations against observations, we created a luminosity histogram as in \citet{Neugent:2020}. Our results are shown in Fig.~\ref{fig:lumhisto} for an MW composition, to be compared with two sets of super-solar M31 observations \citep{Neugent:2020,Massey:2021}. Following the method presented in \citet[][see their Fig.~11c]{Neugent:2020}, we normalized the number of red supergiant stars in our theoretical sample to 1000. We defined red supergiants as log~T$_{\rm eff}$~$<$~3.7 (Neugent 2021, private communication). }

{As Fig.~\ref{fig:lumhisto} attests, our models reproduce the observations of red supergiant luminosities as closely as other published sets of stellar models. Because we worked only with massive stars, our stellar models have a lower mass limit at 9~M$_{\odot}$, meaning that our data do not contain any assymptotic giant branch stars. Therefore our theoretical histogram does not reach below L~$\sim$~4.2. } 

\section{Numerically problematic late phases of stellar evolution}\label{sec:extrapol}

\subsection{Role of the Eddington limit}\label{sec:eddington}

{The maximum luminosity that can be transported by radiation
while maintaining hydrostatic equilibrium is called the Eddington luminosity \citep{Eddington:1926}.
However, in the envelopes
of massive stars, where the density is low, changes in the elemental opacities can cause the \textit{local} radiative luminosity to exceed this Eddington luminosity \citep{Langer:1997,Sanyal:2015}. To maintain hydrostatic equilibrium, density and pressure inversions develop in such envelopes. 
In the absence of efficient convection (which is also typical for the low-density envelopes of massive stars, \citealt{Grassitelli:2016}), this can lead to numerical difficulties in 1D stellar evolution codes \citep{Paxton:2013}, meaning in practice that the
time-steps become exceedingly small, preventing further computation of the model.
While less massive and less metal-rich stars are only affected by this issue in their late evolutionary phases, more massive and high-metallicity stars (at $\gtrsim$\,40\,M$_{\odot}$ for solar composition) can exceed the Eddington limit inside their envelopes already during the core hydrogen-burning phase \citep{Graefener:2012,Sanyal:2015}.}

{Stellar evolution codes employ pragmatic solutions to avoid or overcome the above-mentioned numerical difficulties \citep{Agrawal:2021a}. For example, in the PARSEC stellar models, an artificial limit is set for the temperature gradient \citep[see Sec. 2.4 of][as well as \citealt{Alongi:1993}]{Cheng:2015}, which ensures that the density gradient never becomes negative and thus inefficient convection is prevented. Additionally, a mass-loss enhancement following \citep{Vink:2011} is applied whenever the \textit{total} luminosity of the star approaches the Eddington luminosity.} 
{On the other hand, in the MIST stellar models (\citealt{Choi:2016}) computed with the MESA code (\citealt{Paxton:2013}), density inversions are suppressed through the MLT++ formalism: the actual temperature gradient is artificially reduced to make it closer to the adiabatic temperature gradient whenever the radiative luminosity exceeds the Eddington luminosity above a predefined threshold. Radiative pressure at the surface of the star is also enhanced. This approach again increases the convective efficiency, helping the stars to overcome density inversions.} 
{Yet another pragmatic solution is employed in the GENEVA models \citep[see Sec. 2.3 of][]{Ekstroem:2012}: the mixing length is set to be comparable with the density scale
height, which helps avoid density inversions \citep{Nishida:1967,Maeder:1987}, while the mass-loss rates are increased by a factor of three whenever the local luminosity in any of the layers of the envelope is higher than five times the local Eddington luminosity.} 

{Most of the stellar models in this work are not effected by the above-mentioned numerical difficulties, and are thus computed with the Bonn code without interruption until their core helium  is exhausted. However, the very massive high-metallicity models do develop density inversion regions due to their proximity to the Eddington limit. Instead of artificially avoiding or surpassing these density inversions, we employed another solution.}

\subsection{Another solution: Inflated envelopes and the direct extension method}

{If the density and pressure inversions are not avoided in some pragmatic way (as above), the envelope grows \citep[or inflates, ][]{Graefener:2012,Sanyal:2015,Sanyal:2017}, and the star may become a core hydrogen-burning cool supergiant \citep[cf. Sect. 5 of][]{Szecsi:2015}. These cool supergiants might explain globular cluster formation \citep{Szecsi:2018,Szecsi:2019}. Because we wish our models to be applicable in this field of research, and because for these masses, the available observational constraints cannot exclude the existence of stars with such inflated envelopes, we continue the approach of the previously published Bonn models, and let the envelope of all our stars inflate as well.
This makes our BoOST models special amongst other stellar evolutionary models, but it comes at a cost.
}

{In some models (those with very high mass and high metallicity), the time-steps become exceedingly small and the computation is halted before the end of core helium-burning is reached. These models need to be further treated in postprocessing (we call this the direct extension method, see below),} to make them ready to still serve as a proxy for the chemical yields and radiation of the remaining lifetimes. {In terms of stellar populations, the missing phases comprise less than 3\% of the stellar lifetimes because the Eddington-limit proximity only influences the most massive models, which live shorter lives. Moreover, the lower the metallicity, the less relevant this effect: for example, in our lowest metallicity grid called dwarfE, all the models, even the inflated ones, are properly computed until the end.} In Figure~\ref{fig:HRD}, crosses mark the models that were postprocessed with the direct extension method, and the tables in Appendix~\ref{sec:flagcount} provide some quantitative summary. 
{Eight of the 90 models have evolved past the end of the main sequence, but stopped short of completing core helium-burning, and 27 of the models stopped at the end of the main sequence. These are therefore the models for which we developed the method below. The remaining 58 models are complete.}

\subsection{The method}\label{sec:method}

A {1D} stellar evolutionary model sequence consists of consecutive stellar models (structure models, {or profiles}) belonging to a certain age. One such structure model consists of about~1500 grid points (layers) between the core and the surface. For every layer, physical variables such as local temperature and density are computed in the code, including the chemical composition of the layer. {When the computation was halted due to the above-mentioned numerical difficulties, we postprocessed the data by} removing mass from the surface layer by layer from the last computed structure model. This approach allows us to predict the composition of the material ejected by the stellar winds even during the phases for which the code did not converge. 

We continued to remove layers until the projected lifetime of the star ended. 
The projected lifetime was estimated as follows. 
If the model had already burned away at least 2\% of helium when the simulation stopped, the remaining lifetime was calculated by linearly extrapolating the central helium abundance as a function of time until it reached zero. If the model had not burned that much helium, but stopped before this (e.g., at the terminal-age main sequence), the remaining lifetime was simply defined as 10\% of the main-sequence lifetime. {In five cases (cf. Table~\ref{sec:flagcount})} was the terminal-age main sequence not reached: here first we established the projected main-sequence lifetime by quadratically extrapolating the central helium mass fraction as a function of time, and then again assumed that the post-main sequence lasts for 10\% as long as the main sequence. We ensured that the whole process, including the quadratic extrapolation, provided a good estimation for the projected lifetimes by testing it on existing models. For stars that lose much mass, however, the process may only provide a lower limit because the lifetime of stars is inversely proportional to the actual mass.

{\textsf{Mass-loss rates.} During the direct extension phase, the same mass-loss rate prescriptions were applied as in the stellar evolution code. Namely, \citet{Nieuwenhuijzen:1990} rates were applied for supergiants with T$_{\rm eff}$~$<$~22500 and X$_{\rm surf}$~$>$0.45, and \citet{Hamann:1995} scaled by ten (which is consistent with \citealt{Nugis:2000} and is representative of Wolf--Rayet stars) otherwise. In both cases, a metallicity dependence of $\sim$Z$^{0.86}$ was included, following \citet{Vink:2001}.}

\begin{figure}[!t]
        \includegraphics[height=\ratioA\columnwidth,angle=270]{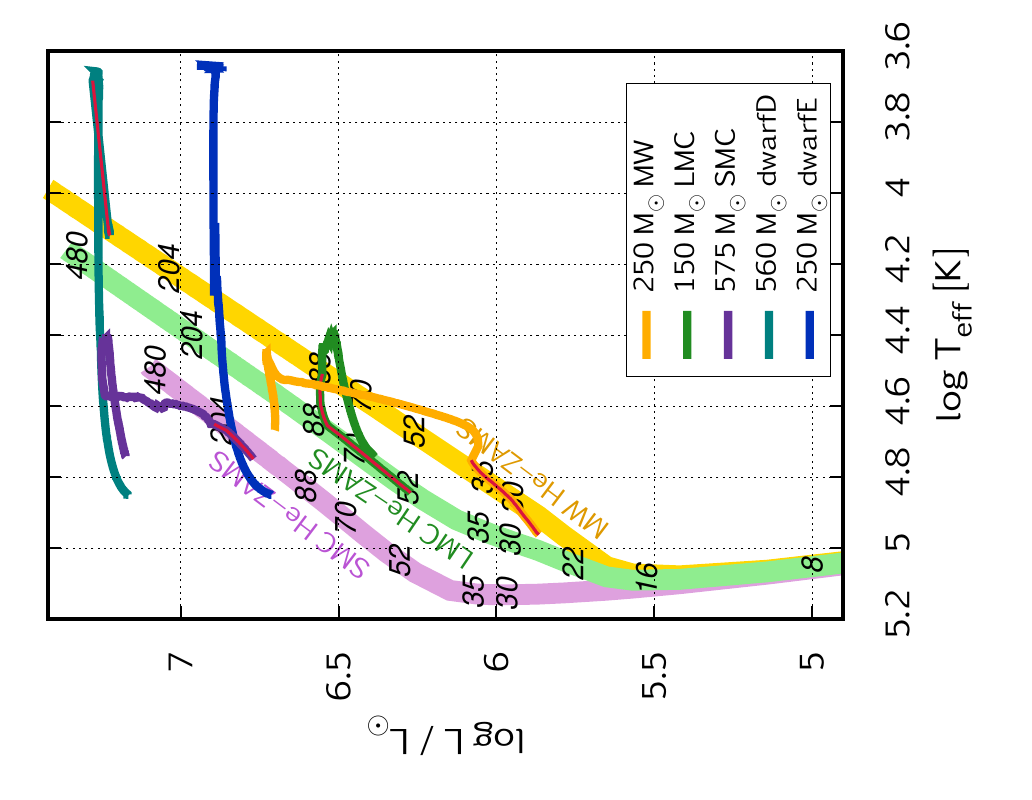}
        \caption{
                Example of the results of the direct extension method (Sect.~\ref{sec:extrapol}). The complete evolution of five stellar models (see the key legend) is plotted. The direct extension phase (typically lasting for 9\% of the lifetime of these particular models) is marked with red in four of them (MW, LMC, SMC, and dwarfD); the fifth model (dwarfE) was properly computed without interruption, and is shown here as reference. During the direct extension phase, the models converge to their corresponding helium ZAMS position while loosing mass from their surface layers (cf. Sect.~\ref{sec:method}).
                Helium ZAMS positions in the HR diagram are shown for MW, LMC, and SMC compositions. The details of constructing these lines are given in Sect.~\ref{sec:HeZAMS}. Numbering indicates stellar masses (in M$_{\odot}$). 
        }\label{fig:HeZAMS}
\end{figure}

\textsf{HR~diagram and radius. \ } If mass layers are removed from a stellar model, this is expected to change the stellar structure and thus influence surface temperature and luminosity,  and hence not only the radius of the star, but also the total ionizing radiation, which is important for stellar feedback predictions. We accounted for this by causing these quantities to transition towards the so-called \textit{helium zero-age main sequence} (helium ZAMS, cf. Sect.~\ref{sec:HeZAMS}) in the HR~diagram. To achieve smoothness, we took the surface temperature and the surface luminosity to be log-linear functions of the surface helium mass fraction. The stellar radius was calculated from these according to the Stefan–Boltzmann law.
When a layer was removed, the helium mass fraction of the next layer was used as a weight to find the new surface temperature and luminosity values between the old ones and those of the helium ZAMS given the new total mass of the star. In this way, the stellar models in the HR diagram converged smoothly and directly towards the helium ZAMS while losing mass from the surface layers. Examples are shown in Fig.~\ref{fig:fixpoints3} and Fig.~\ref{fig:HeZAMS}. 

\subsection{Caveats}\label{sec:caveats}

Numerical difficulties in stellar modeling are a known issue \citep{Aarseth:2008}; {we have summarized some of the solutions offered by modelers in Sect.~\ref{sec:eddington}. The common feature of all these solutions is that they are pragmatic: \textbf{technical maneuvers} usually need to be employed for models to evolve without interruption. In the absence of a homogeneous sample of massive stars \textbf{observed through} various metallicities and various evolutionary phases, based on which 1D stellar evolutionary models could be constrained, \textbf{none of the solutions can be established as preferable over the others}.}
{Our solution, in which the inflated envelope develops and the remaining life is continued by carefully postprocessing the models when numerical
issues are encountered, is just one possible way to solve the common issue in stellar evolution modeling pragmatically. The inflated-envelope phase has special astrophysical applications \citep{Moriya:2015,Szecsi:2015,Szecsi:2018,Szecsi:2019} that make our BoOST models especially useful in globular cluster research, for example.} 

While our direct extension method is quite robust and provides an acceptable approximation for the late phases of very massive stars, 
there necessarily are some caveats. 
Removing layers from the last computed structure model involves the assumption that no more mixing occurs during the remaining lifetime. This may not be true.
The evolution of stars near the Eddington limit is unclear, and the projected lifetimes ignore remaining changes in the model until the end of helium burning.
As no nuclear reactions are simulated for this phase, the central quantities in our output files are simply kept the same as in the last simulated structure model. We add a flag to all our published tables to indicate when this is the case. All these caveats should be kept in mind when using BoOST populations. 

Helium stars are expected to evolve away from the helium ZAMS during their nuclear burning lifetime. In the absence of properly computed helium star evolutionary models in a sufficiently wide mass range (as explained in Sect.~\ref{sec:HeZAMS}), we approximate the post-main-sequence position of our helium stars simply by the helium ZAMS position. This is still within the error bars of massive star physics. In other words, the fact that the mass-loss rates of these stars so unconstrained, for example, introduces a larger uncertainty in \textit{any} grid of stellar models than our simple treatment of the direct extension towards the helium ZAMS.

We emphasize that the direct extension method did not need to be applied for most of the published data ({in terms of stellar lifetimes}, $>$97\% of the published data are properly computed with the code; also cf. Fig.~\ref{fig:HRD}, where crosses indicate when the method is applied). 
Even at high metallicities, only the latest phases of the tracks are influenced; and at low metallicies, we hardly had to apply it at all, as shown by the tables in Appendix~\ref{sec:flagcount}. The reason for this is that envelope inflation starts at higher masses when the metallicity is low \citep{Sanyal:2017}. 
In our lowest-metallicity grid with 1/250~Z$_{\mathrm{MW}}$, even the highest-mass star of 560~M$_{\odot}$ is computed with the Bonn code without any numerical issues until the end of core-helium exhaustion, making the grid called dwarfE perfectly complete; {and for applications in which the upper mass in the population is chosen to be 150~M$_{\odot}$, the IZw18 grid can be used without problem as well because up to this mass, it is not influenced by the DEM treatment.}

\section{Final-fate predictions}\label{sec:core}

\subsection{Final core-mass {estimates}}\label{sec:coremass}

The core mass of a pre-supernova star before the collapse (defined either as the helium-rich or the carbon-oxygen-rich central region) {is often used as} an estimated upper limit for the mass of the compact remnant \citep{Belczynski:2002,Belczynski:2008,Fryer:2012}. {While our modeling stops before carbon burning begins, we provide a proxy for these pre-supernova core-mass values based on the status of the model at the end of core-helium burning.} Our models develop a carbon- and oxygen-rich core in this phase already because helium mainly burns into carbon and oxygen. We define the He core mass to be the mass coordinate where the mass fraction of everything but helium drops below 12\%, and the CO core mass where the mass fraction of everything but carbon and oxygen drops below 12\%. 

During the remaining evolution, the inner regions of the core would undergo nuclear processing, {and the core mass would \textbf{change somewhat:} for a naked helium star because mass is lost in the hot wind, \textbf{and} for a hydrogen-rich star because shell helium-burning replenishes the core with more carbon and oxygen.} However, because core carbon-burning and the subsequent burning phases last for about 1\% of the life of a massive star, {these changes are expected to be minor enough for an order-of-magnitude estimate.}
{For example, \citealt{Chieffi:2013} reported that in their Solar metallicity models (computed with the FRANEC code, see their Table~1), stars in the mass range of 13$-$40~M$_{\odot}$ change their CO core mass during the post-helium-burning phases by merely 1$-$4\% in terms of the initial mass; and stars in the mass range of 60$-$120~M$_{\odot}$ do so typically by 3$-$8\%.}
({The situation is further complicated by the technical difficulty of defining core masses throughout the life of a star;} we refer to Sect.~3.3 of \citealt{Kruckow:2018} for further discussions and references.)
{We therefore suggest that the pre-carbon-burning core mass values we provide for our models can serve as an order-of-magnitude estimate for the pre-supernova values.}

In the case of the models for which the direct extension method was applied (Sect.~\ref{sec:extrapol}), the definitions above do not always hold because the core composition may not yet have reached the required amount of carbon and oxygen. Therefore, we define the CO core mass for these models as 0.8~times the He core mass. We chose this value because we found that the CO cores of models that are not treated with the direct extension method are about 0.7--0.8~times as massive as their corresponding He cores.

\subsection{Assigning supernova types to stellar models}\label{sec:supernova}

Associating supernova types with stellar models is a complex task, hence simplifying assumptions are often made. In star formation studies, for example, when the feedback from supernova explosions is included, it is commonly assumed that all massive stars explode as core-collapse supernovae \citep[e.g.,][]{Gatto:2017}. Moreover, the kinetic energy of the explosion is sometimes simply taken uniformly to be 10$^{51}$~erg.
We suggest caution with this approach, especially when our very massive BoOST models are applied, for the following reasons.

In the case of very massive CO cores, stellar models are known to undergo pair-creation related instability during their oxygen-burning phases \citep{Burbidge:1957,Langer:1991,Heger:2003,Langer:2007,Kozyreva:2014}. This is in fact what would occur in our very massive models as well if their simulations were continued after the helium burning. Thus, simply associating such a stellar model with a core-collapse supernova explosion is inaccurate.
Instead, very massive models are expected to undergo one of the following final fates (relying on the work of \citealp{Heger:2002,Woosley:2007} and \citealp{Chatzopoulos:2012}).

For core masses higher than 130~M$_{\odot}$, the star will directly collapse into a black hole without a supernova explosion. 

For core masses between 65--130~M$_{\odot}$, the model will explode as a pair-instability supernova. This will completely disrupt the whole star, leaving no remnant. The brightness of such a supernova depends strongly on the amount of nickel that is synthesized \citep{Herzig:1990,Dessart:2013}, but according to the analysis of \citet{Kasen:2011}, some of these supernovae should be observable out to large distances. The total explosive energy in such a stellar model is about 10$^{51}$--10$^{52}$~erg.

For core masses between 40--65~M$_{\odot}$, the model will undergo pair instability but should not explode in a pair-instability supernova explosion. Such a model may be associated with large pulsations leading to mass ejection and flashes of emitted light, which is called a pulsational pair-instability supernova \citep{Woosley:2007,Sukhbold:2016,Stevenson:2019}. However, these models will continue their evolution until an iron core forms, and will then explode as a core-collapse supernova. 

For core masses below 40~M$_{\odot}$, iron core collapse is expected, which may lead to a regular core-collapse supernova. 

Since research on core-collapse supernovae, pair-instability supernovae, and pulsational pair-instability supernovae is currently ongoing, we suggest that the user investigate the related literature for further developments before assigning supernova types to our models. No core-collapse type supernova explosions should be assigned to models with cores above 130~M$_{\odot}$ in any case: these certainly do not explode, as explained above. We also suggest to refrain from assigning remnant masses to models with cores between 65-130~M$_{\odot}$: these do explode, but leave no remnant. 

The situation for our chemically homogeneously evolving grid IZw18-CHE may become even more complicated due to the fast rotation of these models. Models with core masses of $\sim$12--30~M$_{\odot}$ may be progenitors of long-duration gamma-ray bursts in either the collapsar or the magnetar scenario, as explained in \citet{Szecsi:2017long}; also cf. Chapter~4.7 of~\citet{Szecsi:2016}. For the pair-instability processes in them, we refer to \citet{AguileraDena:2018}.

\section{Discussion}\label{sec:discussion}

\subsection{Comparison to other  synthetic populations based on the Bonn code}\label{sec:Bonnpop}

The four Bonn grids that existed before the BoOST project (MW, LMC, SMC, and IZw18) have already been applied in population synthesis codes. Examples include \textit{Bonnsai} \citep[][in which the ages of observed massive stars can be established]{Schneider:2014b} and \textit{ComBinE} \citep[][in which predictions of gravitational wave event rates are made]{Kruckow:2018}. While the basic method of creating a synthetic population out of a stellar grid is the same in these two projects and in ours, the details tend to depend on many things. For one, the mass range of the models depends on the actual scientific question: \textit{Bonnsai} and \textit{ComBinE} worked with stars only up to 100~M$_{\odot}$. 
We intend the BoOST models for {astrophysical applications such as} star formation studies in clusters and galaxies in which the contribution from very massive stars may be important, therefore we extended the grids to 500~M$_{\odot}$. In order to study the early universe (e.g., high-redshift galaxies and the origin of globular clusters), a good metallicity coverage including sub-IZw18 metallicities is achieved here. 

\ftextbf{BoOST populations also differ from earlier population synthesis studies in that the models and the interpolated tracks are optimized for smoothness and consistency. For example, in \textit{Bonnsai,} only the main-sequence phases of the massive single-star models were included,} while for star formation studies, the main-sequence phase and the post-main-sequence phase should be included (i.e., the role of massive supergiants and pure helium stars may be relevant). 

\subsection{Future updates}\label{sec:future}

We foresee possible future updates of BoOST models as follows. The first possible update concerns stellar rotation. In the present work we uniformly set all rotational velocities to a 100~km~s$^{-1}$ initial value; however, massive stars rotate with various rotational rates that can to some extent change the predictions in terms of chemical yields, radiation, final core mass, etc. While the 100~km~s$^{-1}$ value we use here is typical for massive stars \citep[as observed in the MW and the Magellanic Clouds, cf., e.g.,][]{Hunter:2008,Dufton:2013,RamirezAgudelo:2013,RamirezAgudelo:2017}, a possible future update of BoOST may be carried out by including stellar models with various rotational velocities. 

Similarly, the initial composition of the BoOST grids could be refined. While the current version of the grids covers a broad range in metallicities from Galactic down to 1/250~lower, we provide only eight metallicity values. This constitutes quite a discrete binning. One possible future endeavor is to simulate grids with a much better resolution in metallicity than this, for instance, by either computing stellar models with metallicities in between or performing interpolation between the grids published here. 

The same is true for the post-helium-burning phases of our models: while they were omitted in the present version for consistency (and for not being relevant for stellar feedback related applications), a possible update may include these phases.

We certainly plan to provide in an updated version of BoOST the chemical yields retained in the stellar envelope (to be released by a stripping of the envelope due to binary interaction). This would allow combining binary population synthesis studies with star formation studies in a powerful way. 

Finally, we encourage future studies in the direction of solving the convergence issues in the inflated envelope near the Eddington limit in a reliable and physically consistent way in the Bonn code and in other stellar evolution codes such as MESA (cf. \citealt{Agrawal:2021b}). While our method of direct extension for the phases in which the models are numerically unstable is quite robust and produces an acceptable result, it is of course not free of caveats (as discussed in Sect.~\ref{sec:caveats}). Therefore, when stellar evolutionary models become available in which these inflated phases are reliably computed, we will update our interpolated tracks and synthetic populations accordingly.

\section{Conclusions}\label{sec:summary}

The BoOST project covers the mass-metallicity parameter space with an unprecedented resolution for the first time. We presented nine grids of massive stars between Galactic and 1/250~lower metallicities, including interpolated tracks and synthetic populations. 
They are available \href{http://boost.asu.cas.cz/}{under this link} as simple tables. The stellar models were computed with the Bonn evolutionary code and were post-processed with methods optimized for massive and very massive stars. Interpolated tracks and synthetic populations were created by our newly developed stellar population synthesis code \textsc{synStars}. Eight~of the grids represent slowly rotating massive stars with normal or classical evolution, while one grid represents fast-rotating, chemically homogeneously evolving models. In addition to the common stellar parameters such as mass, radius, surface temperature, luminosity, and mass-loss rate, we present stellar wind properties such as the estimated wind velocity and kinetic energy of the wind. Additionally, we provide chemical yields of 34 isotopes, and the mass of the core at the end of the stellar lifetimes. 

The BoOST models (grids, tracks, and populations) are thus suitable for further scientific applications, for example, in simulations of star formation in various environments. 
Future updates are planned in terms of adding models with various rotational rates, and with various initial compositions (i.e., a better resolution in metallicity). Post-helium-burning phases will also be added in future work, as will chemical yields retained in the envelope.

In the future we plan to apply the BoOST grids to study the formation and early evolution of globular clusters and young massive clusters in a metallicity-dependent way \citep[following][]{Wunsch:2017,Szecsi:2019}. Beyond this, however, BoOST models open the door for testing the effect of stellar metallicity in many astrophysical contexts in a simple and straightforward way. By optimizing the models for an easy application by the user, our BoOST project harvests the full scientific potential of the Bonn stellar evolutionary code and will contribute to a new era of studying massive stars and their roles in various fields of astrophysics. 

\section*{Acknowledgment}
This research was funded in part by the National Science Center (NCN), Poland under grant number OPUS 2021/41/B/ST9/00757. For the purpose of Open Access, the author has applied a CC-BY public copyright license to any Author Accepted Manuscript (AAM) version arising from this submission.
D.Sz.~was supported by the Alexander von Humboldt Foundation. 
P.A. acknowledges the support from  the  Australian  Research  Council  Centre of  Excellence  for  Gravitational  Wave  Discovery  (OzGrav), through project number CE170100004. 
R.W. acknowledges the support from project 19-15008S of the Czech Science Foundation and institutional project RVO:67985815. 
The authors thank I.~Brott, V.~Brugaletta, M.~Kruckow, I.~Mandel, {K.~Neugent,} D.~Sanyal, F.~Schneider, H.~Stinshoff, A.~Vigna-Gomez, S.~Walch-Gassner {and the anonymous referees} for valuable discussions and kind advice.

\bibliographystyle{aa}
\bibliography{References}


\begin{appendix}
        
\section{Columns in the BoOST data files}\label{sec:units}

BoOST data files are available \href{http://galaxy.asu.cas.cz/pages/boost}{under this link}. The authors welcome feedback from the community, in particular, if there are physical quantities that a next version of our BoOST stellar models and populations should provide in order to serve the community's scientific goals better. Our aim is to provide a flexible model set that can be used in several astrophysical applications.

The current version (v1.3) of BoOST tables contain the columns described below. 

\subsection{Stellar models}\label{sec:appstellar}

All data files contain 608 lines. The $n$th line in one model data file has a comparable evolutionary interpretation in another model data file. Columns are

\begin{enumerate}[1.]
        \item Time [yr]
        \item Actual mass [M$_{\odot}$]
        \item Surface temperature [K]
        \item Surface luminosity [log\,L$_{\odot}$]
        \item Radius [R$_{\odot}$]
        \item Mass-loss rate [log M$_{\odot}$~yr$^{-1}$]
        \item Surface gravity [log cm~s$^{-2}$]
        \item Surface rotational velocity [km~s$^{-1}$]
        \item Critical velocity (assuming an Eddington factor for pure electron scattering) [km~s$^{-1}$]
        \item Eddington $\Gamma_e$ factor calculated for pure electron scattering
        \item Flag marking whether the phase is simulated [0] or approximated with the direct extension method [1] (cf. Sect.~\ref{sec:extrapol})
        \item -- 24. Surface abundances of elements (by summing the abundance of all corresponding isotopes): $\epsilon$(H), $\epsilon$(He), $\epsilon$(Li), $\epsilon$(Be), $\epsilon$(B), $\epsilon$(C), $\epsilon$(N), $\epsilon$(O), $\epsilon$(F), $\epsilon$(Ne), $\epsilon$(Na), $\epsilon$(Mg), $\epsilon$(Al), where $\epsilon$(X)~$=$~N$_{\rm X}$/N$_{\rm H}$+12, and N$_{\rm X}$ is the number fraction of element~X
        \setcounter{enumi}{24}
        \item Helium-core mass [M$_{\odot}$]
        \item Carbon-oxygen core mass  [M$_{\odot}$]
        \item -- 60. Surface mass fraction of isotopes: $^{1}$H, $^{2}$H, $^{3}$He, $^{4}$He, $^{6}$Li, $^{7}$Li, $^{7}$Be, $^{9}$Be, $^{8}$B, $^{10}$B, $^{11}$B, $^{11}$C, $^{12}$C, $^{13}$C, $^{12}$N, $^{14}$N, $^{15}$N, $^{16}$O, $^{17}$O, $^{18}$O, $^{19}$F, $^{20}$Ne, $^{21}$Ne, $^{22}$Ne, $^{23}$Na, $^{24}$Mg, $^{25}$Mg, $^{26}$Mg, $^{26}$Al, $^{27}$Al, $^{28}$Si, $^{29}$Si, $^{30}$Si, and $^{56}$Fe.
        \setcounter{enumi}{60}
        \item -- 94. Core mass fraction of isotopes: $^{1}$H, $^{2}$H, $^{3}$He, $^{4}$He, $^{6}$Li, $^{7}$Li, $^{7}$Be, $^{9}$Be, $^{8}$B, $^{10}$B, $^{11}$B, $^{11}$C, $^{12}$C, $^{13}$C, $^{12}$N, $^{14}$N, $^{15}$N, $^{16}$O, $^{17}$O, $^{18}$O, $^{19}$F, $^{20}$Ne, $^{21}$Ne, $^{22}$Ne, $^{23}$Na, $^{24}$Mg, $^{25}$Mg, $^{26}$Mg, $^{26}$Al, $^{27}$Al, $^{28}$Si, $^{29}$Si, $^{30}$Si, and $^{56}$Fe.
\end{enumerate}

\subsection{Interpolated tracks}\label{sec:apptracks}

The table contains 1856 tracks, and all tracks contain 608 lines. Thus, this data file has about 1.1\,M lines (file size: $ \sim$~800\,MB). Tracks are marked by their initial mass values before their record starts (in M$_{\odot}$ and in cgs units). The following columns are provided:

\begin{enumerate}[1.]
        \item Initial mass [cgs units]
        \item Time [cgs units]
        \item Actual mass [cgs units]
        \item Mass-loss rate [cgs units]
        \item Wind velocity [cgs units]
        \item Kinetic energy generation rate of the wind [cgs units]
        \item Luminosity [cgs units]
        \item Stellar radius [cgs units]
        \item Surface temperature [K]
        \item Mask [integer]
        \item Type of interpolation [integer]
        \item Surface rotational velocity [km~s$^{-1}$]
        \item Critical rotational velocity [km~s$^{-1}$]
        \item Eddington factor (see column~10. in Sect.~\ref{sec:appstellar} above)
        \item Flag marking whether the phase includes the direct extension method [1] or not [0] (cf. Sect.~\ref{sec:extrapol})
        \item Helium-core mass [M$_{\odot}$]
        \item Carbon-oxygen core mass  [M$_{\odot}$]
        \item -- 50. Surface mass fraction of isotopes: $^{1}$H, $^{2}$H, $^{3}$He, $^{4}$He, $^{6}$Li, $^{7}$Li, $^{7}$Be, $^{9}$Be, $^{8}$B, $^{10}$B, $^{11}$B, $^{11}$C, $^{12}$C, $^{13}$C, $^{12}$N, $^{14}$N, $^{15}$N, $^{16}$O, $^{17}$O, $^{18}$O, $^{19}$F, $^{20}$Ne, $^{21}$Ne, $^{22}$Ne, $^{23}$Na, $^{24}$Mg, $^{25}$Mg, $^{26}$Mg, $^{26}$Al, $^{27}$Al, $^{28}$Si, $^{29}$Si, $^{30}$Si, and $^{56}$Fe.
\end{enumerate}

\subsection{Synthetic populations}\label{sec:apppops}

These files contain 608 lines (inherited from the stellar models, see above) plus 50 extra lines to ensure the interpolated quantities behave well. The following columns are provided:

\begin{enumerate}[1.]
        \item Time [cgs units]
        \item Mass lost from all massive stars in the form of stellar wind [cgs units]
        \item Mass lost from all massive stars in the form of dynamical ejection [cgs units] (set to zero for all BoOST models in this paper)
        \item Kinetic energy in the winds [cgs units]
        \item Integrated bolometric luminosity of the population [cgs units]
        \item Integrated UV~flux [cgs units]
        \item Total mass released in stellar winds (accumulated) [cgs units]
        \item Total mechanical energy produced in the stellar winds [cgs units]
        \item -- 56. Mass fraction of isotopes in the winds: $^{1}$H, $^{2}$H, $^{3}$He, $^{4}$He, $^{6}$Li, $^{7}$Li, $^{7}$Be, $^{9}$Be, $^{8}$B, $^{10}$B, $^{11}$B, $^{11}$C, $^{12}$C, $^{13}$C, $^{12}$N, $^{14}$N, $^{15}$N, $^{16}$O, $^{17}$O, $^{18}$O, $^{19}$F, $^{20}$Ne, $^{21}$Ne, $^{22}$Ne, $^{23}$Na, $^{24}$Mg, $^{25}$Mg, $^{26}$Mg, $^{26}$Al, $^{27}$Al, $^{28}$Si, $^{29}$Si, and $^{30}$Si. 
        
\end{enumerate}

        \onecolumn

\section{Technical details of the direct extension method}

\subsection{Establishing the helium zero-age main sequence}\label{sec:HeZAMS}

To be able to apply the direct extension method (Sect.~\ref{sec:extrapol}), we need to establish the point toward which the stellar models converge in terms of surface temperature and luminosity (and thus radius) when they lose mass in their late evolution. This point is the helium ZAMS, that is, the position in the HR~diagram at which homogeneous, thermally adjusted helium stars are predicted to be by simulations.

The position of the helium ZAMS is well known for massive stars, but it is less well known for very massive stars. Model grids were computed with the Bonn code only for MW and SMC metallicity up to masses of 25 and 109~M$_{\odot}$, respectively. They are presented in Fig.~19 of \citet{Koehler:2015}. In this figure, the 35~M$_{\odot}$ point of the MW helium ZAMS was obtained by linearly extrapolating the data above 25~M$_{\odot}$ (D. Sanyal 2019, private communication). No helium ZAMS data for the LMC (and for sub-SMC metallicities) are available.

Some of our very massive models in the MW grid retain as much as 44~M$_{\odot}$ when the simulation stops and those in the SMC grid as much as 170~M$_{\odot}$. Additionally, we have models with LMC and sub-SMC metallicities.
This means that even if we rely on the helium-ZAMS data in Fig.~19 of \citet{Koehler:2015}, we still have to supplement it to cover our parameter space.

The helium-ZAMS models from Fig.~19 of \citet{Koehler:2015} themselves are unpublished but have kindly been provided by D.~Sanyal (2019, private communication). We extrapolated the MW and SMC data in terms of $M$, logT$_{\rm eff}^{(M)}$ , and log(L/L$_{\odot}^{(M)}$) \textbf{up to 500~M$_{\odot}$}. 
We also interpolated between these two data sets using log(Z) as the interpolation parameter to obtain data for the LMC. The results are shown in Fig.~\ref{fig:HeZAMS}. 

When supplementing the MW and SMC data, we applied linear extrapolation and also reduced all quantities by a factor of 3. We chose this number for the following reasons. First, the helium-ZAMS models are initial models with a homogeneous composition, while our models would in reality become helium-ZAMS stars with a relaxed composition (i.e., nuclear burning would be ongoing inside). The surface temperature of such relaxed models is typically somewhat higher, while the surface luminosity is somewhat lower than those of unrelaxed models. Second, extrapolating linearly above 100~M$_{\odot}$ using the original data leads to highly unphysical results. For example, a 200~M$_{\odot}$ helium-star in the LMC grid would have a surface temperature of $\sim$3300~K with an extrapolation like this. This is hardly physically possible because massive stars (including very massive stars) are not expected to have a surface temperature lower than $\sim$4000~K \citep[cf. the Hayashi line,][]{Kippenhahn:1990}. Third and most conclusively, the helium-ZAMS lines we would obtain by a simple linear extrapolation of the original data lie at higher surface temperatures than the lines in the last computed stages of some of our most massive stellar models. During the late phases of evolution, our models should evolve to higher and not lower surface temperatures when they lose mass (and thereby uncover helium-rich layers). For all these reasons and after testing several values, we decided that including a reduction of a factor of 3 in our linearly extrapolation of the helium-ZAMS lines in the HR diagram gives the most physically consistent result. Examples are presented in Fig.~\ref{fig:HeZAMS}.

\subsection{Fraction of the data that includes the direct extension method}\label{sec:flagcount}

 Tables~\ref{tab:fc} provide the percentage of the lifetime influenced by the treatment of direct extension method (DEM) {and the central helium mass fraction Y$_{\rm cen}$ of the last computed structure model} (cf. Sect.~\ref{sec:extrapol}). An asterisk denotes if the value is reached during the main sequence (as opposed to the post-main-sequence, which is the most common case). 
The corresponding lines of the published data files are flagged with [1] in the relevant column (Column~11, cf. Sect.~\ref{sec:units}). 

\small

\begin{table*}
        \caption{Fraction of the data that includes the direct extension method, cf. Sects.~\ref{sec:extrapol} and~\ref{sec:flagcount}
        }\label{tab:fc}

\begin{multicols}{3}

\begin{tabular}{|rcl|}
        \hline                  
\oldtextbf{\textit{MW}} & `DEM' & Y$_{\rm cen}$ \\
        \hline                  
9~M$_{\odot}$ & 0.0\% & 0.00 \\ 
12~M$_{\odot}$ & 0.0\% & 0.00 \\ 
15~M$_{\odot}$ & 0.0\% & 0.00 \\ 
25~M$_{\odot}$ & 1.3\% & 0.17 \\ 
40~M$_{\odot}$ & 8.8\% & 0.99 \\ 
60~M$_{\odot}$ & 8.9\% & 0.99 \\ 
80~M$_{\odot}$ & 9.0\% & 0.99 \\ 
120~M$_{\odot}$ & 9.0\% & 0.99 \\ 
250~M$_{\odot}$ & 9.0\% & 0.99 \\ 
500~M$_{\odot}$ & 9.0\% & 0.99 \\ 
\hline
\end{tabular}

\begin{tabular}{|rcl|}
                \hline                  
\oldtextbf{\textit{LMC}} & `DEM' & Y$_{\rm cen}$ \\
\hline
9~M$_{\odot}$ & 0.0\% & 0.00 \\ 
12~M$_{\odot}$ & 0.0\% & 0.00 \\ 
19~M$_{\odot}$ & 0.0\% & 0.00 \\ 
30~M$_{\odot}$ & 0.0\% & 0.01 \\ 
40~M$_{\odot}$ & 3.0\% & 0.41 \\ 
70~M$_{\odot}$ & 8.6\% & 1.00 \\ 
100~M$_{\odot}$ & 8.5\% & 1.00 \\ 
150~M$_{\odot}$ & 9.7\% & 0.98 \\ 
260~M$_{\odot}$ & 8.6\% & 1.00 \\ 
500~M$_{\odot}$ & 8.5\% & 1.00 \\
\hline
\end{tabular}

\begin{tabular}{|rcl|}
                \hline                  
\oldtextbf{\textit{SMC}} & `DEM' & Y$_{\rm cen}$ \\
\hline                  
9~M$_{\odot}$ & 0.0\% & 0.00 \\ 
12~M$_{\odot}$ & 0.0\% & 0.00 \\ 
15~M$_{\odot}$ & 0.0\% & 0.00 \\ 
30~M$_{\odot}$ & 0.0\% & 0.00 \\ 
40~M$_{\odot}$ & 0.9\% & 0.10 \\ 
55~M$_{\odot}$ & 3.2\% & 0.40 \\ 
100~M$_{\odot}$ & 8.8\% & 0.99 \\ 
150~M$_{\odot}$ & 8.4\% & 1.00 \\ 
250~M$_{\odot}$ & 11.2\% & 0.97$^{*}$ \\ 
575~M$_{\odot}$ & 8.3\% & 1.00 \\ 
\hline
\end{tabular}

\begin{tabular}{|rcl|}
                \hline                  
\oldtextbf{\textit{dwarfA}} & `DEM' & Y$_{\rm cen}$ \\
\hline  
9~M$_{\odot}$ & 0.0\% & 0.00 \\ 
12~M$_{\odot}$ & 0.0\% & 0.00 \\ 
19~M$_{\odot}$ & 0.0\% & 0.00 \\ 
30~M$_{\odot}$ & 0.0\% & 0.00 \\ 
40~M$_{\odot}$ & 0.0\% & 0.00 \\ 
55~M$_{\odot}$ & 0.9\% & 0.07 \\ 
100~M$_{\odot}$ & 8.7\% & 0.99 \\ 
150~M$_{\odot}$ & 8.2\% & 1.00 \\ 
250~M$_{\odot}$ & 9.4\% & 0.99 \\ 
560~M$_{\odot}$ & 15.1\% & 0.92$^{*}$ \\ 
\hline
\end{tabular}

\begin{tabular}{|rcl|}
                \hline                  
\oldtextbf{\textit{dwarfB}} & `DEM' & Y$_{\rm cen}$ \\  
        \hline                  
9~M$_{\odot}$ & 0.0\% & 0.00 \\ 
12~M$_{\odot}$ & 0.0\% & 0.00 \\ 
19~M$_{\odot}$ & 0.0\% & 0.00 \\ 
30~M$_{\odot}$ & 0.0\% & 0.00 \\ 
40~M$_{\odot}$ & 0.0\% & 0.00 \\ 
55~M$_{\odot}$ & 0.0\% & 0.00 \\ 
70~M$_{\odot}$ & 0.0\% & 0.00 \\ 
150~M$_{\odot}$ & 5.0\% & 0.56 \\ 
250~M$_{\odot}$ & 15.3\% & 0.92$^{*}$ \\ 
560~M$_{\odot}$ & 10.4\% & 0.97$^{*}$ \\ 
\hline
\end{tabular}

\begin{tabular}{|rcl|}
                \hline                  
\oldtextbf{\textit{IZw18}} & `DEM' & Y$_{\rm cen}$ \\   
\hline                  
9~M$_{\odot}$ & 0.0\% & 0.00 \\ 
17~M$_{\odot}$ & 0.0\% & 0.00 \\ 
20~M$_{\odot}$ & 0.0\% & 0.00 \\ 
26~M$_{\odot}$ & 0.0\% & 0.00 \\ 
45~M$_{\odot}$ & 0.0\% & 0.00 \\ 
77~M$_{\odot}$ & 0.0\% & 0.00 \\ 
100~M$_{\odot}$ & 0.0\% & 0.00 \\ 
150~M$_{\odot}$ & 0.0\% & 0.00 \\ 
257~M$_{\odot}$ & 6.1\% & 0.68 \\ 
575~M$_{\odot}$ & 10.0\% & 0.98 \\ 
        \hline  
\end{tabular}

\begin{tabular}{|rcl|}
                \hline                  
\oldtextbf{\textit{dwarfD}} & `DEM' & Y$_{\rm cen}$ \\  
        \hline                  
9~M$_{\odot}$ & 0.0\% & 0.00 \\ 
12~M$_{\odot}$ & 0.0\% & 0.00 \\ 
19~M$_{\odot}$ & 0.0\% & 0.00 \\ 
30~M$_{\odot}$ & 0.0\% & 0.00 \\ 
40~M$_{\odot}$ & 0.0\% & 0.00 \\ 
55~M$_{\odot}$ & 0.0\% & 0.00 \\ 
80~M$_{\odot}$ & 0.0\% & 0.00 \\ 
100~M$_{\odot}$ & 0.0\% & 0.00 \\ 
250~M$_{\odot}$ & 0.0\% & 0.00 \\ 
560~M$_{\odot}$ & 18.9\% & 0.88$^{*}$ \\ 
\hline
\end{tabular}

\begin{tabular}{|rcl|}
                \hline                  
\oldtextbf{\textit{dwarfE}} & `DEM' & Y$_{\rm cen}$ \\  
\hline                  
9~M$_{\odot}$ & 0.0\% & 0.00 \\ 
12~M$_{\odot}$ & 0.0\% & 0.00 \\ 
19~M$_{\odot}$ & 0.0\% & 0.00 \\ 
30~M$_{\odot}$ & 0.0\% & 0.00 \\ 
55~M$_{\odot}$ & 0.0\% & 0.00 \\ 
70~M$_{\odot}$ & 0.0\% & 0.00 \\ 
80~M$_{\odot}$ & 0.0\% & 0.00 \\ 
100~M$_{\odot}$ & 0.0\% & 0.00 \\ 
250~M$_{\odot}$ & 0.0\% & 0.00 \\ 
560~M$_{\odot}$ & 0.0\% & 0.00 \\ 
\hline
\end{tabular}

\begin{tabular}{|rcl|}
                \hline                  
        \oldtextbf{\textit{IZw18-CHE}} & `DEM' & Y$_{\rm cen}$ \\       
\hline                  
9~M$_{\odot}$ & 0.0\% & 0.00 \\ 
13~M$_{\odot}$ & 0.0\% & 0.00 \\ 
20~M$_{\odot}$ & 0.0\% & 0.00 \\ 
30~M$_{\odot}$ & 0.0\% & 0.00 \\ 
51~M$_{\odot}$ & 0.0\% & 0.00 \\ 
77~M$_{\odot}$ & 0.0\% & 0.00 \\ 
150~M$_{\odot}$ & 0.0\% & 0.00 \\ 
294~M$_{\odot}$ & 0.0\% & 0.00 \\ 
388~M$_{\odot}$ & 0.0\% & 0.00 \\ 
575~M$_{\odot}$ & 8.0\% & 1.00 \\ 
        \hline  
\end{tabular}
\end{multicols} 

\end{table*}

\end{appendix}

\end{document}